# The Effects of Hypoxia, Metabolic Restriction and Magnetic Fields on Chromosome Instability and Karyotype Contraction in Cancer Cell Lines

A thesis submitted to McGill University in partial fulfillment of the requirements for the degree of Ph.D. in Occupational Health

by

**Ying Li**

January 2012

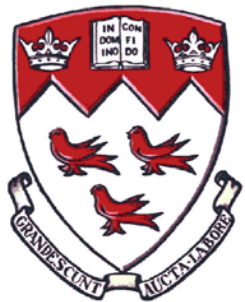

Department of Epidemiology, Biostatistics and Occupational Health
Faculty of Medicine
McGill University
Montreal, Québec, Canada

# ACKNOWLEGEMENTS

It is a pleasure to thank the many people who made this thesis possible. There are many individuals to whom I owe my sincerest and greatest appreciation, as without them this thesis would not have been possible.

I would like to take this opportunity to express my sincere gratitude to my supervisor. I can honestly say that without you, this thesis would not exist. Thank you for being a role model and inspiring me into pursuing an education in the field of Occupational Health. Dr. Héroux runs the lab with his own money, and provided excellent supervision, guidance, constructive criticism and encouragement throughout the period of my PhD. I am deeply touched and inspired by his sincere love, his devotion to science and his independant way of thinking.

I am very grateful to my PhD Comprenhensive and Protocol committee, Drs. Heather Durham, Jim Hanley, and Wayne Wood for their inspiration, support and guidance. I also want to thank Dr. Rebecca Fuhrer for the Principal's Award funding, and the Royal Victoria Hospital's Department of Surgery for providing laboratory space. I am grateful to my former supervisor, Dr. Jim Nicell, for his guidance and inspiration in research.

I also want to thank Janet Moir and Lorne Beckman of the Royal Victoria Hospital for laboratory support, Igor Kyrychenko and Donaldas Guedrikas of the InVitroPlus Laboratory for contributions to the development of the automated systems, and Michel Bourdages, Institut de Recherche d'Hydro-Québec, for equipment contributions.

I would especially like to thank my parents for their love and support. Finally, I dedicate this work to my grandparents who are no longer with us. I miss you very much.

# ABSTRACT

A biological assay based on chromosome counts in human cancer cells was developed as an index of metabolic state. The assay is then used to describe the action of a variety of metabolic agents: oxygen, melatonin, vitamin C, the drugs oligomycin and imatinib, as well as extra-low frequency (ELF) magnetic fields (MFs). This led us to uncover a basic mechanism of interaction between ELF MFs and biological materials. The action of MFs is through an alteration in the structure of water, originally described by Russian physicists at Lomonosov University in Moscow, in the early 1980s.

Our work started as an investigation of the effects of oxygen on cancer cells in culture, later expanded to other metabolic agents. Chromosome counts above 46 are observed in the majority of human tumours. But while real tumours grow in low oxygen and nutrient restricted environments, cultured cancer cells are provided with 21% oxygen and generous nutrition. We studied the chromosome counts of cancer cells as they were metabolically altered, observing that five metabolic restrictors induced chromosome losses in five hyperploid cancer cell lines. These karyotype contractions (KC) allow cancer cells to support fewer chromosomes, increase their proliferation rate and acquire the phenotype of a stable, growing tissue similar to stem cells. Hyperploid cancer cells can expand or contract their karyotypes through rapid mechanisms of endo-reduplication or chromosome loss. These fast meta-genetic mechanisms may explain the surprising adaptability of tumours to changing micro-environments and therapeutic interventions. Furthermore, KC may provide a basis for the previously observed carcinogenic action of some anti-oxidants, positioning metabolic restriction as a meta-genetic mechanism of tumour promotion.

Biological effects of ELF MFs have lacked a credible mechanism of interaction between fields and living material. The karyotype changes produced by 6-day exposures to ELF MFs between 25 nT and 5 µT were evaluated in our five human cancer cell lines. Similar to the chemical metabolic restrictors, all cancer cells lines lost chromosomes from all MF exposures, with a mostly flat dose-response. Continued MF exposures for three weeks allow a rising return to the baseline, unperturbed karyotypes. From this point, small MF increases or decreases are again capable of inducing KCs.

Our data suggests that the KCs are caused by MF interference with mitochondria's ATP synthase (ATPS), compensated by the action of AMP-activated protein kinase (AMPK). The effects of MFs are similar to those of the ATPS inhibitor oligomycin. They are amplified by metformin, an AMPK stimulator, and attenuated by resistin, an AMPK inhibitor. Over environmental MFs, KCs of various cancer cell lines show exceptionally wide and flat dose-responses, except for those of erythro-leukemia cells, which display a progressive rise from 25 nT to 0.4 µT.

These observations lead us to uncover a subtle mechanism of interaction between MFs and human metabolism. MFs cause an alteration in the structure of water that impairs the flux of protons in ATPS hydrophilic channels, with many downstream biological effects. Although the connection between MFs and ATPS inhibition through increased proton impedance is fairly clear, the consequences of typical human MF exposures on AMPK and metabolism should be more complex to unravel. This mechanism may be environmentally important, in view of the central role played in human physiology by ATPS and AMPK, particularly in their links to diabetes, cancer and longevity. Our work provides a defensible mechanism to explain the action of MFs on biological materials.

# RÉSUMÉ

Un test métabolique utilisant des cellules humaines cancéreuses a été développé pour décrire les effets d'une variété d'agents: oxygène, mélatonine, vitamine C, les drogues oligomycine et imatinib, ainsi que les champs magnétiques (CM) de basse fréquence (BF).

Basé sur le comptage de chromosomes (CC) de cellules cancéreuses, il nous amené à découvrir un mécanisme d'interaction entre CM BF et le matériel biologique qui passe par une altération de la structure de l'eau décrite par des physiciens russes au début des années 1980.

Nos recherches ont débuté par une investigation des effets de l'oxygène sur des cellules en culture, subséquemment étendue à d'autres agents. Des CC supérieurs à 46 sont observés dans la majorité des tumeurs humaines. Alors que les tumeurs réelles se développent dans un milieu pauvre en oxygène et nutriments, les cellules cancéreuses en culture sont entourées par un taux d'oxygène de 21% et une nutrition généreuse. Nous rapportons des réductions de CC suite à une activité métabolique restreinte chez cinq types de cellules cancéreuses hyperploïdes. Ces contractions permettent aux cellules de supporter moins de chromosomes, d'augmenter leur prolifération et d'acquérir le phénotype d'un tissu stable en progression, semblable aux cellules souches. Les cellules hyperploïdes peuvent augmenter ou réduire leur karyotype par des mécanismes rapides d'endo-reduplication ou de pertes chromosomiques. Ces mécanismes méta-génétiques rapides pourraient expliquer l'adaptation surprenante des tumeurs aux environnements variables et aux interventions thérapeutiques. La contraction des karyotypes pourrait fournir une base à l'action carcinogénique préalablement observée de certains antioxydants, présentant la restriction métabolique comme un mécanisme méta-génétique de promotion cancéreuse.

Les effets biologiques des CM BF n'avaient pas de mécanisme crédible d'interaction entre champ et matériel vivant jusqu'à maintenant. Cinq lignées de cellules cancéreuses exposées aux CM BF pendant 6 jours dans la plage de 25 nT à 5 µT ont montré des changements de karyotype, tout comme les restricteurs métaboliques, avec une dose-réponse essentiellement plate. Une continuation de l'exposition sur trois semaines permet un retour progressif au karyotype original de base. De ce point, de petites augmentations ou décroissances de CM sont à nouveau capables d'induire des contractions de karyotypes.

Nos croyons que les contractions sont causées par une interférence des CM avec l'ATP synthase (ATPS) des mitochondries, compensée par l'action de la protéine kinase activée par l'AMP (AMPK). Les effets des CM sont similaires à ceux de l'olygomycine, un inhibiteur de l'ATPS. Ils sont amplifiés par la metformine, un stimulateur de l'AMPK, et atténués par la résistine, un inhibiteur de l'AMPK. Sur la plage des CM environnementaux, les contractions de karyotypes de diverses lignées cancéreuses montrent des doses-réponses plates, sauf pour celles des cellules erythro-leucémiques, qui montrent une augmentation progressive de 25 nT à 0.4 µT.

Ces observations nous ont mené à découvrir un mécanisme subtil d'interaction entre CM et le métabolisme humain. Les CM causent une altération de la structure de l'eau qui réduit le flux de protons dans les canaux hydrophiles de l'ATPS. Bien que la connexion entre CM et l'inhibition de l'ATPS par l'augmentation de l'impédance aux protons soit raisonnablement claire, les conséquences des expositions magnétiques humaines typiques sur l'AMPK et le métabolisme sont plus complexes à dégager. Ce mécanisme d'interaction pourrait être important pour l'environnement, en vue du rôle central joué dans la physiologie humaine par l'ATPS et l'AMPK, particulièrement dans leurs liens avec le diabète, le cancer et la longévité. Nos travaux fournissent un mécanisme crédible pour expliquer l'action des CM sur le matériel biologique.

# CONTRIBUTIONS OF AUTHORS

**Li Y, Héroux P, Kyrychenko I (2010). Cytotoxicity Testing with Anoxic K-562. P201-009: In Vitro Testing Methods. IUTOX 2010 (XII International Congress of Toxicology), 19-23 July 2010, Barcelona, Spain.**

Dr. Paul Héroux and I designed this study. I performed the experiments and statistical analyses in the lab. Dr. Igor Kyrychenko helped in the design of the automated systems, the RSF1 medium, and provided biological consultation.

**Héroux P, Li Y (2010). Cytotoxicity Variables in High-Throughput Computer-Vision Studies. P201-074: In Vitro Testing Methods. IUTOX 2010 (XII International Congress of Toxicology), 19-23 July 2010, Barcelona, Spain.**

Dr. Paul Héroux and I designed this study. Dr. Héroux wrote the computer program. I performed the experiments and statistical analyses in the lab.

**Li Y, Héroux P, Kyrychenko I (2012). Metabolic Restriction of Cancer Cells in vitro causes Karyotype Contraction - an indicator of Cancer Promotion? Tumor Biology 33(1), pp. 195-205. doi: 10.1007/s13277-011-0262-6.**

Dr. Paul Héroux and I designed this study. I performed all the experiments and statistical analyses in the lab. Dr. Igor Kyrychenko helped in the design of the automatic system, the RSF1 medium and provided biological consultation.

**Li Y, Héroux P. Extra-Low-Frequency Magnetic Fields alter Cancer Cells through Metabolic Restriction. Submitted to *PLoS ONE*.**

Dr. Paul Héroux and I designed this study. I conducted the experiments and statistical analyses in the lab.

# TABLE of CONTENTS

# List of Figures

Fig. 3.3: *Crest Doubling Frequency* differences between adjacent Oxygen concentrations (0-2 %, 2-5 %, 5-10 %, 10-21 %) in simultaneous tests, based on 8 tests. Bars represent ± 95 % CI of 8 tests. Differences in *Crest Doubling Frequency* between adjacent oxygen concentrations are always positive, confirming that the *Crest Doubling Frequency* varies with oxygen concentration as depicted in Fig. 3.2.

Fig. 3.4: Histogram of object Footprints as a function of Oxygen level (1,430,000 cells/curve). Data averages 3 sets of experiments. Higher oxygen levels stimulate the appearance of larger objects exhibiting macrophage-like size. 50 to 300 μm² are "normal cells", and 300 to 600 μm² are macrophage-like. 0 % oxygen curve has 5.71 % of macrophage-like counts, 2 % has 7.29 %, 5 % has 8.78 % and 21 % has 11.37 %.

Fig. 3.5: Rapid inflammatory response to oxygen in K562 is illustrated by time-series histograms of cell shape quantified as *Roundness*. The anoxic gas phase of a continuously monitored cell culture is replaced by atmoxic gas at time zero, triggering the appearance of the "Roundness 2" peak in this 3D mesh plot. The total cell number climbed from 22,000 to 46,000 from the near to the far edge of the graph.

Fig. 3.6: Apobodies and Necrobodies per Cell (%) in K562 cultures under 0 % and 21 % Oxygen. 0 % oxygen (top curves) shows more apobodies and necrobodies than 21 % oxygen (bottom curves).

Fig. 3.7: Apoptosis rates measured on Oxic and Anoxic transitions in K562 cultures.

Fig. 3.8: K562's enhanced clustering under anoxia, a sign of reduced metastatic potential. *Hex-distance* histograms are compiled at 1, 4, 7, 10, 18, 29, 40, 51, 62, 73 hours. Each curve shows averages ± 1 σ of Hex-*distance* for 5,000 to 32,000 cells. Among the different oxic environments in the legend, cyto-adhesion is strongest for anoxic (red) and anoxic transition (blue) cells, while atmoxic cells are more easily shed.

Fig. 3.9: The anoxic histogram (blue) is averaged from our data and from 3 repeats by *Plate-Forme de Cytogénétique* of Maisonneuve-Rosemont hospital (108 metaphases). The atmoxic histogram (red) is averaged from our data. Similar atmoxic data is published by ATCC.

Fig. 3.10A: G-banding and SKY results from 20 anoxic K562 cells, detected in all but 4 (absence of i(14)(q10): normal chromosome 14). Chromosomal abnormalities are related to the 3n (69 chromosomes) level, according to ISCN 2009: -2, dup(2)(q?): duplicated long arm segment localized on p, -3, -4, +der(5)t(5;6)(q?;p?), dup(6)(p?), +der(6)t(6;20), +inv(7)(p?p?), -8, -9, del(9)(p12), dup(9)(q?): duplicated long arm segment localized on 9p, der(10)t(3;10), +der(10)t(3;10;17), -11, der(12)t(12;21), -13, der(13)t(9;13), -14, der(14)t(2;14), i(14)(q10), -15,der(17)t(9;17)x2, ?del(18)(q?), der(18)t(18;22), der(18)t(3;18), -20, -21, dup(22q?q?), -X, der(X)t(X;8), dup(X)(q?): duplicated long arm segment localized on Xp.
FISH results. BCR (green)-ABL (orange)-ES translocation probe results. The fused green-orange (white spot) reveals the Philadelphia chromosome, detected in 95 % of CMLs. From Plate-Forme de Cytogénétique, Maisonneuve-Rosemont Hospital.

Fig. 3.10B: Naumann et al.'s (2001), M-FISH karyotype of a K562 67-karyotype, found in 15 of 19 cells investigated.

Fig. 3.11: Four days after Atmoxic transition (0 % > 21 %), intermediate karyotypes (63 to 67) bridge the gap between the Anoxic mode of 62 and the Atmoxic mode of 69. 157 metaphases from 3 different tests in our laboratory. 11 % of atmoxic transition karyotypes lie both below and above the graph, but such outliers are almost entirely absent from the anoxic transition.

Fig. 3.12: Output example of karyotype evolution simulation program. EndoReDuplication and asymmetric segregation events are applied with specific probability levels (two dimensional scan) to the anoxic distribution of Fig. 3.9. The simulation result are then compared with the actual 4-day atmoxic transition karyotype of Fig. 3.11.

Fig. 3.13: K562 chromosome counts produced by 60-Hz, 1 µT MFs applied either horizontally or vertically in three 6-day assays. The baseline T-12 culture vessel contains anoxic cells at less than 4 nT (60-Hz), with an average of 61.5 chromosomes (horizontal line), and a very narrow distribution (at left). Box plots show median (solid), average (dotted), 25 and 75 % limits (box), 10 and 90 % limits (whiskers), and outside values (dots). 56 (Assay 1), 50 (Assay 2) and 51 (Assay 3) metaphases karyotyped in each orientation. Inside the box plots are average chromosome losses, with standard deviation. The Student's t-test results quantify the probability that the horizontal and vertical results are identical.

Fig. 3.14: K562 chromosome counts as a function of 60-Hz Magnetic Flux Density. 6-day assays with, in sequence, 65, 28, 50, 77, 46, 33, 65, 102, 56 and 50 metaphases. Approximate ranges for domestic, commercial and occupational exposures are indicated.

Fig. 3.15: Average chromosome losses in erythro-leukemia, breast, lung and colon cancer cells as a function of 60-Hz Magnetic Flux Density. The references ("0") for naïve cells (< 4 nT) are: 66 (HEL), 74 (MCF7), 57 (NCI-H460) and 54 (COLO 320DM) chromosomes. 6-day assays with, in sequence, 32, 22, 29, 32; 19, 22, 19, 21; 29, 22, 24; 22, 34 and 46 metaphases. HEL, NCI-H460 and COL 320DM assays used 21 % oxygen, rather than anoxic conditions, as some anoxic karyotype modes are too close to 46.

Fig. 3.16: K562 chromosome counts return to baseline after 3 weeks of continuous 1 µT MF exposure. 65, 102, 50 and 37 metaphases.

Fig. 3.17: K562 Chromosome counts obtained after 6 days by altering baseline MFs of 0.1 µT and 1 µT. For 0.1 µT, 20, 31, 37 (baseline), 31, 35 metaphases. For 1 µT, 28, 28, 37 (baseline), 28, and 28 metaphases. Although the symmetry of the chromosome counts is strong, there is more cell decay with increased than with reduced fields.

Fig. 3.18: Object Diameter histograms for 6-day anoxic exposures of K562 cultures to 0.4 µT MF at 60-Hz and oligomycin at 2.5 ng/ml. The lower 4 curves are: imatinib (0.04 µg/ml) in blue, resistin (40 ng/ml) in violet, metformin (0.01 mg/ml) in light green and melatonin-vitamin C (0.3 µg/ml, 26 µg/ml) in dark green. Exposed cultures are adjusted to a common small particle count maximum.

Fig. 3.19:  NCI-H460 cell number ratios between initial and 4-day counts under Baseline (<4 nT), 0.05, 0.4 and 5 µT MFs.

# List of Symbols and Abbreviations

| | |
|---|---|
| AMPK | AMP-activated protein kinase |
| ATPS | ATP synthase |
| CIN | Chromosome Instability |
| CIN-END | Chromosome Instability- Endo-Reduplication |
| CML | Chronic Myelogenous Leukemia |
| $D$ | Diffusion of oxygen |
| ELF | Extra-Low-Frequency |
| EMF | Electro-Magnetic Field |
| END | Endo-Reduplication |
| FISH | Fluorescence In Situ Hybridization |
| $IC_{50}$ | Inhibitory Concentration for 50 % of maximum proliferation |
| $J$ | The rate of oxygen consumption of K562 |
| KC | Karyotype Contraction |
| M-FISH | Multiplex Fluorescence In Situ Hybridization |
| MF | Magnetic Field |
| NIM | Nickel-Iron-Molybdenum alloy |
| ROS | Reactive Oxygen Species |
| SKY | Spectral Karyotyping |
| $x$ | Depth within the culture medium |
| $\Phi$ | Oxygen concentration |

# List of Tables

---------------------------------------------------------------------------------------

# 1.0 INTRODUCTION

---------------------------------------------------------------------------------------

# 1.1 Chromosome Instability and Metabolic Restriction of Cancer Cells

## 1.1.1 Chromosome Instability and Karyotype Contraction

Cancer cell lines have disorganized phenotypes and karyotypes, and a pivotal question in oncology has been whether this disorganization is an essential or a peripheral characteristic of cancer (Lengauer et al., 1998). One prominent characteristic of a cancer cell is its chromosome count. Although increased chromosome counts are routinely observed in the majority of tumors, little physiological meaning has been associated with this variable. If human tumor cells are cultured *ex vivo*, chromosome counts may be increased even further, thus displaying a basic cancer cell characteristic labeled chromosome instability (CIN) (Yoon et al., 2002; Paulson et al., 2009).

In the clinic, unstable phenotypes and karyotypes are often observed in the progression of tumors through stages, and many cancer cell lines cultivated *in vitro* also change because of CIN, often into hyperploid karyotypes. CIN represents the ability of cancer cells to spontaneously alter genetically, either in chromosome structure or in chromosome counts. Therefore, *in vivo* as well as *in vitro*, the majority of cancer cells display hyperploid as well as unstable karyotypes. Beyond their state at a point in time, tumors are also thought to evolve relatively slowly, using mechanisms of clonal expansion based on the genetic instability arising from the initial cancer lesion (Nowell, 1976).

Some perspective on the meaning of chromosome count increases can be gained by reviewing the four classes of genetic alterations classically recognized in neoplasms: gene amplification, sequence changes, translocations (such as BCR:ABL in K562), and *chromosome count* *alterations* (Lengauer et al., 1998). These classes testify to a plurality of mechanisms in cancer, but all create hyperplasia, suggesting that these lesions find a common ground in metabolic stimulation (Jones and Thompson, 2009). The term *Karyotype Contraction* was introduced by us to describe the loss of chromosomes by a cancer cell, but with the understanding that the loss is reversible. Contraction does not affect the carcinogenicity of the cell line, and a cell lineage could go through repeated contractions and expansions as the environment of the cells changes.

### 1.1.2   Anoxia and Metabolic Restriction

The development of cancer cells is widely believed to be limited by oxygen and nutrients, occulting their truly critical property, the ability to thrive in diverse sites.  Because of their metabolic flexibility, cancer cells survive both in the highly hypoxic tumor cores and in the blood. However,  very little oncological research has been conducted using hypoxic and metabolically restricted models.

### Anoxia and Hypoxia in Normal Cells

The percentage of oxygen in most mammalian tissues varies widely,  ranging from 1 % to 6 %, which is much lower than the 21 % oxygen conventionally found in the incubators used to culture cells. Even within a single tissue such as the blood, oxygen levels vary, depending on the location within the circulation (Ward, 2008). Oxygen pressure around the cells is lower than in the systemic capillaries, as oxygen is being consumed by the cells. Cellular levels range from 0.66 % to 5.3 %, with an average measured in the tissues of lower animals of 3%. Consequently, cells cultured under 21% oxygen actually are exposed to a hyperoxic environment (Wright and Shay, 2006).

ROS are formed intracellularly in proportion to oxygen concentration (Freeman et al., 1982). Cells maintained under hyperoxic conditions suffer from oxygen toxicity due to increased free radical generation (Halliwell, 1981). Although most of the oxygen consumed by cells is metabolized to water, a small portion is univalently reduced to free radical intermediates (Balin et al., 2002). Free radicals and particularly ROS are deeply influential on cell behavior. Oxidative stress is not only considered a cause of aging and cell death, but is also recognized as affecting molecules relevant to cell signaling pathways and cell development processes (Covarrubias et al., 2008). They underlie such pivotal phenomena as cancer, aging and the toxic effects of metals (Harman, 1956; Pourahmad et al., 2003; Valko et al., 2007). As the concentration of ROS increases, a temporary arrest of growth can occur through paralysis of gene expression, leading to cellular replicative senescence and apoptosis (Davies, 1999). High levels of ROS activate apoptosis, followed at higher levels by type 2 cell death, autophagy (Scherz-Shouval and Elazar, 2007). The highest ROS concentrations cause necrotic cell death (Bras et al., 2005). Because of damage to fatty acids, proteins and DNA, high and sustained production of ROS would prove quickly fatal to unprotected cells.

A corollary of the previous argument on oxygen toxicity is that cells growing under anoxic conditions should provide a sensitive system to assess the impact of toxic agents. And stable *in vitro* cellular models with low ROS levels should allow more accurate determinations in research studies.

## Anoxia and Hypoxia in Cancer cells

The situation is even more extreme in the case of cancer cells. 82 % of oxygen readings taken in solid tumors are less than 0.33 % (Kizaka-Kondoh et al., 2003), and stem cells are hosted in niches that are very low in oxygen (Hill et al., 2009). Cancer cells within cancer patients are therefore exposed to a wide range of oxic conditions, from very hypoxic to richly oxygenated, as cells from cancer niches metastasize through blood vessels to the periphery. Many studies have demonstrated that hypoxia on a variety of cancer cell types alters cell behavior.  The work available in the literature on cancer cell hypoxia typically reaches down to 1 % oxygen, and there are scant studies on true anoxia. Cell survivals and proliferation rates reported by these deep hypoxia studies vary widely (Anderson et al., 1989; Cipolleschi et al., 1993; Cuvier et al., 1997; Papandreou et al., 2005; Papandreou et al., 2005). One interesting study on a wide range of bone marrow samples compared proliferation at 19, 3 and 1 % oxygen, showing that the large majority of samples generated more cells, and increased colony-forming units under deepening hypoxia (Thompson et al., 2007).

The oxygen and nutrient gradients (Jezek et al., 2010; Matsumoto et al., 2008) between the core and cortex of a tumor influence both the phenotype and genotype of cancer cells.  Karyotypes of cancer cells react to reduced oxygen. It is of historical interest to note that in its early history, K562 sported 45-46 chromosomes (Lozzio and Lozzio, 1975). After 3.5 years in *atmoxic* culture (topped by natural atmospheric gas), it acquired the karyotype widely known today as "pseudo-triploid" because its mode of 69 coincides with 3n, although individual chromosomes are variously present as doublets, triplets or quadruplets (Naumann et.al, 2001).

If the investigator is looking to monitor the behavior of cancer cells in the stages of the original process of carcinogenicity, anoxia appears a better simulation of the natural environment of tumors than the 21 % oxygen favored in routine cell culture.

## Metabolic Restriction and Restrictors

The process of tumor evolution over time occurs in an environment limited in nutrients, which cancer cells overcome by using their ability to thrive in diverse sites, which may be their truly critical property. Because of their metabolic flexibility, cancer cells can survive both in the nutrients-limited tumor cores and in the blood. Cellular metabolism has two important segments, glycolysis and mitochondrial oxidation. ATP can be produced by three distinct cellular processes: (1) glycolysis in the cytosol, and, in mitochondria, (2) the citric acid cycle/oxidative phosphorylation and (3) beta-oxidation. Cancer cells have the redundancy to circumvent the absence of what is apparently an essential energy source, oxygen, and sustain their proliferation rate by relying on glycolysis.

While assessing cell behavior in our assays, we were aware that attempts to influence metabolism could be defeated by metabolic pathway redundancy. We used in our work five chemical methods, *metabolic restrictors*, to slow down or restrict metabolism by impairing oxidative phosphorylation, ATP synthesis or ATP use, challenging the cells to restore homeostasis. Two of the 5 *metabolic restrictors* we used act specifically on mitochondria (Schultz and Chan, 2001), while 3 act more generally within the cell, but all effectively induced KCs in our tests.

The first is anoxia, well tolerated by tumors because of their use of glycolysis (Warburg, 1956), and the second is oligomycin, a specific ATPS inhibitor (Masini et al., 1984). The other three agents, two anti-oxidants and an inhibitor, reduce oxygen consumption. The neuro-hormone melatonin is a direct scavenger of $O_2^-$ (Poeggeler et al., 1994), while vitamin C generally acts as an anti-oxidant. The last, imatinib, a cancer drug used to treat Chronic Myeloid Leukemia and other malignancies (Takimoto and Calvo, 2008), is a competitive inhibitor of the BCR-ABL enzyme's ATP-binding site.

The clearest mechanistic example of metabolic suppression, through quenching of oxidative phosphorylation, is oligomycin, which inhibits ATPS and triggers AMP-activated protein kinase (AMPK), a sensitive regulator of ATP levels. The activation of AMPK suppresses bio-synthesis and stimulates catabolism, which in cancer cells afflicted by CIN leads to chromosome losses in later cell divisions.

The basic lesion that makes the cells cancerous, generally a lesion targeting metabolism, also makes the chromosome count of these cells unstable. Because of CIN, cancer cells are therefore ideal test objects to assay for changes in metabolism since these changes, through the action of AMPK, are reflected in chromosome counts in the next cell division. We have therefore used chromosome count as a reporter of cellular metabolic state.

# 1.2 Magnetic Fields

Since the 1979 Wertheimer and Leeper (Wertheimer and Leeper, 1979) article relating MFs, or more accurately, wire codes, with childhood cancer, the link between cancer and power-frequency MFs has been under investigation. There have been hundreds of epidemiological studies attempting to link MFs with cancer. Some of these studies were apparently successful. Milham (1982) reported an increase in deaths (proportional mortality ratio) due to leukemia in 486,000 men whose occupations were associated with electric or MFs.

Feychting of the Swedish Karolinska Institute reported (1992, 1993) a rate of leukemia close to four times the expected rate in a case-control study conducted in children exposed to higher MFs in their homes. Tynes and Andersen (1990) reported a significant 2.1 relative risk of breast cancer in Norwegian men potentially exposed to EMF. Thériault (1994) reported that workers with the highest cumulative exposure to MFs had an elevated risk (1.95) of brain cancer. Armstrong (1994) reported an association between exposure to pulsed EMFs and lung cancer in electric utility workers in Quebec, Canada, and France, with odds ratios rising to 3.11 in the highest exposure group. The dossier shows that the majority of studies find a weak association between EM fields and cancer (Coleman, 1990).

But epidemiological links between EMF and adult leukemia and brain cancer did not give completely clear or unanimous results (Kheifets, 1995). The exception is childhood leukemia (Albohm et al., 2000), leading the International Agency for Research on Cancer to attach the class 2B carcinogen designation to MFs in June 2001 (World Health Organization, 2002).

There have also been reports of alterations in cell behavior under excitation by EMFs of various frequencies and non-thermal intensities (Goodman et al., 1995). Consistent evidence from a number of different laboratories, that ELF fields interfere with breast cancer cells MCF-7 near 1.2 µT, has been reported (Ishido et al., 2001). Many have also detected a diversity of effects above 2.5 µT, higher than common environmental exposures. These include lengthened mitotic cycle and depressed respiration (Goodman et al., 1979), increased soft agar colony formation (Phillips et al., 1986), inhibition of differentiation with increased cell proliferation (Chen et al., 2000), as well as DNA breaks with apoptosis and necrosis (Lai and Singh, 2004).

Although epidemiology and *in vitro* studies on the health effects of ELF MFs have proceeded for decades, the process of understanding the biological effects of MF has been difficult and controversial. It has been argued that environmental 60-Hz MFs, certainly within the class of non-ionizing radiation, and furthermore incapable of raising tissue temperatures, could not have significant impacts on cells, because of the absence of a mechanism. The activation energy available from MFs seemed inferior to the background thermal noise typical of biological systems. In this context, population, *in vivo* and *in vitro* studies failed to provide a link strong enough to convince many investigators.

Since we had in our laboratory a highly stable and sensitive model capable of detecting alterations in the metabolism of  cancer cells based on their phenotype and karyotype, it seemed logical to use it to investigate this question. The unique quality of our model is determined by the use of anoxia which, beyond its representativity of the conditions in the core of tumors, has the virtue of stabilizing the chromosome range of hyperploid cultures (Li et al. 2012) to very restricted values, giving excellent sensitivity. Preliminary work showed that we could, with simple materials (structural steel), attenuate environmental ELF MFs to values lower than 4 nT, much lower than in a typical environment, adding the opportunity for clear exposure conditions.

# 1.3 Turning Points and Research Orientation

Before proceeding further into the document, we want to highlight three elements that we consider pivotal in determining the path of the work presented.

The first important decision was to perform cell assays in various levels of oxygen, and ultimately under anoxia, using boxes flushed with gas.

Since anoxia is a better simulation of the environment of tumors than the 21 % oxygen favored in routine cell culture, we needed to determine how our main cellular model, K562, reacts to various levels of oxygen, and we experimented with the full range of oxygen concentrations (0 to 21 %). From automated microscopy observations, we attempted to determine the oxygen concentration most favorable (normoxic condition) for K562, and essentially found that the lower, the better, as we shall see below. The sensitivity of  the K562 phenotype and karyotype to oxygen did not come as a surprise, as it has been known for a while that hematopoietic progenitor cells similar to K562 react favorably to reduced oxygen concentrations (Sandstrom et. al, 1994). But the ability of all of our cancer cell lines to thrive without *any* oxygen would surprise most biologists, because anoxia has been traditionally viewed at a lethal condition for living cells.

The second decision was to go beyond the investigation of cell phenotypes and into their genotypes, which was strengthened by the solidity of our computer vision-based data on cell phenotypes. The dependability and time-course of the phenotype changes caused by anoxia and *metabolic restrictors* lead us to suspect that corresponding karyotype changes must also occur. After the first cancer cell chromosome counts were confirmed to change with metabolism, we attempted to find a general connection between metabolic activity level and chromosome counts.

Later, we attempted to normalize chromosomes numbers in cancer cells using *metabolic restrictors*. The cancer cell lines used in this study were deliberately chosen because of their hyperploidy, which means that their chromosome counts are substantially increased beyond the human norm of 46.

Since our cancers cells subjected to anoxia or metabolic restriction did not revert back to normal tissue, but could, rather, expand their karyotype back to their original numbers when the metabolic conditions were restored, we realized that karyotype expansion and instability is a peripheral effect of the metabolic enhancement caused by the initial cancer lesion. Our view is that the *chromosome count alterations* capable of initializing cancer may be fundamentally different from the later karyotype changes (CIN). Our follow-up experiments attempted to develop the concept of KC and its pathological significance.

The conclusion of our mechanistic and metabolic work was that, somewhat paradoxically, metabolic restriction may act as a meta-genetic mechanism of tumor promotion. This conclusion allowed us to provide a hypothesis for previously unexplained observations in epidemiology, the carcinogenic action of some anti-oxidants, such as vitamin A, among cancer patients.

A third turning point occurred when we confirmed that MFs could induce KCs. The similarity of action between MFs and oligomycin led us to uncover the mechanism linking MFs and ATPS, using the previous work of Russian physicists on water. In view of the widespread occurrence of MFs in the general environment, we felt it was important to document effects in the range from 25 nT to 5 µT on the chromosome counts of our five cancer cell lines. From our observations, chromosome losses could occur in cancer cells with only discrete, but still observable, phenotypical changes under the microscope. Therefore, investigators looking for MF effects *in* *vitro* could have been deceived by the relatively normal appearance of the cell cultures, showing only small phenotype variations, while substantial genetic changes were occurring. The transient nature of these changes would compound the difficulty of detection.

Our research therefore reached two specific objectives, one related to basic tumor biology, and the second related to the biological action of  MFs. In basic tumor biology, we studied the effects of metabolic restrictors on cancer cell KC, and related the phenomenon to the surprising adaptability of tumor cells. In our ELF-MF work, we documented the ability of MFs to suppress metabolism in cancer cells, and described and explored the biological mechanism linking MFs and increased cancer  rates.

---------------------------------------------------------------------------------------------

# 2.0 MATERIALS and METHODS

---------------------------------------------------------------------------------------------

When this project started, the InVitroPlus Laboratory at the Royal Victoria Hospital was specialized in computer-based, real-time monitoring of *in vitro* human cell cultures. Experiments on human cells were largely automated. Cells contained in T-25 flasks or 96-well plates were placed on a motorized stage so that they could be moved automatically. With computer control and the scrutiny of a microscope camera, multiple cell cultures could be observed for as much as a week at a time. The system harvested large numbers of images that were interpreted by computer vision software to provide, for example, growth curves over time. Experiments could be monitored remotely through the Internet (Héroux et al., 2004). Round suspension cells like K562 were particularly convenient for the purpose of automated computer-based object recognition, because their generally simple shapes were easy to identify. The automated assaying produced accurate results designed to simultaneously compare the evolution of multiple cell cultures. From sequential images gathered automatically, a number of variables were analyzed, such as proliferation rate, cell size, expression of macrophages (objects 3 to 5 times the average K562 size) and cytoadhesion. The laboratory strived to produce stable biological models with high sensitivity for assaying various agents. To support this goal, cells were grown in synthetic medium in order to avoid the irregularities associated with changes in the composition of serum. The work reported here benefited from these methods because they provided solid documentation of the effects of oxygen and *metabolic restrictors* on cell phenotypes. However, when the emphasis of the project turned to karyotyping and genetics, the implementation of the work went from computer-based to classical laboratory work, as karyotyping methods are essentially manual, with little hope of being automated. This meant that we had to invest considerable time and effort into the refinement and execution of classical methods of karyotyping for various cell lines. This involved hundreds of hours of slide preparation as well as thousands of hours of microscope observation.

Consequently, this work, at the methodological level, is a cross between new techniques of computer-based automated observations and traditional biological techniques.

## 2.1 Cancer Cell Lines

Cancer cells are commonly used in *in vitro* work, not only because they are representations of disease, but also because they can propagate indefinitely without substantial changes. New immortal non-cancerous cellular models that have been transfected with human telomerase reverse-transcriptase (hTERT) are just becoming available. But in the end, the instability of our cancer cells, derived from human tumors, turned out to be a benefit in our investigations, because it allowed us to sensitively detect metabolic changes.

To investigate the effects of the various agents studied in this thesis, we used five cancer cell types: K562 and HEL 92.1.7 (erythro-leukemias), NCI-H460 (lung cancer), COLO 320DM (colon cancer) and MCF7 (breast cancer). They represent the major types of cancers leading to high death rates, and were obtained from the ATCC (http://www.atcc.org/). The particular cell lines within each type were chosen because they are hyperploid, which means they have chromosome numbers higher than normal human cells, at 46 chromosomes. The spread in chromosomes numbers allowed us to obtain statistically significant data more easily.

The cells are maintained at 5 % $CO_2$, 90 to 100 % humidity and 0 %, 2 %, 5 % or 21 % oxygen as needed. When received, the cells were immediately grown in T-75s and aliquots frozen into vials for conservation in liquid nitrogen. Two generations of cells were frozen as 5 aliquots to insure that 25 cell renewals would be available in the course of the project. Cells adapted to synthetic RSF1 (see 2.2 below), a process which takes many months, were conserved in the same way. Adaptation of cells to 0 % oxygen and < 4 nT MF are rapid in terms of obtaining a surviving culture, although metabolic adjustment takes longer.

## 2.1.1 Erythroleukemia: K562 and HEL 92.1.7

Erythroleukemia is characterized by the myeloproliferation of erythrocyte and leukocyte precursor stem cells.

K562 was the first continuous human Chronic Myelogenous Leukemia (CML) cell line, established by the Lozzios from a female patient in terminal blast crisis (Lozzio and Lozzio, 1975). It has a modal chromosome count of 69. K562 is a stem cell that can develop characteristics similar to early-stage erythrocytes, granulocytes and monocytes. It expresses the critical BCR:ABL protein (Philadelphia chromosome), detected in 95 % of CML cases. Many of the changes in CML cells are believed to be controlled by this oncoprotein. Its unregulated tyrosine kinase activity is thought to activate a number of cell cycle-controlling proteins and enzymes, inhibit DNA repair, cause genomic instability and maybe blast crisis. K562's properties have been established in detail, it is easy to recognize through image analysis, and is sensitive to toxic attack.

HEL 92.1.7 is a human erythroleukemia cell line from a male patient with Hodgkin's disease who later developed erythroleukemia, obtained by Martin (Martin and Papayannopoulou, 1982). It spontaneously and inducibly expresses globins. It has a modal chromosome count of 66. We chose K562 and HEL 92.1.7 because they are well documented human leukemia cell lines.

## 2.1.2 Lung cancer: NCI-H460

NCI-H460 was established by Gazdar in 1982 from the pleural fluid of a patient with large cell cancer of the lung. It has a modal chromosome count of 57. The NCI-H460 lung cancer cell line was chosen because it is one of the most fatal cancer types, and because there have been reports of an association between lung cancer and exposure to MFs.

### 2.1.3 Colon cancer: COLO 320DM

COLO 320DM is a human colon carcinoma cell line derived from a patient with colorectal adenocarcinoma, with a modal chromosome count of 54 (range 49 to 61). It was established by Quinn because of its unusual characteristics (Quinn et al., 1979). It is morphologically different from most colon cell lines with unusual cell products, double minutes, and homogeneously staining regions. This unusual cell line is valuable for studies of apudomas of the colon. It was chosen because of the high fatality rate of this cancer.

### 2.1.4 Breast cancer: MCF7

MCF7 is an adenocarcinoma cell line derived from the pleural effusion of a human breast cancer metastatic site (Soule et al., 1973). The modal chromosome count with normal incubation is 82, ranging from 66 to 87. It retains the ability to process estradiol via cytoplasmic estrogen receptors and expresses the WNT7B oncogene. The stemline chromosome counts ranged from hyper-triploidy to hypo-tetraploidy, with the 2S component occurring at 1 %. Substantial *in vitro* work using MFs has been done with this cell line (Ishido et al., 2001).

## 2.2 Synthetic Cell Culture Media

The choice of culture medium is critical in the assessment of toxicity, which is usually attenuated by peptide and protein plasma components such as albumin. Serum-based media might best represent physiological conditions for blood cells, but may not be best for the cancer cells resting within solid tumors cores.

Synthetic media allow the exclusion or controlled dosing of purified protein, such as albumin, insulin and transferrin, that are critical to many cells. Use of synthetic media is however associated with a number of practical problems. The ingredient concentrations in most commercial synthetic media are kept secret, due to company business practices. Many of the secret and other ingredients are included to achieve cell proliferation rates comparable to those of serum-based media for specific cell types.

To have full knowledge of medium composition, we decided to develop our own synthetic medium based on the classical RPMI 1640. The RSF medium was developed in our laboratory by Dr. Igor Kyrychenko. A revised version, RSF1, was later implemented with insulin levels divided by 10, closer to plasma values, and as shown in Table 1.

Insulin is still provided at concentrations much higher than physiological levels in RSF1. High insulin decreases the activity of SKN-1/Nrf2 (Tullet et al., 2008; Jasper, 2008), throttling down detoxification enzymes, and leaving cells less protected. Extra copies of the gene for SKN-1 extend the life of *C. elegans* by 25 to 30 percent (Tullet et al., 2008). This means that the high insulin levels in RSF1 may provide a sensitive model for the effects of ROS, for example, and possibly a useful model to display accelerated chronic toxicity.

There is no iron salt in RPMI 1640, but iron is necessary to insure sustained survival of a cell line. The iron in RSF1 is provided as transferrin saturated with iron (see Table 1), and there may also be some small amount of iron bound to the added albumin.

| RPMI-1640 with l-glutamine<br>(Sigma 61-030-RM) |
|:---:|
| Sodium bicarbonate, 2 g / l<br>(Sigma S-6014) |
| Sodium selenite, 20 nM<br>(Sigma S-5261) |
| Bovine Serum Albumin, 4 g/ l<br>(Sigma A3311). |
| Bovine Transferrin (iron saturated), 25 mg/ l<br>(Sigma T1408) |
| Bovine Insulin, 1 mg/ l<br>(Sigma I5500) |

**Table 1: RSF1 medium composition.**

Transferrin appears indispensable for DNA synthesis of most growing cells, and is often referred to as a growth factor because proliferating cells express high numbers of transferrin receptors (CD71). Transferrin also acts as a cytokine, and has roles that may not be related to its iron-carrying capacity. Transferrin is used at a low level of 25 mg/l in RSF1, partly because of cost. It is present in normal plasma at 2500 mg/l, 100 times more (about 3.4 % of total blood protein). In RSF1, transferrin concentration is 100 times below blood levels, but is 100 % rather than 30 % saturated. This translates to an iron concentration of 0.03 mg/l, initially bound to transferrin in RSF1.

K562 can divide for a short period of time in unsupplemented RPMI 1640, probably from the action of residual protein. This represents a very artificial situation (extremely low protein concentrations), which could be used to enlighten specific mechanisms. As supplemented in Table 1, RSF1 is competitive in terms of proliferation rate with serum-based RPMI-1640 for the erythroleukemia cells (K562 and HEL), since it was optimized for them, but the other cell lines divide more slowly without serum, and could not be sustained over very long periods. The same composition was used for all, in the interest of uniformity.

## 2.3 Culture Vessels and Incubation

### 2.3.1 Culture Vessels

Depending on the particular determination (proliferation, apoptosis, etc), the seeding density and the duration of the test, various formats are used for short-term and chronic experiments that are automatically monitored by computer vision over time. The formats are T-12, T-25 and microplate. Each format can be executed at different levels of oxygen. The T-12 format, as we implemented it, provides four aliquot and simultaneous results from four separate vessels, each with its separate exposure. A T-25 provides a single result at a single oxygen level, for example, but with a large sampling surface, appropriate for detection of rare events or cell types. The microplate provides 77 individual results at a single oxygen level. All the 96 wells of the microplate cannot be used, because of mechanical limitations of the motorized stage (Fig. 2.6). These formats can provide a maximum of 12, 25 and 0.32 $cm^2$ of cell culture surface per determination, respectively, which, with seeding density, will determine the precisions of the determinations (see 2.9.3). The cells are routinely incubated in vented T-25 flasks, and transferred to either T-12s (Falcon 353018), T-25s (Sarstedt 83.1810.502) or various microplate models, as different vessels are used in various experiments.

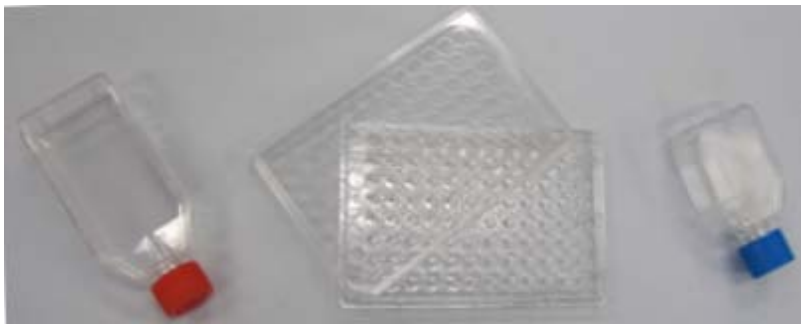

**Fig. 2.1: Cell Culture Vessels: T-25, 96-well Microplate with cover, and T-12.**

Although T-25s provide the most surface, T-12s are preferred in experiments involving MFs because their smaller area insures a more uniform MF density within the longitudinal profile of the magnetic shields (see Fig. 2.3).

The small size of the T-12s is also indispensable for *differential* experiments where cells are simultaneous seeded into four T-12s to assess different agents applied to each T-12.

For example, the effect of four different oxygen levels can be assessed accurately in the same time frame in this manner. The four T-12s are welded together in a square, using lines of epoxy glue along their edges. This insures identical mechanical history and stable confocality when the array is positioned on the motorized microscope stage for chronic computer-vision observations. The four different oxygen concentrations are implemented by connecting the vented cap of each T-12 with a long flexible latex tube to much larger vessels (1.6 liters) containing the required levels of oxygen and carbon dioxide. To compensate for $CO_2$ diffusion through the tubes, the incubator itself maintains $CO_2$ conditions during the test.

The cell densities are adjusted initially from automated cell counting of microscope fields acquired using camera and custom vision software (ImageProPlus macros) in early tests, and later using an automated cell counter from Millipore, the Scepter.

## 2.3.2 Incubation

We use two Forma 3310 as well as one Baxter (WJ 501 S) incubator equipped with HEPA filters, stabilized for temperature, $CO_2$ and humidity. The incubation conditions ($CO_2$, temperature and humidity) are constantly monitored and logged at 1 min interval by our data acquisition system.

The automated image acquisition hardware (see section 2.6), including camera, motorized stage and microscope is entirely contained within the incubator to maintain cell-friendly conditions.

Some modifications were made to the hardware to make them temperature and humidity tolerant.

## Low Oxygen Boxes

Our laboratory has developed the capability of exploring multiple oxygen levels simultaneously by using multiple environmental boxes placed in a single incubator. Specific oxygen concentrations are secured by gas-tight containers (Star Frit Lock & Lock) large enough (1.6 liters, polypropylene with silicone gasket) to stabilize oxygen levels over the time of the assays in the T-12s or T-25s that they contain (Wright and Shay, 2006). Many boxes can be fitted in a single incubator, but they are large enough to act as buffers, passively maintaining atmospheric conditions over the course of even 10-day tests.

The containers are slowly filled from the bottom with 4 to 5 box volumes of the cold medical gas mixtures supplied through a 0.22 μm filter by specially ordered gas tanks (5 % $CO_2$, 95 % nitrogen). Bubble tubing is used for connections. Pressure measurements and water immersion

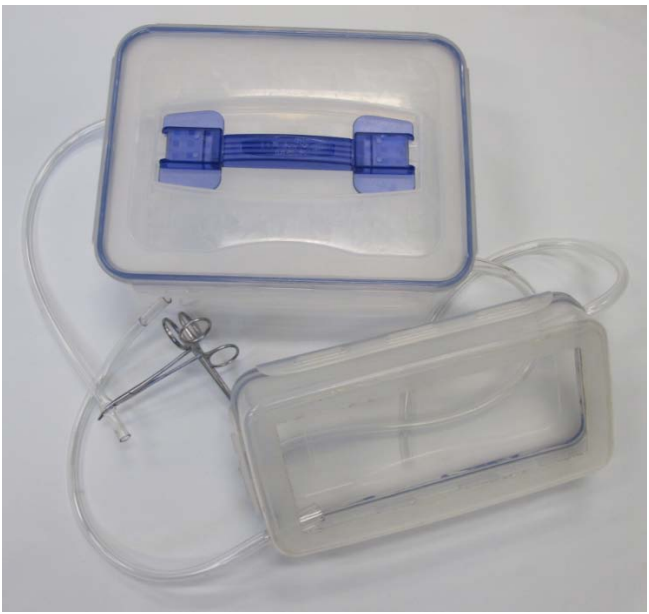 tests were made to insure that the configuration provides effective sealing.

**Fig. 2.2: Low oxygen boxes. The larger size at top can accommodate the two inner magnetic shields. The lower one is fitted with thin glass on the bottom to allow microscopy observations. Bubble tubing, hemostats and silicone gaskets provide sealing.**

To minimize exposure to atmospheric gases during manipulations in the laminar flow hood, the culture vessel's ventilation holes are plugged right after box opening, and the culture vessels are flushed with the gas mixture using sterile Pasteur pipettes.

Some boxes were modified to be useable over the microscope. Particularly thin glass was glued to a cut-out in the bottom of the polypropylene containers to allow observations in the T-12s or T-25s on inverted microscopes. This allows multiple observations without the need to repeat the gas flushing procedure.

Oxygen levels intermediate between 0 % and 21 % can be obtained by including in the box, together with the culture vessel, other vessels of various sizes equilibrated with air with a porous plug. Over a short period, the atmosphere within all containers averages out.

## 2.4  Metabolic Restrictor Concentrations

The five metabolic restrictors used in our tests were chosen based on their metabolic effects on various aspects of cell metabolism, are described in section 1.1.2. Oligomycin and imatinib were used at sub-toxic levels: serial dilutions allowed the  determination of concentrations suppressing proliferation to 50 % of normal ($IC_{50}$), still adequate for karyotyping. Melatonin-Vitamin C were optimized for maximum chromosome drop through serial dilutions, and these levels were subsequently found to be physiological, as they matched those in bone marrow (Zhuang et al., 1998) and plasma.

Oligomycin (O4876), metformin (D150959)  and melatonin (M5250)  were obtained from Sigma and resistin from Prospec Protein Specialists, East Brunswick, New Jersey, USA.

# 2.5 Magnetic Field Hardware

## 2.5.1 Magnetic Shielding

**Fig. 2.3: The three layers of magnetic shielding. The Narda EFA-300's magnetic probe is in place of the culture vessel. Magnetic field coils are below, but not in contact with the two smaller shields, insulated from the outer shield by a layer of rigid foam or acrylic.**

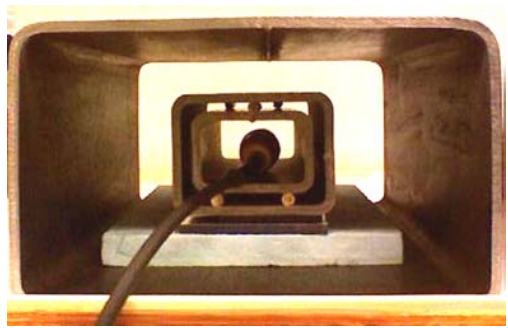

*Naïve* cells must be kept in a low 60-Hz MF environment, as laboratory and incubator fields are too high. Inside a low-field (0.4 µT average) incubator, Forma 3310, shielding is provided by three cylindrical layers of 6.3 mm thick structural steel. The three shields have outside dimensions 15.2 x 25.4 x 36, 7.6 x 10.2 x 20, and 5.1 x 7.6 x 20 cm (Fig. 2.3).

The shields reduce 60-Hz fields by a factor of 144, estimated from the attenuation of 8.8 µT to 61 nT. Our measuring instrument, the Narda EFA-300 (see below), has a floor of 5 nT, above the *naive* level achieved by the three-layer shielding, estimated at 3 nT. The inner shielding layer may be removed for particular experiments, depending on the field magnitude and frequency. When keeping cells under no-field and anoxic conditions, the two inner shields with the culture are placed inside boxes which fit inside the third external shield.

## 2.5.2 Magnetic Field Measuring Instruments

Two triaxial data-logging MF meters were used (Fig. 2.4): a Field Star 1000 (Dexsil Corporation, Hamden, CT. USA) and a Narda EFA-300 (Narda Safety Test Solutions, Pfullingen, Germany). The Field Star is capable of recording MF density in the X, Y and Z axes, and their vectorially integrated RMS value each second, within a narrow bandwidth: 55 to 65 dB ±3 dB.

The Narda EFA-300 comes with an optional MF probe connected with a cable. It can measure MF spectra between 5 Hz and 32 kHz and field strengths from 100 nT to 32 mT within that bandwidth. The bandwidth on the Narda is selectable (for example, 5 Hz to 2 kHz). In the narrow-band power-frequency mode, the measurement floor is 5 nT.

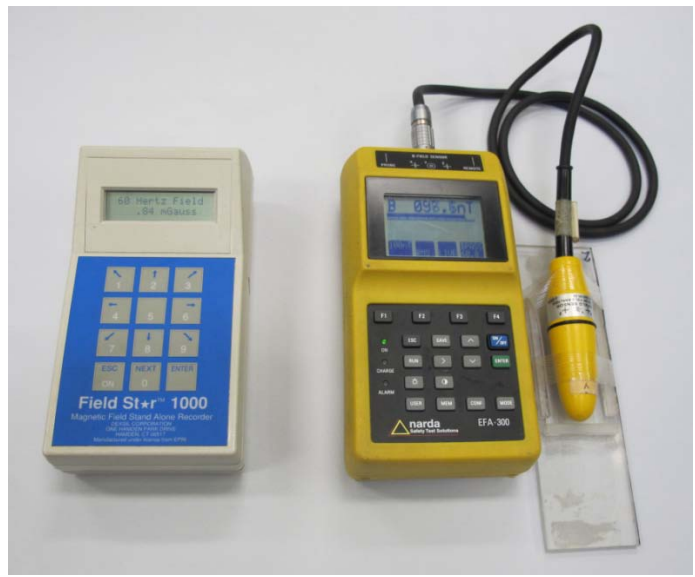

Miniature coils were also developed to perform uniaxial MF measurements in the restricted space of a multiwell dish.

**Fig. 2.4: Magnetic Field Instruments, the Field Star 1000, and the Narda EFA-300 with MF probe. The probe is fixed to a jig allowing precise positioning within the magnetic shields.**

## 2.5.3 Applied Magnetic Fields

MFs are applied by rectangular coils (19 x 25.6 cm) with 20 to 50 turns of #25 AWG varnished copper wire. The sources are either variable transformers (Variacs) connected to the 60-Hz network or various audio amplifiers fed by computer-based sinewave generating software for other frequencies. When applied MFs were particularly low (~25 nT), it was necessary to add some passive filtering (28 µF) in parallel with the coils (8 Ω) to reduce high-frequency parasites on the waveform, which was controlled using an oscilloscope.

The MF applied to cell cultures is not uniform, because it is generated by a coil located outside the two inner shields, but inside the outer shield. The coil's field geometry is stretched, but not substantially disturbed in shape by the internal shields, and so is minimum at the coil center.

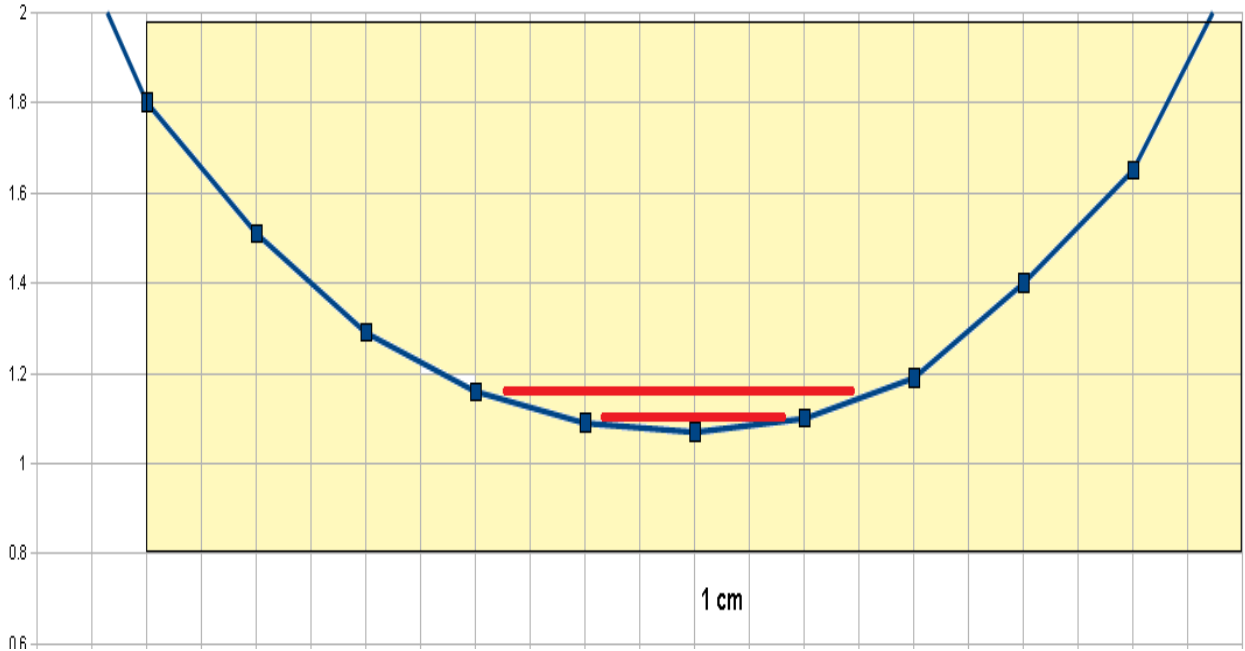

**Fig. 2.5: Magnetic field density (µT) generated by an exposure coil vs longitudinal distance inside a magnetic shield pair. The two red lines show the extent of T-25 and T-12 culture vessels, and the yellow rectangle is the shield outline.**

The red lines in the figure represent the length of T-12 (3.35 cm) and T-25 (6.3 cm) culture flasks. These results indicate that a T-25 is accurate to better than 20 %, while a T-12 is accurate to better than 10 %.

The coils generating MFs were wound around plastic forms, as shown in Fig.2.3. There is no contact between the coils and the magnetic shields. The electrical currents required to produce MF exposures in all out tests were small enough that the electrical wiring was not warm to the touch.

After a clinical thermometer was carefully read inside the incubator, the same thermometer was used to read the temperature within cell culture medium after a 24 hour exposure at 5 µT, and the column did not shown any detectable difference from the incubator baseline.

The cells used in our experiments need to be occasionally withdrawn from the incubators for passaging, which means that they spend some short intervals of time outside of volumes within which MFs are well controlled. The effects we have monitored (KC) in our data occur over many generations of cells and many days, so we do no expect that a 20 min hiatus will have a detectable influence on the outcome of our 6-day tests. Therefore, we do not believe that these short transient exposures were a significant factor in our results.

# 2.6 Data Acquisition Hardware

## 2.6.1 Microscope, Camera, Imaging Software

Images used in our work were produced using Diavert (Leitz) or Laborlux D (Leica) microscopes, in brightfield illumination, at magnifications of 20 to 2.5x, with Infinity X (21 Mpixels) or Infinity Lite (1.5 Mpixels) CMOS cameras (Lumenera). Basic object recognition functions were provided by Media Cybernetic's ImagePro Plus.

## 2.6.2 Stage

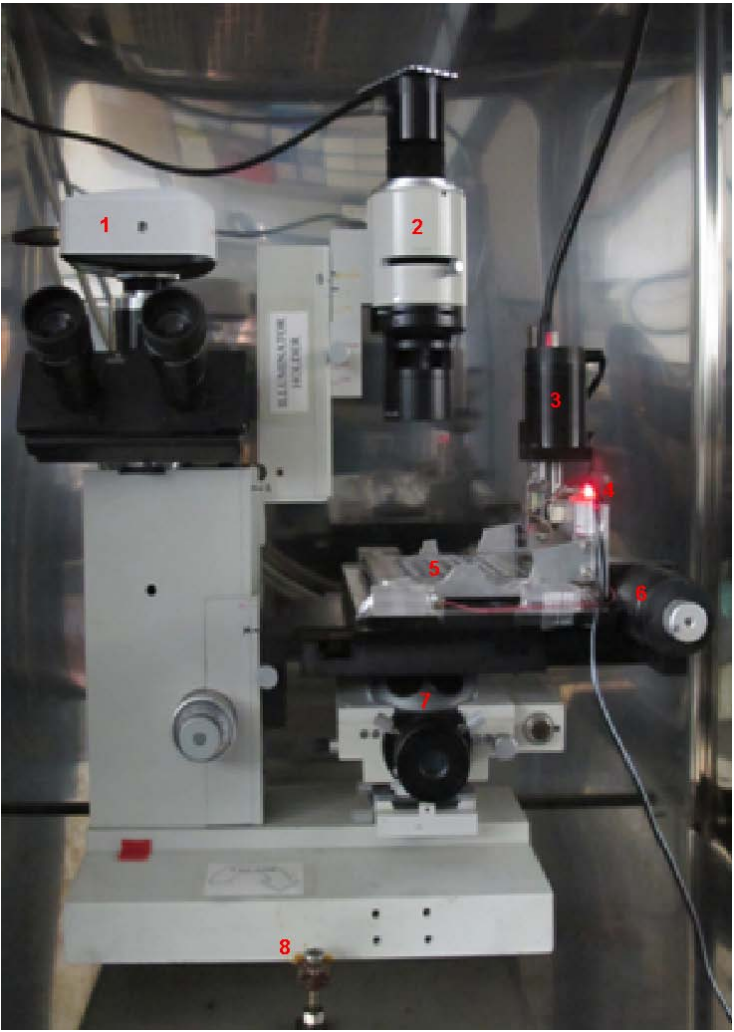

In *manual assays*, T-25s were successively placed at time intervals on the motorized stage (Marzhauser MCL) mounted on the microscope (Diavert) installed inside the incubator.

**Fig. 2.6: The automated data acquisition system housed inside one of the two Forma 3310 incubators.**

**1: CMOS camera.**
**2: Illuminator.**
**3: Z-movement motor.**
**4: LED, end-of-line indicator.**
**5: microplate.**
**6: motor.**
**7: turret and optics.**
**8: leveling.**

In *automated assays*, T-25s, microplates or sets of four T-12s were placed on the stage at the beginning of the test, and were scanned over the full duration of the test, while images were acquired automatically. In order not to disturb the cells during the tests, the stage acceleration was limited to 0.34 cm/s², which, as was observed, was barely enough to produce a vibration of settled cells in a T-12 filled with 3 mm of the medium.

Automatic focusing software developed in our lab maintained image quality over time. This arrangement insured reproducible and identical conditions for all four T-12s investigated simultaneously. Automated operation provided the large number of images necessary to reduce sampling errors (section 2.9.3). Stage control, image acquisition process, analysis and first line data compilation procedures were realized by custom software developed in our laboratory (Héroux et al., 2004).

# 2.7 Cell Phenotype Variables

Cell *Proliferation*, *Footprint*, *Roundness* and *Hex-Distance* were studied to monitor cell behavior. The implementation of these variables is discussed below.

## 2.7.1 Proliferation

*Proliferation* is often represented in the literature by growth curves quantifying relative cell numbers, population doubling times (hours) or growth constants. Cell proliferation rate is an important functional cellular variable. Basal proliferation levels vary according to the particular cell line, culture medium (a range of serum-based and synthetic media were tried) and incubation characteristics, among which, atmospheric oxygen.

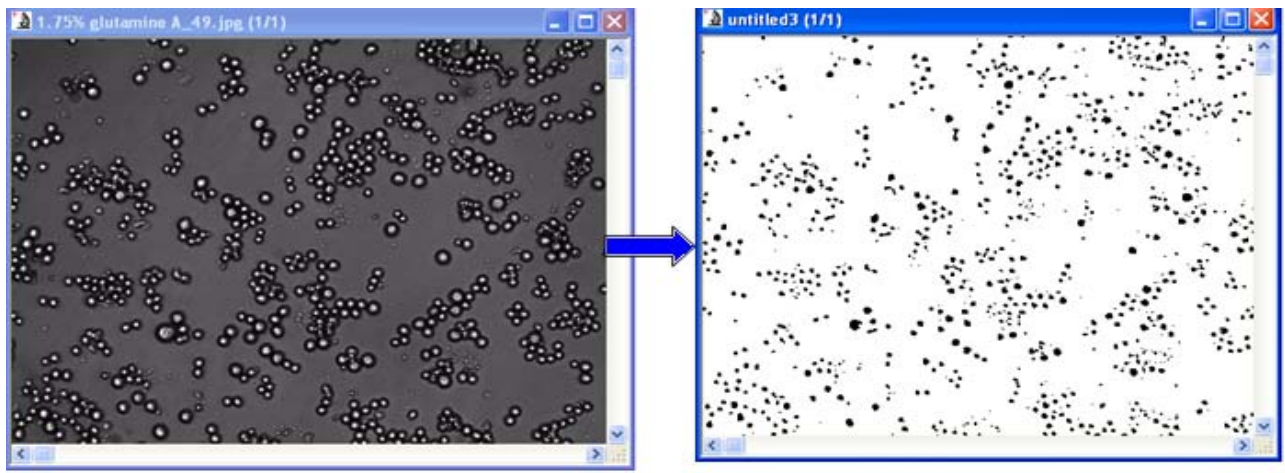

**Fig. 2.7: Cell Proliferation counting procedure. The cells coalescing together in the culture are segmented by an algorithm written in Visual Basic, using ImageProPlus functions.**

The pictures of cells acquired either manually or automatically with a digital camera are analyzed using ImageProPlus software (Media Cybernetics), yielding cell counts. The purpose of the algorithm is to optimize image segmentation of cell clusters which form most of the image surface as cell numbers increase. Based on the different illumination level of the cell body, cell edges and background, individual cells are separated by adaptive software within each image. Even under very unfavorable conditions, the algorithm works with a recognition accuracy of 81 % (Héroux et al., 2004).

We quantified cell multiplication as a relative rise in cell numbers calculated from successive images typically 1 hr apart. The *doubling time* is frequently used, but the ***doubling frequency*** (the inverse of the doubling time) has the advantage of being proportional to proliferation speed.

## 2.7.2 Footprint

Although most authors refer to "cell size" when discussing the apparent area of objects settled on a culture surface, as seen through a digital camera and microscope, we use the term *Footprint* (pixels or $\mu m^2$) as distinct from actual cell size in tridimensional suspension.

The cell *Footprints* may be measured using IPP-based software with a special focal plane sweeping procedure designed to reduce any possible bias in the footprint values obtained.

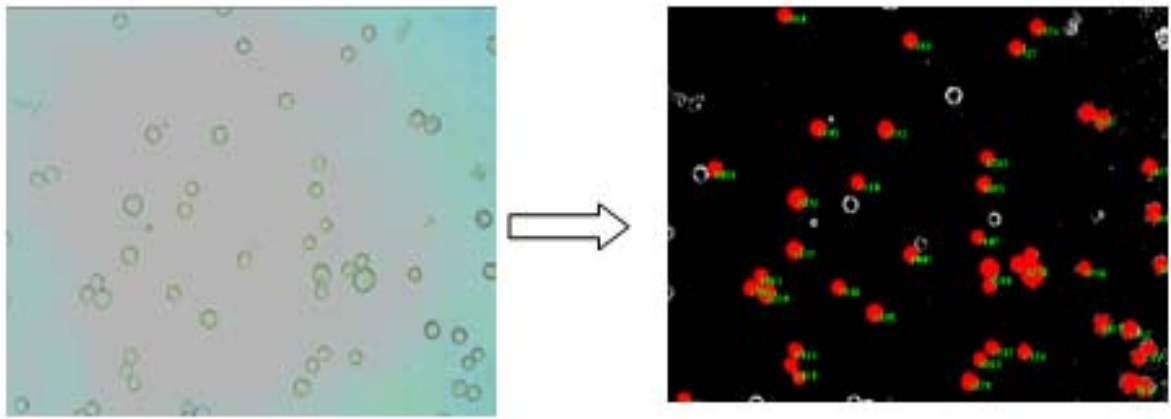

**Fig. 2.8: Computer perception of objects in culture, with their calculated *Cell Footprints* displayed. Cell Footprint is processed and measured with an algorithm written in Visual Basic, using ImageProPlus functions.**

How flat a cell lays on the bottom of a dish is dependent on the integrity of its internal cytoskeleton. It has been documented that short-term hypoxia produces alterations in the spectrin cytoskeleton (Glantz and Morrow, 1996) in many studies of necrosis that sometimes involved reperfusion. Other studies point to alterations of the cytoskeleton, for example in cardiac myocytes (Ganote and Van der Heide, 1987). It is known that ROS affect cell shape through actin cytoskeleton toxicity in mussel haemocytes (Gomez-Mendikute and Cajaraville, 2003). And neuron cells have been observed as slightly smaller after being incubated in 1 % of oxygen compared to 20 % oxygen (Xie et al., 2005). But no detailed studies about cell size changes or karyotype changes under various levels of oxygen have been reported in the literature.

### 2.7.3 Roundness

*Roundness* can be accessed from images using algorithms similar to those developed for *Footprint* detection. *Roundness* provides information about cell state that is different from that provided by *Footprint*.

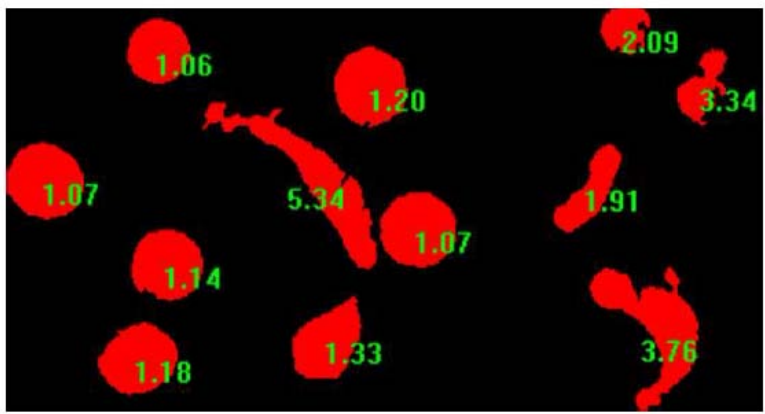

Fig. 2.9: Computer perception of objects in culture, with their calculated *Roundness* displayed. The 1.91 and 3.76 objects are macrophages. The 2.09 and 3.34 objects are dead cells. *Roundness* can be revealing of cell type and function.

*Roundness* is the dimensionless ratio between the imaged cell perimeter, and a circle of identical area. A circle has a *Roundness* of 1; all other shapes yield higher numbers. For suspension cells, which are normally close to circular, *Roundness* quantifies taxis of the sub-population of macrophage-like cells within the culture. The variable is useful to quantify the expression of macrophages and their activity within a cell population.

### 2.7.4 Apoptosis

Software to detect apoptosis automatically has been developed in our laboratory, and is effective as long as images for a given site are acquired within a reasonable frequency (30 min or less) and also if cell density does not exceed a certain maximum. Apoptosis occurs rapidly and its detectable result, a group of apobodies, is rapidly reabsorbed by neighboring cells, possibly losing events if the frequency of image acquisition is too low. If the cell population is too high, the complexity and density of objects increases, making the detection of apoptosis, which is optimal for a relatively isolated group of apobodies, more difficult.

Reactions to the presence of toxicants can span days (Héroux et al., 2004). It has been shown, for example, that excess iron in the body induces apoptosis in some cell types such as macrophages and HeLa cells (Pirdel et al., 2007; Cozzi et al., 2003).

### 2.7.5 Hex-Distance

*Hex-Distance* is a variable developed in our laboratory, quantifying the stickiness of cells to one another. As cells multiply and move, they may form different patterns, depending on their level of self-adhesion.

*Hex-Distance* is quantified as 360° divided by the sum of the angles shadowed on a cell's surface by its 6 nearest neighbors. In ideal two-dimensional packing of identical spheres (Table 2), dimensionless *Hex-Distance* is equal to 1. It rates 5 to 7 for cells newly seeded at 9,000 to 7,000/cm$^2$.

Hex-Distance = $$\frac{\pi}{\sum_1^6 \sin^{-1} \frac{1}{D_i} \sqrt{\frac{A_i}{\pi}}} \quad (1)$$

where $D_i$ are center-to-center distances between one cell and its 6 nearest neighbors of area $A_i$

| Hex-Distance Value | 1 | 2 | 4 | 6 |
|---|---|---|---|---|
| Geometrical | 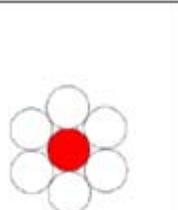 | 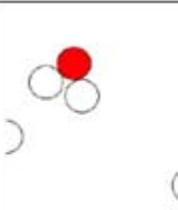 | 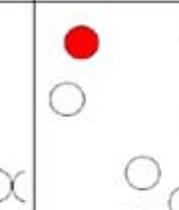 | 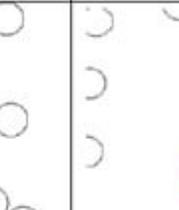 |
| Cell Culture | 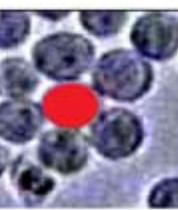 | 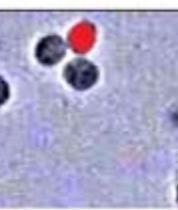 | 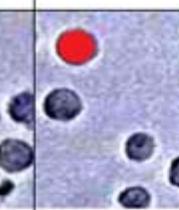 | 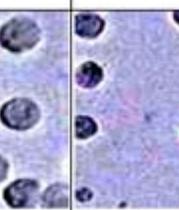 |

**Table 2: Hex-Distance examples illustrated in ideal geometry, and in actual K562 cell culture micrographs.**

# 2.8 Cell Genotype Variables

The genetic structure of cells exposed to metabolic restrictors or MFs was generally determined using karyotyping, as chromosome counts and G-banding. It is difficult to identify and group uniformly stained chromosomes, but G-banding allows 'bands' to appear on chromosomes, which are the same on the homologous chromosomes, easing identification.

More advanced techniques such as Spectral Karyotyping (SKY) and Fluorescence In-Situ Hybridization (FISH) were also used, but had to be contracted out, as the equipment to perform these procedures is only available at a single hospital in Quebec.

## 2.8.1 Chromosome Counts and G-banding

Metaphase preparation and cytogenetic analysis were performed according to standard cytogenetic procedures on the five cancer cell lines with a trypsin-Giemsa banding technique (Freshney, 1994). For simple chromosome counts, the non-adherent cell lines K562 and HEL 92.1.7 were first exposed in culture flasks until they reached a certain cell density. At conclusion of exposure, colcemid was added for 2 hours, after which the cells were resuspended in hypotonic solution. 12 minutes after, the hypotonic solution is removed, and ice-cold acetic acid-methanol is added. The preparation is dripped unto inclined slides which are placed in a 60°C incubator for 24 hours, after which they are stained with Giemsa. The procedures for the adherent cell lines, NCI-H460, COLO 320DM and MCF7, are more or less the same, except that they need to be trypsinized for cell separation.

The slides are evaluated using a Laborlux D upright microscope with 100X magnification under oil immersion. Metaphase images are captured using a monochrome Infinity X digital microscope camera with a low-noise 21 Mpixels CMOS sensor (pixel-shifting technique). Chromosomes are counted visually with the aid of ImageProPlus software.

SKY (Applied Spectral Imaging, Sky Vision) and FISH protocols of anoxic K562 cells were performed on metaphase spreads at the Banque de Cellules Leucémiques du Québec (Centre de Recherche, Hôpital Maisonneuve-Rosemont, Montreal, Quebec, Canada).

# 2.9 Precision of Determinations

Accurate determination of cell culture properties depends on control of assay conditions, but also on sufficient sampling (cell numbers or culture area in the case of cell counts, or number of metaphases in the case of chromosome counts).

The precision of individual determinations is related to the cells, the medium, test conditions, sampling and image interpretation. Our objective is to provide reproducible and precise *in vitro* determinations, so that fine effects as well as accurate thresholds can be determined.

## 2.9.1 Culture State

Assay errors may relate to the state of the cells at the beginning of a test. The cells may be brought to a certain state by changing the medium at a fixed time before the test, insuring a certain culture density and proliferation rate, avoiding contamination (mycoplasma) and checking for cell parameters (such as cell roundness) that signal the dynamism of cell division. A more troublesome aspect is the accommodation that occurs in a cell population when they are changed from one culture medium to another. For best reproducibility, cell medium variations should be avoided, which we have controlled using synthetic media.

To provide identical starting points for tests on effects of oxygen or metabolic restriction, cell lines were kept under anoxia in RSF1. Later, the same conditions included alternating MFs lower than 4 nT as experiment series focussed on effects of MFs.

## 2.9.2 Environmental Conditions

Although incubator conditions (temperature, $CO_2$ and relative humidity) are tightly controlled and recorded continuously to file (1 min intervals) by a computer, manipulations performed on mechanically delicate cells are difficult to reproduce perfectly in time intervals and mechanical characteristics. Test manipulation conditions are inevitably variable, if conducted by a human. The availability of our automatic robotized system to perform data acquisition over as much as 10 days without human intervention gives us a considerable advantage in precision. Therefore, if initial setup for a test is rigidly standardized, excellent control is achieved on the overall test procedure.

### 2.9.3 Sampling Surface

Since our system is robotized, the acquisition and quantification of large numbers of images, which would exhaust any human technician, is possible to reduce sampling errors. Our laboratory is equipped with large (backed-up) hard disks to acquire and keep permanently large number of images (we already have more than 1.5 million images in our archive).

The need for acquisition over substantial surfaces to reach specific accuracies in the case of a measurement of cell proliferation is illustrated below. Our investigation determined that 34 images from our setup (a total sample of 5.5 cm²) reduced the standard deviation on cell density estimates to less than 4 %. Our Monte Carlo type procedure for reaching that conclusion is illustrated in a graph (Fig. 2.10) generated by a Visual Basic program (Appendix A). Starting from a T-25 culture in expansion, where local cell density deviations are at their maximum, we use our automatic system to acquire 168 images. The software randomly picks 100 images from the 168, counts the cells within them, and compiles this data. The program simulates under-sampling of the 100-strong data set (Monte Carlo method), and estimates the error that would be associated with counting the cells in fewer images to obtain a reading. Assuming that the 100-image average is the true value, we then plot the standard deviation as a function of the number of images sampled. From this graph, we can see that using 34 images (5.5 cm$^2$) reduces the standard deviation on cell counts to less than 4 %. Superposed and underneath the graph are scattered pale blue points which represent the convergence of the arithmetic mean of a random mathematical function (random triangular of 10), offset by 55, with increased sample size.

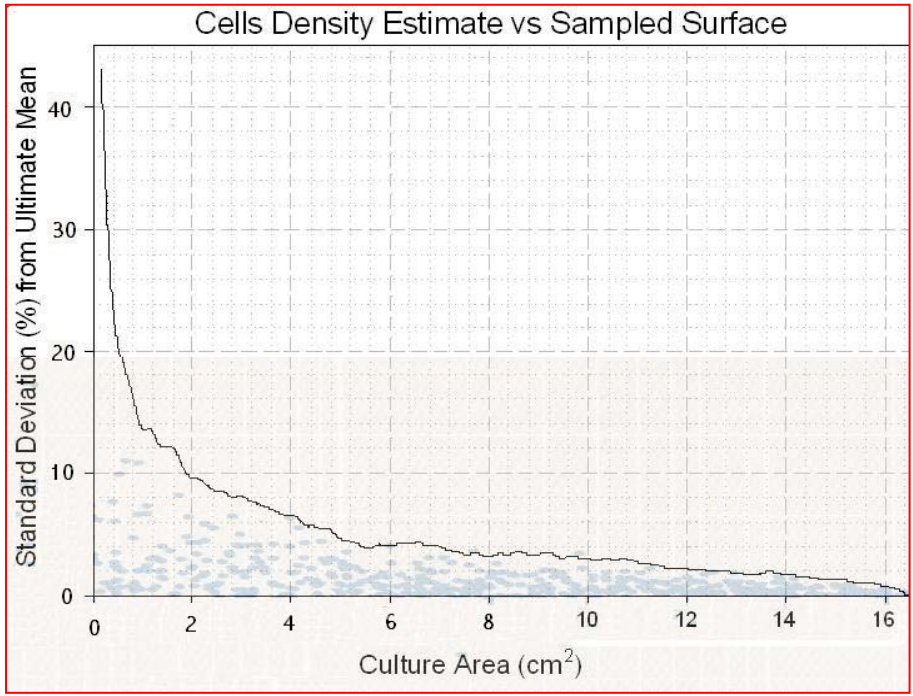

**Fig. 2.10: Accuracy of Cell Density estimate for K562 at 10,000 cells/cm$^2$, as a function of sampled surface. Up to 100 images from a set of 168 images are randomly picked to simulate under-sampling. The standard deviation as a function of the number of images sampled is plotted. 34 images (5.5 cm$^2$) reduce the standard deviation on cell counts to less than 4 %.**

To explain the choices made in sampling strategies for various variables, one can consider that seeding a plate at 10,000 cells / cm$^2$ will yield a certain number of cell counts per cm$^2$ (10,000) which can increase and be followed over time by successive manipulations or by an automated system. However, investigating spontaneous apoptosis in a population of unchallenged K562 cells will yield far fewer events, each lasting for about 30 minutes, in the order of 20 per day per cm$^2$.

Therefore, tests may be carried out at low surface (low $cm^2$, i.e. microplate ) when gross toxicity is expected, but with larger sampling surface as well as more frequent readings when thresholds or more subtle variables are investigated.

Within each experiment, more than one instance of a culture can usually be run. For example, in a microplate, more than one well can be used to represent a single exposure condition, providing multiple measurements within a single dish. When using T-25s, several T-25s with identical content can be run simultaneously. This, of course, is different from running experiments at different times. When experiments are run sequentially over time, as opposed to simultaneously, the main risk of variations is with the cells themselves. Simultaneous experiments would usually harvest cells from a single source and divide them as aliquots, while tests conducted in sequence must use cells that are not aliquots. We believe that if cells are harvested from independent cultures, simultaneous tests are equivalent to sequential tests, as the main source of variance is the cells themselves and possibly the details in handling them, as opposed to variations in incubator conditions. Environmental controls in our incubators are very good, and all environmental data is kept on file by an automated data acquisition system.

Data obtained from two tests are usually compared by means of Student's t-test and comparisons among multiple groups are performed using an analysis of variance. A p-value of 0.05 is considered significant.

### 2.9.4 Accuracy of Chromosome Counts

The most critical variable of this thesis is the observations on chromosome counts. Therefore, the accuracy of this determination is of great importance. The factors that limit this accuracy are connected with the clarity of chromosome spreads. Given an unlimited amount of time, a patient experimenter could work for long periods of time until absolutely clear and ideal chromosome spreads are obtained. This is not realistic in practice. The display of chromosome karyotypes on images is not always ideal. An important factor in determining image clarity is the colcemid time. Colcemid arrests mitotic cultured cells in metaphase and causes the chromosome to condense. The longer the cells are exposed to colcemid, the more cells are arrested, which is desirable, but the shorter the chromosomes become. Long-term exposure to colcemid can lead to very short chromosomes, and these very short chromosomes may become undetectable, leading to counting errors. Shorter chromosomes are harder to identify because no bands can be resolved, which leads to difficulties of counting and analyzing cell karyotypes. The metaphases are therefore the result of a technical compromise.

Due to these technical reasons and the irregularities in the spread of chromosome pairs by the metaphase plate, there are almost inevitable overlaps between chromosomes that may affect the accuracy of chromosome counts. To increase the accuracy of chromosome counting, image screening and accuracy monitoring procedures are routinely used. Screening chromosome images is a procedure widely used in karyotyping, where only 20 % of the images are typically selected among hundreds of images available. The selection criteria are that the chromosomes spread well and are of substantial length.

To monitor the accuracy of our own chromosome counting, we randomly selected 20 images for cell lines COLO 320DM and MCF7. COLO 320DM and MCF7 are both adherent cell lines and tend to grow in clumps, which leads to more difficult karyotyping.

Each of the 20 images was counted 3 times, and the average chromosome count was used as the true chromosome count. The chromosome counting error for each image was calculated based on these 20 images. The counting error for COLO 320DM (chromosome count 46-49) is 1.86 % with a standard error 0.58, and MCF7 (chromosome count 61-75) rates at 1.16 % $\pm$ 0.39. The accuracy for the other cell lines is expected to be higher, because of their more regular dispersion on the plates. This means that in practice, the chromosome counts presented in the following section should be accurate to better than $\pm$ 1.

--------------------------------------------------------------------------------

# 3.0 RESULTS

--------------------------------------------------------------------------------

# 3.1 Phenotype Changes under Various Levels of Oxygen

As many important forms of chronic occupational toxicity are associated with metals, our initial research objective was to assess the toxicity of industrial metals *in vitro*. From our initial experiments and from the scientific literature, we quickly realized that the toxicity of metals is intimately tied to the production of reactive oxygen species (ROS) through the Fenton reaction. Essentially, the Fenton reaction uses metals as catalysts to generate highly reactive hydroxyl radicals. To clarify the role of oxygen tension in transition metal toxicity, we were led to conduct *in vitro* tests under various levels of oxygen. Under the microscope, we observed cellular changes in proliferation rate, cell size and roundness, macrophage action and cell adhesion. Hypoxic tests were followed by anoxic tests, and in our anoxic K562 model, we observed smaller, but rapidly proliferating cells.

The hypoxic center of tumors is widely perceived as in need of more perfusion, because hypoxia in viewed as an inherently pathological state. Our measurements below present a different picture.

## 3.1.1 Reducing Oxygen Increases Proliferation

In four manual and four automatic assays in RSF1 medium, K562 cells were seeded at 7,000 cells/cm², and counted over time. In the manual assays, T-25 flasks were filled with 7.5 ml of culture, and measurements taken at 8 hr intervals at 168 locations within each flask. Scanning was performed with a motorized stage, but the five T-25s corresponding to the different levels of oxygen had to be manually exchanged periodically. With human intervention, suspension cells reposition at each measurement, preventing exact monitoring of the same population of cells over time. In these 4 assays, the 168 evenly distributed images gathered for each data point corresponded to sampling of 22.5 % of the T-25 surface.

In the automated assays, four T-12 flasks were filled with 3.5 ml of culture, set on the stage, and slow scans automatically taken at 1 hr intervals. In these 4 assays, 84 images (evenly distributed) were gathered for each data point, 22.5 % of the T-12 surface being sampled.

The same pattern of proliferation, over time and as a function of oxygen concentration, was observed using both techniques. Fig. 3.1 is a compilation of both manual and automated results.

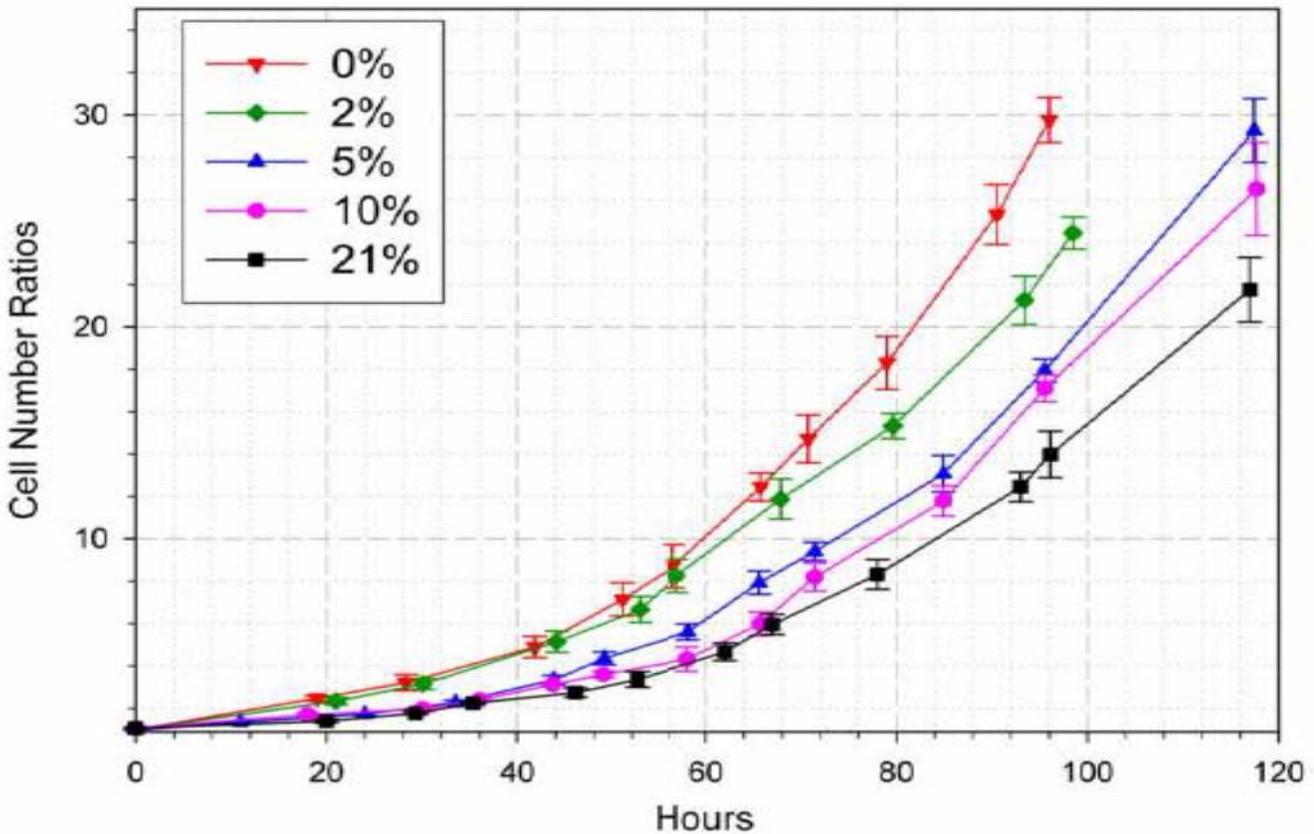

**Fig. 3.1: Classical Cell counts on K562 under various levels of Oxygen. Bars represent ± 95 % CI. The result is a compilation of both manual and automated results.**

The average growth rate (% per hour) based on the curves in Fig. 3.1 are 4.2 % at 0 % oxygen, 4 % at 2 %, 3.71 % at 5 %, 3.61 % at 10 % and 3.48 % under 21% oxygen. K562 cells grow fastest under 0 % oxygen. Our confidence in these results is high because of the data gathering power of joint image analysis and automation. Oxygen from anoxic to atmoxic concentrations is therefore confirmed as smoothly suppressing cell growth (see also Table 3).

## Oxygen Diffusion to Cell level

The diffusion of oxygen in culture medium is limited, so the bottom of a culture flask becomes hypoxic if the cell density is high (Griffith and Swartz, 2006). Because the cell division rate is dependent on oxygen concentration (Fig. 3.1), the cell proliferation rate observed at a given moment depends on gaseous oxygen concentration, medium depth and cell density. Cell-level oxygen concentration in a culture flask can be estimated from a diffusion model based on Fick's First Law, using as parameters oxygen consumption of K562 ($J$), diffusion ($D$) of oxygen in culture medium, 2 x 10$^{-5}$ cm$^2$/s, and the equation

$$J = -D\frac{\partial \Phi}{\partial x}$$

where $\Phi$ is oxygen concentration and $x$ is depth within the culture medium.

The rate of oxygen consumption for K562 ($J$) varies according to different authors, but a single representative value measured under atmoxic conditions would be 28 ng of oxygen / min / 10$^6$ cell (Denis-Gay et al., 1998; Carré et al., 1999;  De Oliveira et al., 2006). The oxygen consumption rate estimate used a cell density of 2 x 10$^5$ cells/cm$^2$, oxygen saturation of 5 ppm at the top of the liquid interface, and culture medium depth of 3 mm.

Since cell proliferation itself alters oxygen levels at the bottom of the culture dish, the cell variables influenced by oxygen will also be altered in the natural course of culture maturation.

## Early and Late Doubling Frequencies

Table 3 goes further than Fig. 3.1 in distinguishing between Doubling Frequency measured "Early" (nominal oxygen condition) and "Late" in the assays (88 % of oxygen remaining). As expected, it shows a stable Doubling Frequency between Early (15.3) and Late (15.24) times under anoxic conditions, but all oxic assays (2 to 21 % oxygen) increase Doubling Frequency by an average of 4.95 % late in the assay, compared to the beginning. The effect is thought to be due to oxygen depletion at the bottom of the flasks, as modeled from the first to the blue columns of Table 3, and this acceleration happens despite the inevitable reduction in the medium's nutrients.

| Nominal Oxygen Concentration (%) | Early[1] Doubling Time ± 95 % CI (hrs) | Early[1] Doubling Frequency ± 95 % CI (µHz) | Late[2] Doubling Frequency ± 95 % CI (µHz) | Late[2] Estimated Oxygen Concentration, bottom of T-25 (%) | Crest Doubling Frequency ± 95 % CI (µHz) | Time Interval used for Crest Doubling Frequency (hrs) |
|---|---|---|---|---|---|---|
| 0 | 18.49 ±1.86 | 15.3 ±1.62 | 15.24 ±1.07 | 0 | 17.58 ±1.94 | 0 to 30 |
| 2 | 20.33 ±2.60 | 14.1 ±1.44 | 14.53 ±1.4 | 1.76 | 15.67± 2.10 | 0 to 40 |
| 5 | 21.70 ±2.45 | 12.87 ±1.25 | 13.00 ±0.92 | 4.4 | 14.63 ±2.06 | 11 to 70 |
| 10 | 24.51 ±2.92 | 11.42 ±1.23 | 11.93 ±1.12 | 8.8 | 13.33 ±1.65 | 24 to 90 |
| 21 | 27.04 ±2.23 | 10.30 ±0.83 | 10.99 ±0.87 | 18.5 | 12.09 ±1.38 | 29 to 96 |
| p –value 0 %-5 % | < 0.001 | < 0.001 | < 0.001 | | < 0.0001 | |

**Table 3. Oxygen Concentration alters Doubling Frequency and Crest Doubling Frequency.**

**1** Early: Cells are less than 12 % confluent (cells occupy less than 12 % of the image's surface), maintaining oxygen levels within 4 % of gas/liquid interface level: from 0 to 50 hours for 0, 2, 5 % oxygen, from 0 to 60 hours for 10 and 21 % oxygen.
**2** Late: 40 % surface coverage by the cells corresponds to 88 % of the oxygen at the gas interface available at the bottom of the flask: 70 hours for 0 % and 2 %, 80 hours for 5 %, 90 hours for 10 % and 95 hours for 21 %.

This data suggests that the acceleration of cell division over time in K562 cultures can be explained by oxygen depletion. K562 oxygen consumption ($J$) is only known as a single, probably atmoxic value (depending on cell density and medium depth), and it is likely that a more precise definition of this variable as a function of oxygen concentration would improve the match between observations and computations.

## Crest Doubling Frequency

The fastest Doubling Frequency within the evolution of a cell culture and for a particular oxygen concentration may carry the most physiological meaning. To obtain the *Crest Doubling Frequency*, a time range is chosen for each oxygen concentration, as shown in the last column of

Table 3. This time interval is early enough so that neither nutrients nor oxygen at cell level are substantially depleted, but late enough to avoid the culture's initial lag phase, if any (more on lag phase below).

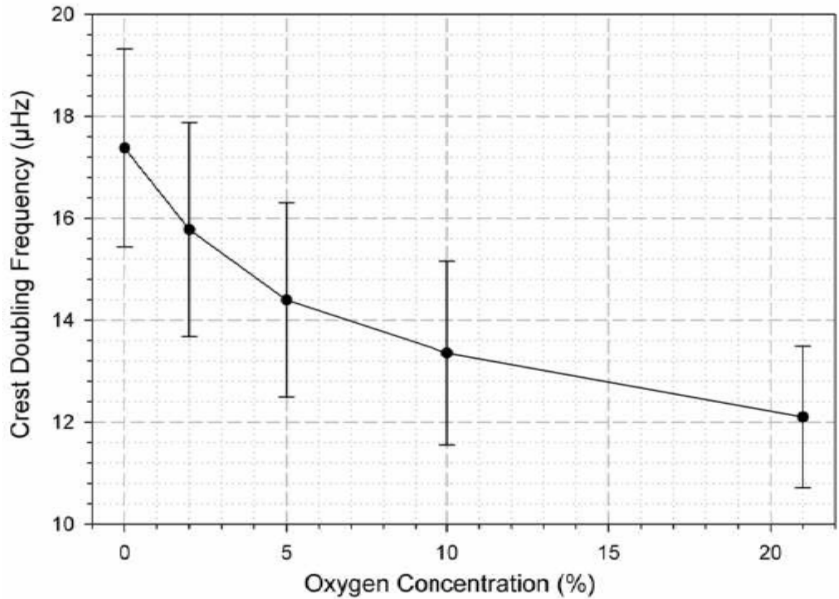

**Fig. 3.2:** *Crest Doubling Frequency* **reduces as Oxygen increases. Bars represent ± 95 % CI of 8 tests. The data dispersion is largely attributable to the initial state of the cell cultures used in independent experiments.**

The simple relation between oxygen level and *Crest Doubling Frequency* is illustrated in Fig. 3.2. The data dispersion is largely attributable to the initial state of the cell cultures used in independent experiments. In individual tests, the *shape* of the *Crest Doubling Frequency* curve as a function of oxygen concentration (Fig. 3.2) is repeatable, as individual curves simply shift vertically as a whole.

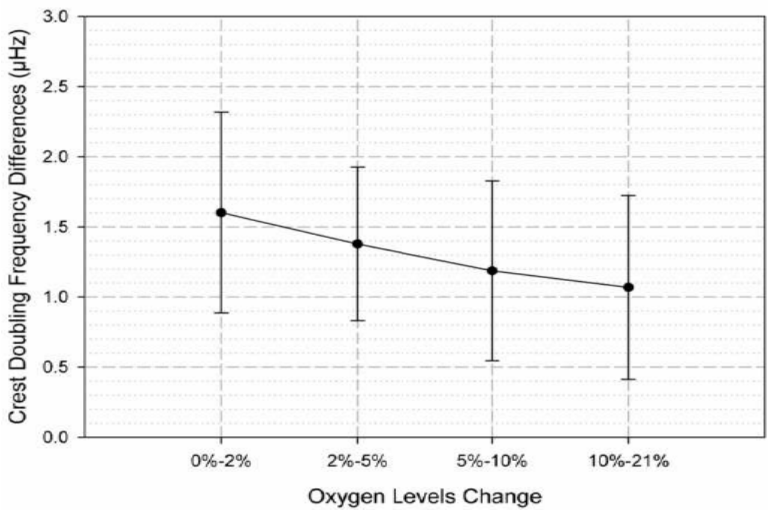

**Fig. 3.3:** *Crest Doubling Frequency* **differences between adjacent Oxygen concentrations (0-2 %, 2-5 %, 5-10 %, 10-21 %) in simultaneous tests, based on 8 tests. Bars represent ± 95 % CI of 8 tests. Differences in** *Crest Doubling Frequency* **between adjacent oxygen concentrations are always positive, confirming that the** *Crest Doubling Frequency* **varies with oxygen concentration as depicted in Fig. 3.2.**

Fig. 3.3, which illustrates the differences in *Crest Doubling Frequency* between adjacent oxygen concentrations within a given test, confirms that the *Crest Doubling Frequency* always varies with oxygen concentration as depicted in Fig. 3.2 (individual curves follow the same shape). The *Crest Doubling Frequency* curves of Figs. 3.2 and 3.3 however integrate important phenomena that occur within the cell culture, and which become apparent if tighter measurements over time are performed. These measurements reveal: (1) a lag phase in cell proliferation on the occasion of any oxidation level change, (2) rapid variations in apparent proliferation rate due to periodic bursts of apoptosis, depending on ROS levels, and (3) among closely spaced groups of cells, cell division bursts that are more or less time-wise spaced apart, again according to ROS level.

## 3.1.2 Reducing Oxygen Changes Cell Appearance and Macrophage Activity

The mutipotential K562 cell line can express erythroid, macrophage and megakaryocyte lineages (Sutherland et al., 1986). Computer-based recognition allows the monitoring of large numbers of objects, and small components of cell populations can be monitored, as long as they can be effectively discriminated. In our tests, objects were assigned to two categories, based on size. Objects between 50 and 300 µm² are classified as "normal cells". This is a reasonable range for a cell of variable karyotype, considering that normal cells usually display a range of 2 (Conlon and Raff, 2003). Objects between 300 and 600 µm² are recognized as macrophage-like. Objects larger than 600 µm² are clumped necrobodies or dead macrophages.

Cell *Roundness* is commonly used as a subjective index of cell health by suspension cell culture technicians, a low value of *Roundness* (good sphericity) indicating rapidly growing cells, and so a healthy culture. Most cell biologists believe that cell membrane synthesis is at least partly determined by intra-cellular pressure (Nierras and Warner, 1999), potentially linking a smooth membrane to rapid cell growth. Whatever the case, a benign and physiological environment is conducive to low *Roundness*. A second group of influences on *Roundness* relate to the sensory or exploratory functions of the cell in relation to its surroundings. For macrophages, this expresses as chemotaxis. It involves multiple focal filopodia contacts on the substrate, coordinated with cytoskeleton and F-actin formation (Kim et. al., 2007). Increased perimeter is related to invasiveness in normal and malignant cells (MacDougall and Kerbel, 1995).

## Reducing Oxygen reduces Cell and Macrophages Footprints

Table 4 compiles *Roundness*, *Footprint* and proportion of macrophage*s* at a cell density of 30,000 cells/cm², but also discriminates for object sub-populations. At each oxygen level, 8 assays (4 manual assays and 4 automatic assays) measured 10,000 objects. The perturbation of re-seeding (5 minutes at 420 g, renewal of the supernatant, and seeding) as well as oxygen depletion in the aerobic cultures were avoided by sampling at 40 to 60 hours post-seeding, depending on cell density.

| % Oxygen | All Objects | | Cells ( 50-300 μm²) | | Macrophages ( 300-600 μm²) | | % Macrophages ± 95 % CI |
|---|---|---|---|---|---|---|---|
| | Roundness ± 95 % CI | *Footprint* ± 95 % CI (μm²) | Roundness ± 95 % CI | *Footprint* ± 95 % CI (μm²) | Roundness ± 95 % CI | *Footprint* ± 95 % CI (μm²) | |
| 0 | **1.52** ± 0.13 | *158* ± *24.2* | **1.41** ± 0.17 | *130* ± *22.2* | **3.29** ± 1.06 | *415* ± *37.1* | 6.85 ± 2.3 |
| 2 | **1.54** ± 0.09 | *166* ± *21.9* | **1.39** ± 0.12 | *133* ± *23.4* | **2.88** ± 0.91 | *422* ± *45.7* | 9.62 ± 5.2 |
| 5 | **1.61** ± 0.12 | *169* ± *22.3* | **1.40** ± 0.17 | *135* ± *19.6* | **2.69** ± 0.81 | *423* ± *34.7* | 11.3 ± 3.4 |
| 21 | **1.73** ± 0.09 | *186* ± *18.3* | **1.47** ± 0.26 | *137* ± *23.7* | **2.72** ± 0.83 | *443* ± *17.0* | 15.93 ± 5.12 |
| p –value 0%-21% | < 0.01 | *< 0.05* | 0.44 | *0.38* | 0.12 | *< 0.05* | < 0.01 |

**Table 4. Average *Roundness* and *Footprint* at various levels of Oxygen.**

The "All Objects" columns of Table 4 show that reducing oxygen makes the detected objects rounder and the *Footprint* smaller. When discriminated according to size as "Cells" or "Macrophages", it can be seen that "Cells" *Roundness* or *Footprint* are fairly indifferent to oxygen, but that "Macrophages" have higher *Roundness* and smaller *Footprint* at low oxygen. "Macrophages" have an irregular shape (high *Roundness*, see Fig. 3.5) when they are actively chemotaxic. The higher *Roundness* of "Macrophages" under 0 % oxygen can be interpreted as higher chemotaxic activity. The differences in adjacent *Roundness* and *Footprint* values as a function of oxygen concentration reported in Table 4 are often small, with overlapping CIs (for example, the "All Objects" *Roundness* series 1.52-1.54-1.61-1.73). The CIs reported in Table 4 are computed for tests widely separated in time. In a specific test, 4 culture aliquots are exposed to 4 oxygen concentrations simultaneously in the same incubator and with the same equipment. We confirm that the tendencies reported in Table 4 for *Roundness* and in *Footprint* are reproduced in individual assays. The CIs of Table 4 reflect slight differences in the state of the root cultures used to make the 4 aliquots, and these differences affect all oxygen levels of a given assay, similar to the situation of Figs. 3.2 and 3.3. It is notable that the tendency for "Macrophage" *Roundness* as a function of oxygen concentration (- 14 %) is reversed compared to "Cells" *Roundness* (+ 4.3 %).

In Table 4, the average anoxic *Footprint* for "All Objects" is 158 μm², compared to 186 μm² at 21 % oxygen (+ 18 %, $p < 0.05$). "Cell" Footprints rise slightly between 0 % and 21 % oxygen (+ 5 %), and "Macrophage" Footprints increase from 415 μm² to 443 μm² (+ 7 %, $p < 0.05$).

## Reducing Oxygen reduces Macrophage numbers

As can be seen in Fig. 3.4, a histogram of cell *Footprints* under various levels of oxygen, higher oxygen levels stimulate the appearance of larger objects exhibiting macrophage-like size. The vertical log scale allows visualization of changes in the number of these larger objects as a function of oxygen level. The cells analyzed in Fig. 3.4 have been cultivated over 3 passages under their specific oxygen concentrations, and their histogram is assumed to be in steady state. The data is based on 3 separate assays, and compiles 4,300,000 objects.

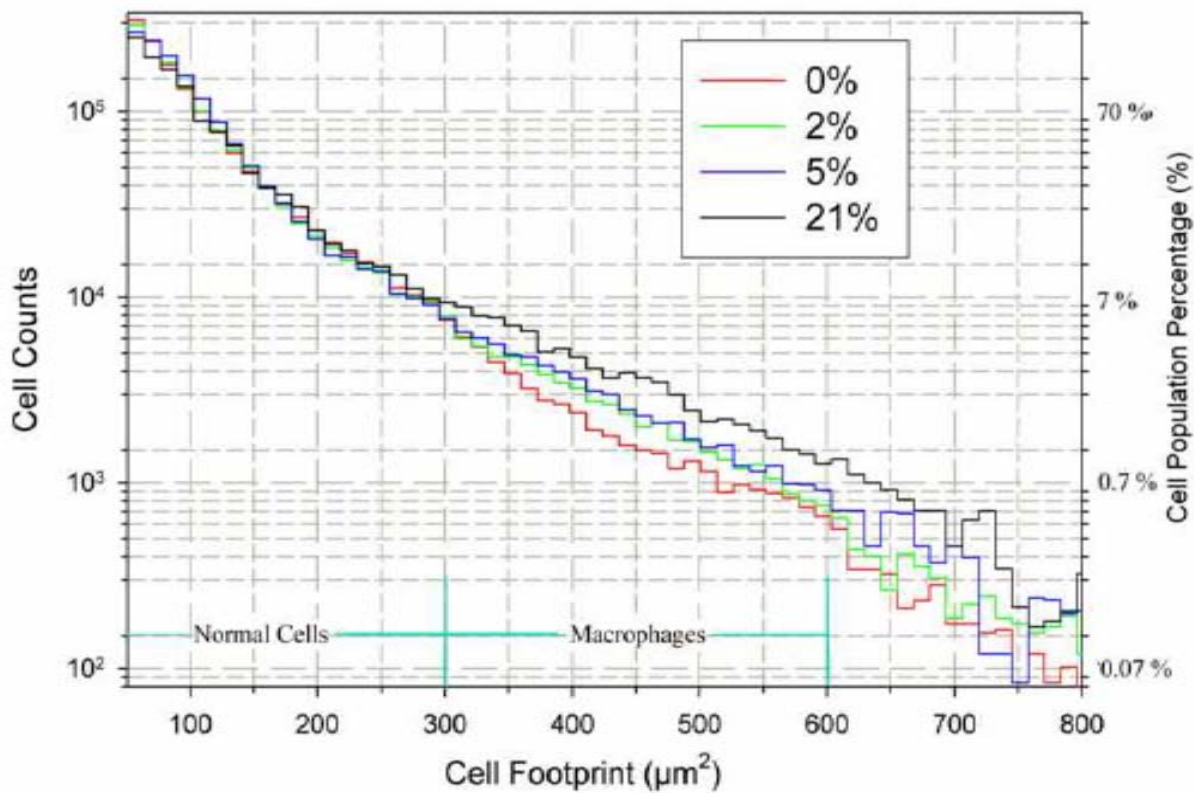

**Fig. 3.4: Histogram of object Footprints as a function of Oxygen level (1,430,000 cells/curve). Data averages 3 sets of experiments. Higher oxygen levels stimulate the appearance of larger objects exhibiting macrophage-like size. 50 to 300 µm² are "normal cells", and 300 to 600 µm² are macrophage-like. 0 % oxygen curve has 5.71 % of macrophage-like counts, 2 % has 7.29 %, 5 % has 8.78 % and 21 % has 11.37 %.**

Over approximately 1,430,000 objects for each histogram, there are 81,653 macrophage-like objects in 0 % oxygen and 162,591 in 21 % oxygen: macrophage-like objects are 99 % more numerous in 21 % oxygen than in 0 % oxygen. In the last column of Table 4, it can be seen that oxygen increases significantly the number of macrophages, an up-regulation of K562's macrophage phenotype.

## Reducing Oxygen increases Macrophage Chemotaxis

The most rapid effect of oxygen we could display is the initial stages of an inflammatory response. Because our techniques gather large number of images over time, it is relatively easy to verify from time-series visual observations on normal or oxic transition assays (assays where cells are taken from one oxygen level to another) that the macrophage sub-population has a more turbulent and short life than the normal cell population. Macrophage life is essentially without reproduction, presents frequent engulfment of particles or other cells, and finally apoptosis. Our data is also compatible with macrophage-like objects being mechanically more fragile (Baba et al., 1991) and likely to be destroyed in a re-seeding process, as larger cells usually are.

We compiled changes in object *Roundness* with level of oxygen in K562 in Table 4. *Roundness* values, the ratio of the perimeter of a cell to that of a circle of identical surface, increase with oxygen concentration, from 1.52 (0 %) to 1.73 (21 %) for the mean and from 1.20 (0 %) to 1.60 (21 %) for the median. The confidence intervals (CIs) for 8 assays are shown in Table 4.

Particularly interesting data can be generated by monitoring the *Roundness* of K562 over time, as the percentage of oxygen overlying the culture is changed. The near edge of the graph in Fig. 3.5 represents a stable *anoxic* culture. Near-circular cells (low *Roundness,* at left) are the most numerous, and the more irregular shapes (high *Roundness,* at right) become monotonously less frequent. At time zero of the figure, the anoxic gas of the culture is replaced by *atmoxic* gas.

Within as little as 4 hours, there is generation of a population of cells at *Roundness* equal to "2",

which solidifies in the 40 hours documented by the graph.

About 30 hours are necessary for *Roundness* adaptations to a new oxygen level to take place.

The R2 peak corresponds to a highly probable morphology of the macrophage population described in Fig. 3.5. The R2 peak is smallest in unstressed, healthily growing anoxic cells. It is expressed under higher oxygen levels, as shown in Fig.3.5, and in any oxic transitions. R2 may be an indicator of culture stress.

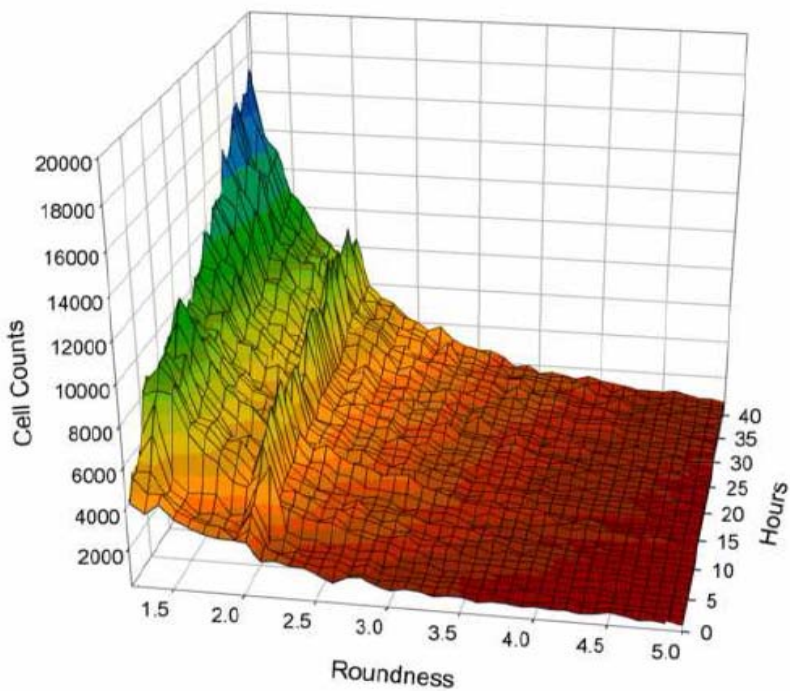

**Fig. 3.5: Rapid inflammatory response to oxygen in K562 is illustrated by time-series histograms of cell shape quantified as *Roundness*. The anoxic gas phase of a continuously monitored cell culture is replaced by atmoxic gas at time zero, triggering the appearance of the "Roundness 2" peak in this 3D mesh plot. The total cell number climbed from 22,000 to 46,000 from the near to the far edge of the graph.**

There is a known molecular mechanism to justify such a fast reaction. In many cells, including

K562, there are specific amino acids in a latent complex, LTGF-β, that allows release of TGF-

β in response to ROS (Barcellos-Hoff and Dix, 1996). TGF-β action has been implicated in a variety of reactive oxygen-mediated tissue processes, particularly inflammation, and in pathologies such as reperfusion injury, rheumatoid arthritis, and atherosclerosis. Macrophages and monocytes respond to low levels of TGF-β in a chemotactic manner (Letterio and Roberts, 1998), explaining the rapid changes displayed in Fig. 3.5.

The inflammatory reaction to ROS displayed in Fig. 3.5 is sustained indefinitely. As well, in cultures chronically maintained at 0, 2, 5, 10 and 21 % oxygen, the proportion of macrophage-like cells (larger than 300 μm²) increases steadily, as shown in the 6th column of Table 4. ROS therefore trigger the appearance of irregular cell shapes (Roundness 2) as in Fig. 3.5, but also of a small population of larger, macrophage-like cells, both of which can be interpreted as early signs of inflammation.

### 3.1.3 Reducing Oxygen Generates more Cell Debris

Routinely, some cells undergo apoptosis or necrosis, introducing debris in the medium. The number of apobodies and necrobodies (the proceeds of apoptosis and necrosis) present in the medium results from a competition between cell turnover or decay, and phagocytosis. These smaller objects, measured in Fig. 3.6, are rarely documented in cell studies, but can conveniently be counted using image analysis. In our high resolution imaging of a single test, we categorized "cells" as objects above 51.3 μm, "apobodies" as objects between 33.4 and 51.3 μm, and "necrobodies" as objects between 12.8 and 33.4 μm. In this T-12 experiment, cells cultured at 0 % and 21 % oxygen concentration were centrifuged and the supernatant renewed. Re-seeding was at 10,000 cells/cm², in the same oxygen concentration. Over 90 hours, 22.5 % of the T-12's surface (84 images) was sampled hourly, a range of 27,000 to 120,000 cells.

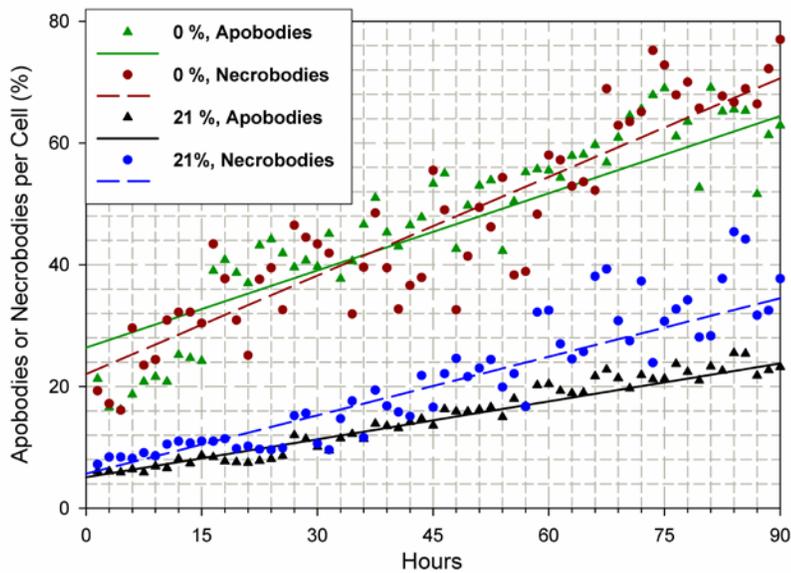

Fig. 3.6: Apobodies and Necrobodies per Cell (%) in K562 cultures under 0 % and 21 % Oxygen. 0 % oxygen (top curves) shows more apobodies and necrobodies than 21 % oxygen (bottom curves).

The readings at the left of Fig. 3.6 start at low values because of the fresh (clean) seeding medium. But the 0 % oxygen condition creates more Apo- and Necrobodies than 21 %, so values start higher after the centrifugation procedure, and continue to diverge upwards from the 21 % debris ratios thereafter. The data of the two curves shows that debris is more common under anoxia than under 21 % oxygen. We have previously shown that anoxic conditions reduce macrophage numbers but increase macrophage activity (Table 4). With the data of Fig. 3.6, we have a picture under anoxia of fewer macrophages (x 0.43) that are more active (*Roundness* changes from 2.72 to 3.29) resulting ultimately in more debris in the medium.

Future investigations may be able to confirm that anoxia provides higher cell turnover (more apobodies, Fig. 3.6) and more discriminating immunity that would be blunted by ROS under more oxic conditions. It may also be that anoxic cells concentrate their resources on heightened proliferation (Fig. 3.1), and reduce the number of phagocytes (Table 4) in what is perceived as a less threatening (oxidizing) environment (Nagai et al., 2004).

Apoptosis was also measured in image series (Fig 3.7) using visual detection ("manual") or custom software (Héroux et al., 2004). The manual measurement shown below includes manipulation of the various culture vessels by the experimenter, as successive placements on the scanning system, and visual rather than software recognition of apoptoses. Automated measurements involve only movement by the motorized stage, but rely on software recognition of groups of apobodies. Because software is more easily confused by complex images than the brain, the automated measurement can only reliably cover about 10 hours of the experiment, as compared to the 50 hours in the manual measurement, when the images also becomes too cluttered even for visual recognition.

Apoptosis is a very delicate variable that is quite sensitive to initial culture state. In spite of this, a hierarchy is consistent between manual and automated measurements. Apoptosis is strongly inhibited by atmoxia, to typical levels of only 0.01 %/hour. The anoxic culture shows slightly higher rates. The transition tests show rates as high as 0.5 % per hour, seemingly increasing with the size of the oxic transition step.

Rises in apoptosis rates coincide with induction of karyotype changes. Inhibition of apoptosis by oxygen in our results bring to mind the "Warburg hypothesis", which posits that mitochondria, the initiating site of apoptosis, in cancer cells are shut down, and would need to restart their apoptosis program for a tumor to regress (Bonnet et al., 2007). The sensing of ROS concentration by mitochondria is probably achieved by the voltage gated family of ROS-$K^+$ channels (Michelakis et al., 2004).

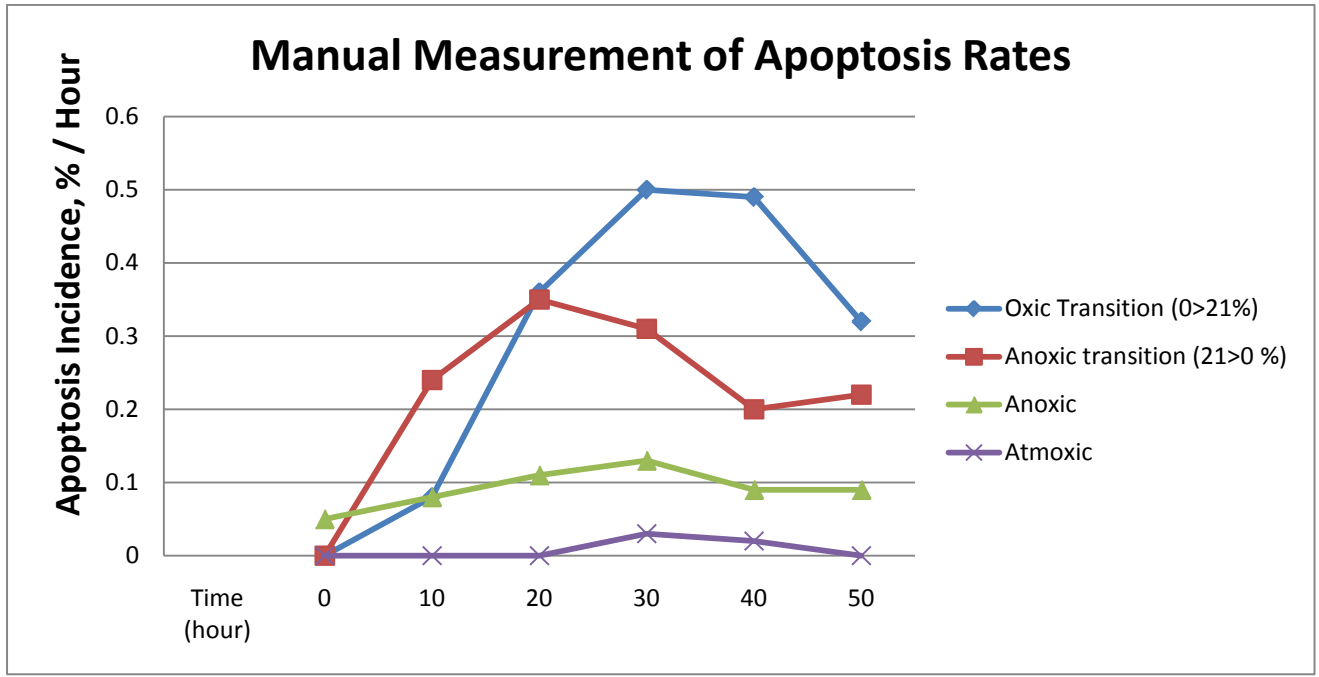

**Fig. 3.7: Manual Measurements of Apoptosis Rates in K562 cultures.**

## 3.1.4 Reducing Oxygen Increases Cell Adhesion

Another cell behavior affected by oxygen is adhesion. Because cell adhesion is lowered late in CML, allowing metastatic release, this variable is particularly relevant to cancer progression. The oxygen gradients found in tumors should support the release of metastatic cells on their periphery, while stabilizing the implantation of metastatic cells in hypoxic environments.

In atmoxic culture, K562 exhibits much less clumping than many other suspension cell lines, perhaps due to down regulation of surface adhesion molecules by the BCR:ABL protein (Shet et. al, 2002). Conversely, hypoxia is reported to suppress the BCR:ABL-dependent leukemogenic signals (Desplat et. al, 2002), allowing expression of adhesion molecules.

Centrifugation for 3 minutes at 250 g, renewal of the supernatant, and seeding of a suspension should produce a random distribution of single K562 cells on a T-12's surface, but we detected early in our assays that growth patterns differed according to oxygen level. Our visual impression was confirmed by the *Hex-Distance* data below.

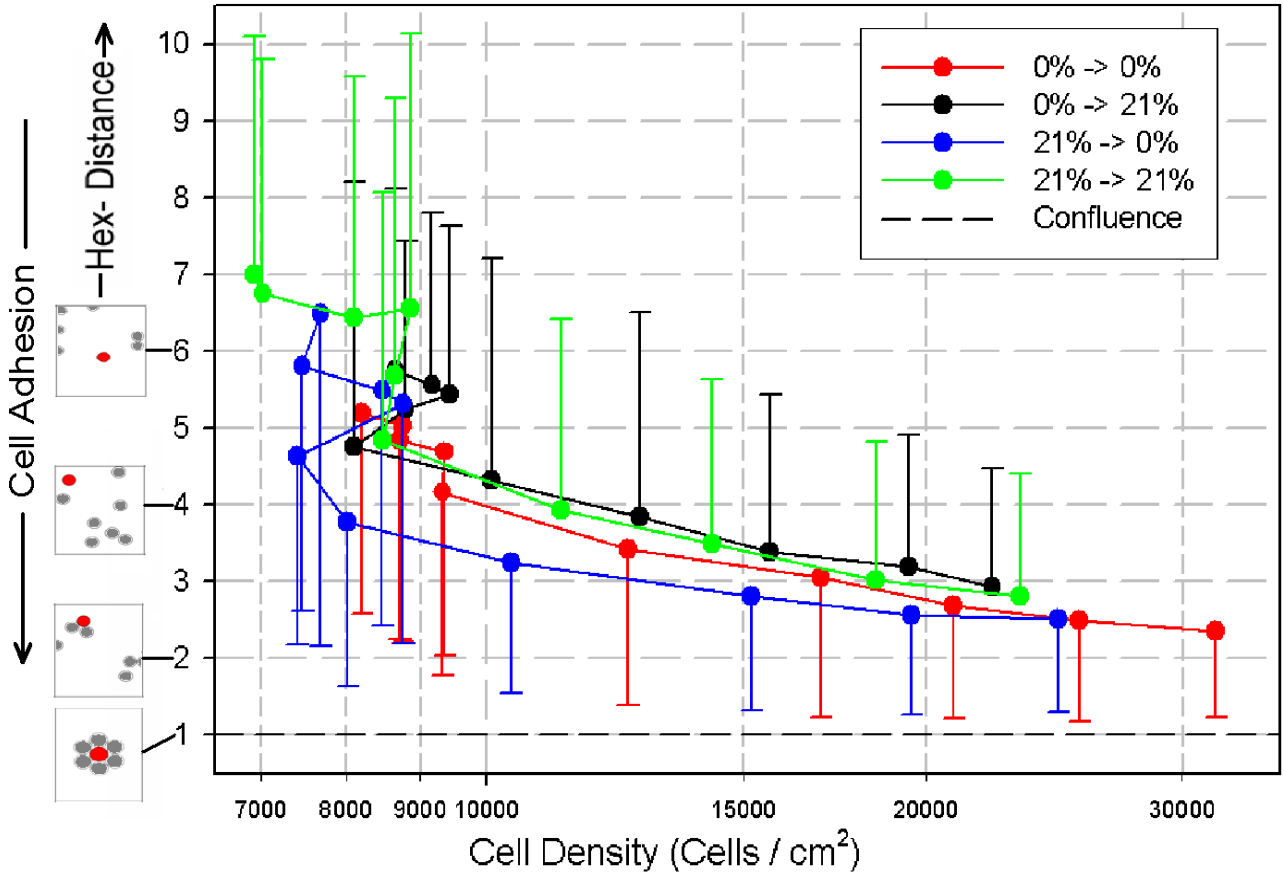

**Fig. 3.8: K562's enhanced clustering under anoxia, a sign of reduced metastatic potential.** *Hex-distance* histograms are compiled at 1, 4, 7, 10, 18, 29, 40, 51, 62, 73 hours. Each curve shows averages ± 1 σ of Hex-*distance* for 5,000 to 32,000 cells. Among the different oxic environments in the legend, cyto-adhesion is strongest for anoxic transition (blue) and anoxic (red) cells, while atmoxic cells are more easily shed.

Looking at the vertical axis at left of Fig. 3.8, one can see thumbnails corresponding to specific values of *Hex-distance*, which is the average distance between a cell (red) and its 6 closest neighbors (gray), expressed in cell diameters. If every cell in a culture had a *Hex-Distance* of 1, this would corresponds to complete cell coverage of the culture surface. When seeded randomly, *Hex-distance* distributions show values compatible with their density (*Hex-distance* of 5 to 7 corresponds to 10,000 cells/cm$^2$).

*Hex-Distance* distributions and how *Hex-distance* values decay towards 1 (= confluence) as cell densities increase over time, quantify the tendency of cells to stick together (lower average *Hex-Distance*) rather than disperse (higher average *Hex-Distance*). To allow a fair comparison of cell clustering between experiments, *Hex-Distance* distributions must be compared against a horizontal axis of cell density rather than time, as is done in Fig. 3.8.

We investigated K562 adhesion in quadruple experiments where anoxic and atmoxic cultures were simultaneously passaged into identical or opposite oxygen environments. The four *Hex-Distance* curves in Fig. 3.8 stop at the right of the graph at the same time (72.5 hours), but at different cell densities, because of different doubling frequencies at different oxygen levels. All curves converge to the lower right of the graph, as cultures become more confluent. The black and blue curves describe *transition assays* in that cells cultured in one oxygen concentration are transferred to a different one, rather than to the same concentration (red and green).

All curves show a change in horizontal direction at the left of the graph. The direction changes correspond to "lags" in the proliferation of cells in the early parts of the curves.

Both the black and blue oxygen *transition* assays exhibit a transient regression in cell numbers at 17.5 hours following the oxygen change, and it takes some time for the new cytoadhesion to establish itself (black curve joining the green at 17.5 hrs, or blue joining the red at 61.5 hrs).

But even the red and green curves (no oxygen transition) experience small proliferation lags around 17.5 hours. The 21 % > 21 % assay (green) is actually a small hyperoxic jump, since the 4-day-old 21 % culture serving to seed the second 21 % culture at 10,000 cells/cm$^2$ is actually somewhat depleted in oxygen (and nutrients). As well, the 0 % > 0 % (red) curve is transiently enriched by the oxygen dissolved in the new medium, which was stored in equilibrium with air. Therefore, all the proliferation lags observed can be associated with oxic transitions.

Concentrating now in Fig. 3.8 to the right of the lag phase due to seeding (10,000 cells/cm$^2$), we see that low oxygen curves (blue and red) have smaller *Hex-Distances* than high oxygen curves (black and green), indicating that the hypoxic cells are more cytoadhesive.

It seems that a hypoxic transition (blue) is most cytoadhesive. This means that atmoxic cells, which are thought to represent a later (metastatic) stage in the evolution of K562, still retain the ability to express self-adhesion when oxygen is withdrawn.

The overall observation is that K562 cells are more cytoadhesive in anoxic culture, and are delayed in proliferation ("lag phase") by any oxic change. Consequently, cells harvested early in a culture, compared to cells harvested late, would not initially proliferate at the same apparent speed when seeded. Because of their habituation to specific oxygen levels, they would experience different lags.

## 3.1.5 Phenotype Summary

The phenotype measurements above converge to present a picture of anoxic K562 as a more dynamic (increased proliferation), stable (less inflammation) and cohesive (more self-adhesion) tissue compared to other levels of oxygen. Table 4 provides specific numerical values to support these assertions, and adds the following elements expressed under anoxia: smoother cell borders (column 4), smaller cell footprints (column 5) and reduced macrophage numbers (column 6) of increased activity (column 7). These changes occur within tens of hours of any oxygen adjustment. Our data suggests a more ordered K562 anoxic phenotype compatible with tissue formation.

Our interest in the anoxic cells is high because they have a different phenotype, represent an extreme case of inhibition of the mitochondrial oxidative metabolism associated with cancer, and also because very little work has been performed on this subject.

Our tests performed at numerous oxygen levels confirmed that the reduction in cell size was a trait generally associated with oxygen removal. The reduced cell sizes caused by anoxia lead us to suspect that karyotype changes may also occur during oxygen withdrawal, as cell size is often connected with chromosome count, particularly in plants.

# 3.2 Karyotype Changes under Anoxia

Does anoxia actually roll back K562, a cell harvested in a late phase of CML, to an earlier stage of the disease? Would a transition from anoxia to atmoxia parallel the transformation witnessed by Lozzio as K562 was stabilized in atmoxic culture? Perhaps more pointedly, are these changes effects of oxygen on BCR:ABL and tyrosine kinase, or an independent effect of ROS on K562 expression? Furthermore, do all these changes apply to other cancer cells? From our observations of the K562 cell phenotype under anoxia, we suspected that the phenotype alterations were shadowed by chromosome changes, and that the cells' more stable phenotype may be mirrored in a karyotype of superior stability.

## 3.2.1 K562 Chromosomes Change in Anoxic Transition

### Anoxic Karyotype

In an attempt to answer these questions, we documented the anoxic and atmoxic karyotypes of K562. Kayotypes are usually established from 50 metaphases or less, but in Fig. 3.9 we compiled 108 and 122 to solidify the clear separation observed between the two karyo-histograms, a modal loss of 7 chromosomes from atmoxia to anoxia. Our anoxic karyo-histogram (blue) is uniquely narrow compared with that of any tumor cultured under atmoxic conditions (Spriggs et. al, 1962), or even of normal human tissue, such as the brain (Yurov et. al, 2007), implying unusual stability.

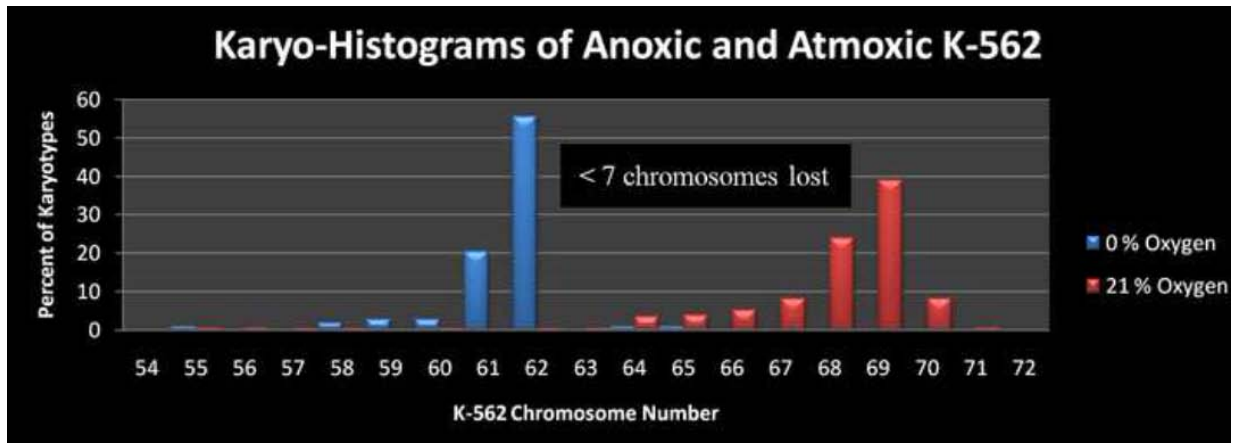

**Fig. 3.9: The anoxic histogram (blue) is averaged from our data and from 3 repeats by *Plate-Forme de Cytogénétique* of Maisonneuve-Rosemont hospital (108 metaphases). The atmoxic histogram (red) is averaged from our data.  Similar atmoxic data is published by ATCC and Dimery (1983).**

Since elimination of oxygen caused chromosome losses in Fig. 3.9, could pleomorphisms and spreads in K562 karyotypes reported by many observers (Dimery et. al, 1983) be explained by oxygen gradients? Oxygen diffuses poorly in culture media (Griffith and Swartz, 2006), so concentrations are lowered at the bottom of a dish (Table 3, column 3), depending on depth and cell density. Cell growth patterns should also influence karyo-histograms, since oxygen concentration at individual cells depends on a cell's position within a cluster (Mathur et al., 2010). Growth-pattern hypoxia is likely responsible for the relatively large lower-side tail at 64 to 67 in the 21 % oxygen distribution of Fig. 3.9.  As would be expected, this lower-side tail is much reduced when oxygen is completely absent (blue in Fig. 3.9). This narrow spread, and the substantial anoxic-atmoxic gap, make K562 an ideal model to test for mechanisms of karyotype transitions induced by oxic changes (Fig. 3.11).

In the case of K562, important karyotype modes across the range of oxygen concentrations are 45-46, 55, 62 and 69. The reduced chromosome count under observed under anoxia is compatible with a lower doubling time, as it takes more time to copy more chromosomes.

## SKY and FISH Studies of Anoxic K562

To document the nature of the chromosome changes associated with anoxia in K562, Spectral Karyotyping and Fluorescence In-Situ Hybridization of 20 anoxic K562 cells was performed. We had questions about the survival of the BCR:ABL and tyrosine kinase lesion in the majority of the anoxic cells, and on the selection of the chromosomes eliminated by anoxia. We also wanted to determine whether the orderly phenotype and narrow karyo-histogram of anoxic K562 had a correspondence in the details of its karyotype.

The 62-chromosome anoxic clone was studied using Spectral Karyotyping and Fluorescence In-Situ Hybridization (Fig. 3.10A). All human chromosomes are represented in the anoxic karyotype. Chromosome 14 shows a Robertsonian translocation, common in Acute Myeloid Leukemia and CML. The lost satellite of chromosome 14 is redundant with microsatellites of 13, 15, 21 and 22. There is also a 20 % incidence of a normal chromosome 14 in the culture. A single normal chromosome 17, reported by Lozzio (Lozzio and Lozzio, 1975), is conserved in both atmoxic and anoxic cultures.

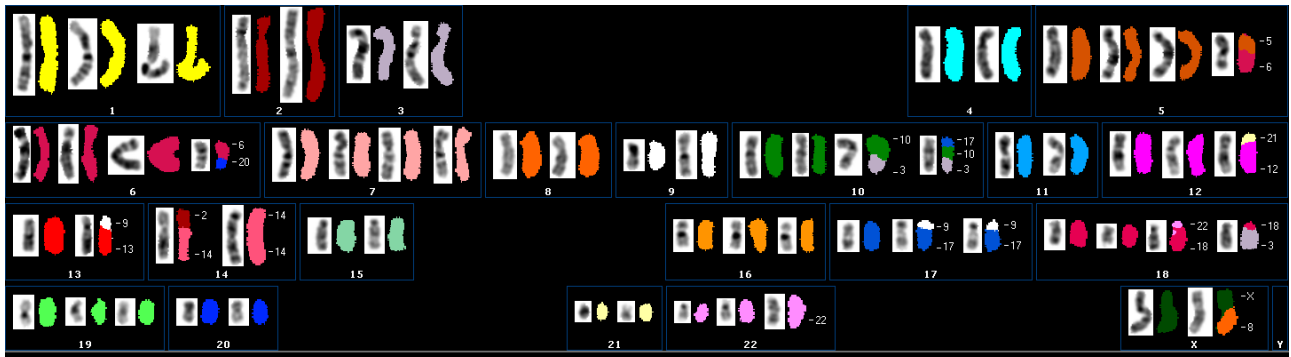

**Fig. 3.10A:** G-banding and SKY results from 20 anoxic K562 cells, detected in all but 4 (absence of i(14)(q10): normal chromosome 14). Chromosomal abnormalities are related to the 3n (69 chromosomes) level, according to ISCN 2009: -2, dup(2)(q?): duplicated long arm segment localized on p, -3, -4, +der(5)t(5;6)(q?;p?), dup(6)(p?), +der(6)t(6;20), +inv(7)(p?p?), -8, -9, del(9)(p12), dup(9)(q?): duplicated long arm segment localized on 9p, der(10)t(3;10), +der(10)t(3;10;17), -11, der(12)t(12;21), -13, der(13)t(9;13), -14, der(14)t(2;14), i(14)(q10), -15,der(17)t(9;17)x2, ?del(18)(q?), der(18)t(18;22), der(18)t(3;18), -20, -21, dup(22q?q?), -X, der(X)t(X;8), dup(X)(q?): duplicated long arm segment localized on Xp.

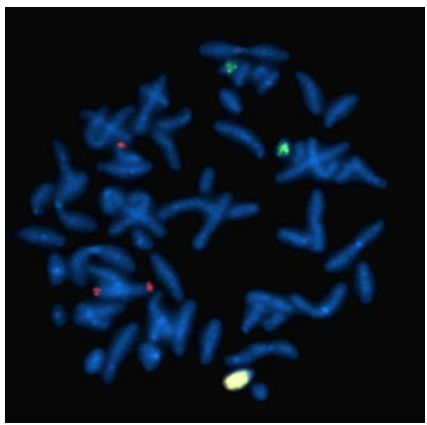

**FISH results.** BCR (green)-ABL (orange)-ES translocation probe results. The fused green-orange (white spot) reveals the Philadelphia chromosome, detected in 95 % of CMLs. From Plate-Forme de Cytogénétique, Maisonneuve-Rosemont Hospital.

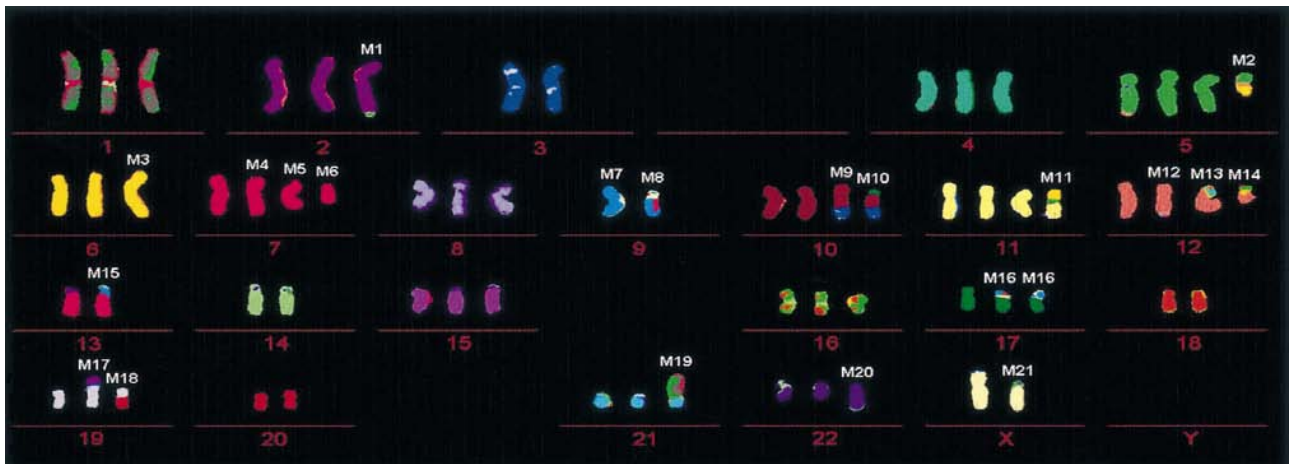

**Fig. 3.10B:** Naumann et al.'s (2001) M-FISH karyotype of a K562 67-karyotype, found in 15 of 19 cells investigated.

The stability of the 62-karyotype (a high of 56 % in Fig. 3.9) created an opportunity for a chromosome-by-chromosome comparison with a documented K562 *atmoxic* 67-karyotype published by Naumann (Fig. 3.10B) (Naumann et al., 2001). Comparison shows (Fig. 3.10) that many markers are lost, and the 62-karyotype has generally moved closer to that of a normal cell. 11 chromosomes move closer to normality, while 2 become more abnormal. Many markers of the 67-karyotype are eliminated in the 62-karyotype.

Fluorescence In-Situ Hybridization (Fig. 3.10A, bottom) showed that the BCR: ABL fusion on chromosome 22, part of the standard *atmoxic* K562 line, is conserved in anoxia. These observations, together with a low rate of apoptosis in anoxic transitions (see below), suggest that anoxia contracts karyotypes by methods that are not degenerative or random, but selective. Although metabolism has a leading role in modulating phenotype and karyotype, the basic cancer lesion, the BCR:ABL gene fusion, is left untouched (Fig. 3.10A, bottom) in the diverse forms of K562.

## 3.2.2 Oxic Karyotype Transitions and CIN

Oxic transition assays were conducted on K562 to further examine the mechanisms of KC and expansion. In *oxic transition assays*, stable anoxic and *atmoxic* cultures are transferred to their opposite oxygen level (21 % > 0 % and 0 % > 21 %). By karyotyping at various time intervals following the transitions, we observed that chromosome losses after anoxia are instantaneous, while gains after *atmoxia* need many cell cycles to complete. The gap (62 to 69 in Fig. 3.9) between anoxic and atmoxic karyo-histograms creates an opportunity to watch the evolution of karyotypes. Four days is an auspicious time to investigate the *atmoxic* transition (Fig. 3.11), as intermediate karyotypes (62 to 69) bridge the gap.

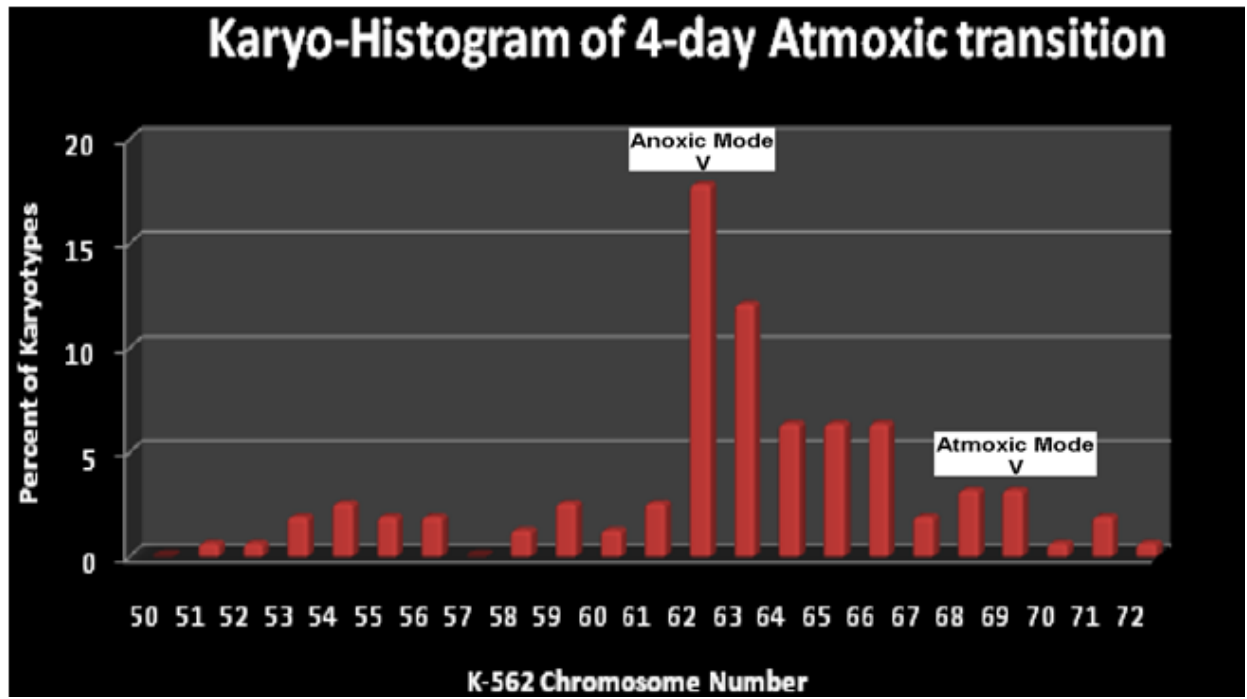

Fig. 3.11: Four days after Atmoxic transition (0 % > 21 %), intermediate karyotypes (63 to 67) bridge the gap between the Anoxic mode of 62 and the Atmoxic mode of 69. 157 metaphases from 3 different tests in our laboratory. 11 % of atmoxic transition karyotypes lie both below and above the graph, but such outliers are almost entirely absent from the anoxic transition.

Our data shows that the traditional mechanism of tumor adaptation, clonal expansion (Nowell, 1976; Zhang et. al, 2001), strains to explain the rapid evolution of karyotypes observed. How can the 63 chromosome karyotype, undetectable under anoxia (blue in Fig. 3.9), appear in Fig. 3.11 at 12 % only 4 days later (~ 4 cell divisions), and disappear afterwards from the long-term *atmoxic* signature (no red 63 in Fig. 3.9)? The results of Fig. 3.11 imply, rather, that CIN (Lengauer et al., 1998; Rajagopalan and Lengauer, 2004) allows chromosome additions to occur with high probability (~21 %, see below) at each cell division, likely under the mechanisms of CIN (Yoon et. al, 2002).

Also, only CIN can explain the rapid chromosomes losses observed in the anoxic transition (21 % > 0 %), where within 1-day, two thirds of metaphases are already within the long-term anoxic envelope. The speed at which the transitions occur far outpaces the conventionally favored mechanisms of genetic change, specifically clonal expansion following random mutation.

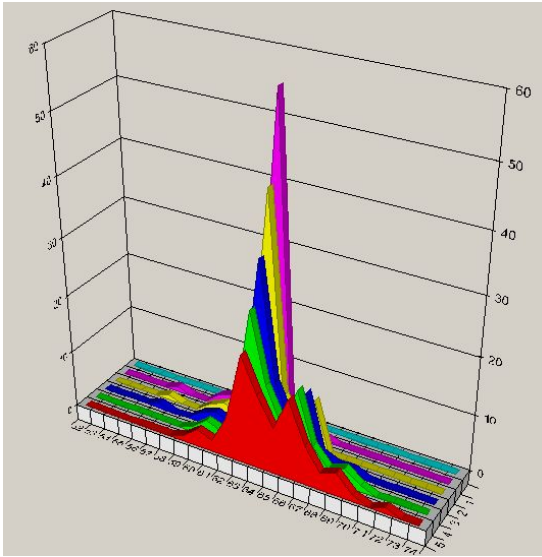

In terms of molecular mechanisms, chromosome counts could increase in the *atmoxic* transition as a result of failed segregation, contributing symmetrically around a mode, or as endo-reduplication (END) (Levan and Hauschka, 1953), a mechanism of unscheduled (extra-mitotic) chromosome duplication. Both mechanisms can be applied numerically using a karyotype evolution simulation program to an initial anoxic karyo-histogram such as the one in Fig. 3.9 by assigning a share of chromosome changes to each mechanism for a certain number of cell divisions, and comparing the output to the karyo-histogram of Fig. 3.11. Computer simulations suggest that the dominant CIN mechanism is END (21 % probability per division), rather than asymmetric segregation (1.5 % probability per division). Too much weight should not be placed on such simulations, because it is difficult to represent accurately uncertain biological mechanisms in software and because for any cell line, all chromosome structures are not equally probable, reflecting islands of stability. But the simulations clarify that in this limited model, END dominates asymmetric segregation, if only because atmoxic transition needs a 7 chromosome mode increase, while asymmetric segregation contributes symmetrical skirts around a mode. Also, KC produces very few unusual karyotypes. Our observations show that tumors can contract and expand their karyotypes with surprising speed as oxygen levels are altered.

Fig. 3.12: Output example of karyotype evolution simulation program. EndoReDuplication and Asymmetric Segregation events are applied with specific probability levels (two dimensional scan) to the anoxic distribution of Fig. 3.9. The simulation result are then compared with the actual 4-day atmoxic transition karyotype of Fig. 3.11.

Following these observations, further questions arose as to the generality of the phenomenon, specifically whether KC can be observed in other cancer cells, and as to whether KC is controlled exclusively by oxygen or is a more general manifestation of reduced metabolism. To elucidate these questions, four other hyperploid cancer cell lines and four different  metabolic restrictors were used to document the extent of the relationship between metabolic restriction and KC.

# 3.3 Anoxic Karyotypes and Karyotype Contraction
# from Metabolic Restriction

At that point, we extended our karyotype investigations to a variety of agents which we knew from our own experiments to be capable of producing size reductions similar to those we observed with oxygen withdrawal in K562. Those agents, melatonin, vitamin C, imatinib and oligomycin we designated collectively as *metabolic restrictors*. Vitamin C, for example, reduces the ROS levels as well as oxygen consumption in cells. Both anoxia and the *metabolic restrictors* in sufficient concentrations can strongly inhibit metabolism. This line of investigation proved very productive when metabolically restricted K562 revealed losses of many more than the 7 chromosomes  lost by anoxia alone, leading almost to normalization of the number of chromosomes in K562. We also expanded our investigations to a number of hyperploid cell lines in an attempt to cover the major tumour types.

KCs on five hyperploid cancer cell lines are presented in Table 5. The "*Atmoxia* Baseline" column lists the chromosome counts observed under standard culture conditions, 21 % oxygen and 5 % $CO_2$. The other columns report the chromosome losses sustained by cell cultures under four strong metabolic restrictions. Anoxia is representative of the deep hypoxia of tumor cores.

"Anoxia alone" (Table 5) results in partial KC, which means that 6 to 8 chromosomes are lost in the 5 cell lines, bringing their totals closer to 46. This contraction remains as long as anoxia persists, but is reversed over a few weeks, if *atmoxia* is restored.

All cell lines in Table 5 have larger *atmoxic* than anoxic count *ranges*. In Fig. 3.9, the left tail of the histograms is larger for *atmoxic* (red, 64 to 67) than anoxic (blue, 58 to 60) distributions. These two observations can be explained if oxygen increases not only chromosome counts, but also chromosome count *spread*. Growth patterns influence karyo-histograms under oxygen, as a cell's position within a group impacts its oxygen exposure. This view is supported in Table 5 by the smaller *atmoxic* chromosome ranges for suspension cells (the two erythro-leukemia types), compared to the adherent lines.

The metabolic restrictors oligomycin and imatinib were used at sub-toxic levels: concentrations ($IC_{50}$) just low enough to allow cell division and karyotyping. Melatonin-vitamin C were optimized for maximum chromosome drop, and these levels were subsequently found to be physiological, as they matched those in bone marrow (Tan et al., 1999) and plasma.

| Cell | Type | Atmoxia Baseline | Anoxia Alone | Atmoxia | | |
|------|------|------------------|--------------|---------|---|---|
| | | Mode (80 % Range) | Mode (80 % Range) | Oligomycin 0.1 µM[4] Mode (80 % Range) | Imatinib 0.08 µM[5] Mode (80 % Range) | Melatonin-Vit C 0.3 µM, 150 µM[6] Mode (80 % Range) |
| K562[1] | Erythro-Leukemia | 69 (64-70) | 62 (58-62) | 48 (46-53) | 47 (45-51) | 48 (45-52) |
| HEL[2] 92.1.7 | Erythro-Leukemia | 66 (62-67) | 59 (57-60) | 47 (46-51) | 48 (47-53) | 49 (46-52) |
| NCI-H460 | Large Cell Lung Cancer | 57[3] (53-65) | 51 (45-52) | 47 (46-49) | 47 (46-50) | 47 (45-51) |
| COLO 320DM | Colo-rectal Adeno-carcinoma | 54[3] (49-61) | 48 (46-49) | 46 (46-48) | 46 (45-48) | 47 (45-49) |
| MCF7 | Breast Adeno-carcinoma | 82[3] (66-87) | 74 (61-75) | 64 (59-66) | 65 (61-68) | 63 59-65 |

**Table 5. Karyotype Contractions in Cancer Cell Lines after 3-day Metabolic Restrictions.**

Number of metaphases for each determination: 122, 108, 30, 25, 25; 50, 50, 30, 25, 20; ATCC, 20, 20, 20, 20; ATCC, 50, 20, 20, 20; ATCC, 35, 20, 25, 30. [1] BRC-ABL positive. [2] BRC-ABL negative. [3] ATCC data. [4] Proliferation IC50 of 0.0125 µM for K562 and HEL IC50 of 0.1 µM for other types. [5] K562 sub-toxic level. [6] The Melatonin-Vitamin C concentrations optimized for K562 chromosome count normalization.

The metabolic restrictors of the last three columns of Table 5 contract karyotypes further than anoxia in all cell lines, whether starting from *atmoxia* or anoxia (not shown). NCI-H460 and COLO 320DM chromosome counts are almost normalized by anoxia alone, possibly indicating their deeper dependence on oxygen metabolism. The apparent failure to reach peri-normal counts in MCF7 may be due to the induction of cell fusions by the metabolic restrictors (Yang et al., 2009).

K562 chromosome count ranges can be expanded beyond the numbers in Table 5. Higher melatonin levels contract them below 46, while hyperoxia (50 % and 95 % oxygen) expands the usually small "2S" (~138 chromosomes) population.

Since anoxia alone reduces karyotypes by 6 to 8 chromosomes while other metabolic restrictors return chromosome counts close to normal, except for MCF7, it can be concluded that hyperploidy in cancer cells is not *essential*, but is circumstantially connected with enhanced metabolism.

# 3.4 Karyotype Contractions from Magnetic Fields

Until now, our phenotype and karyotype results on 5 cancer cell lines has showed that metabolic restriction leads to KC, which has proven to be a rather sensitive assay variable. Anoxic K562, in particular, is not only an ideal model to assess the mechanisms of CIN, as revealed by chromosome count changes (Fig. 3.9), but can also detect metabolic disturbances. Because of a number of experimental precautions, in particular the use of synthetic media and anoxia, our reference K562 cultures are karyotypically and otherwise exceptionally stable. 75 % of the cells have just two chromosome counts, 62 and 61, compared to a wider range under 21 % oxygen (Li et al. 2011). The stability of anoxic K562 chromosome counts has been monitored by periodic controls in our lab for 5 years. These cultures thus provide an extremely precise reference point, as shown in the narrow baselines. From this position, we applied MFs to 5 cancer cell lines to assess possible phenotype and karyotype changes.

## 3.4.1 Induced Currents

Whether biological effects of power-frequency MFs are related to the MF itself, or to the currents induced in tissues by the fields, has been a perennial question. A direct MF interaction supposes an unrecognized component within living tissues that is vulnerable to MFs. Many investigators favor the view that effects occur through potentials produced by magnetically induced currents on the thin membranes within or bordering living cells. Such currents and membrane potentials are familiar to conventional electrophysiology.

In the results illustrated in Fig. 3.13, one aliquot of K562 cell culture is placed in a vertical, and the second in a horizontal MF exposure system. At the same magnetic flux density, the horizontal coil induces currents 6 times larger within the medium. The maximum current density in a culture dish is related to the magnetic flux through its surface, an area 34 x 34 mm for the horizontal coil, compared to 5.8 x 34 mm for the vertical coil. As chromosome losses after 6-days at 1 µT repeatedly came out similarly for both orientations (Fig. 3.13), we concluded that the effect on the main variable of this article, chromosome counts, was dependent on the MF itself.

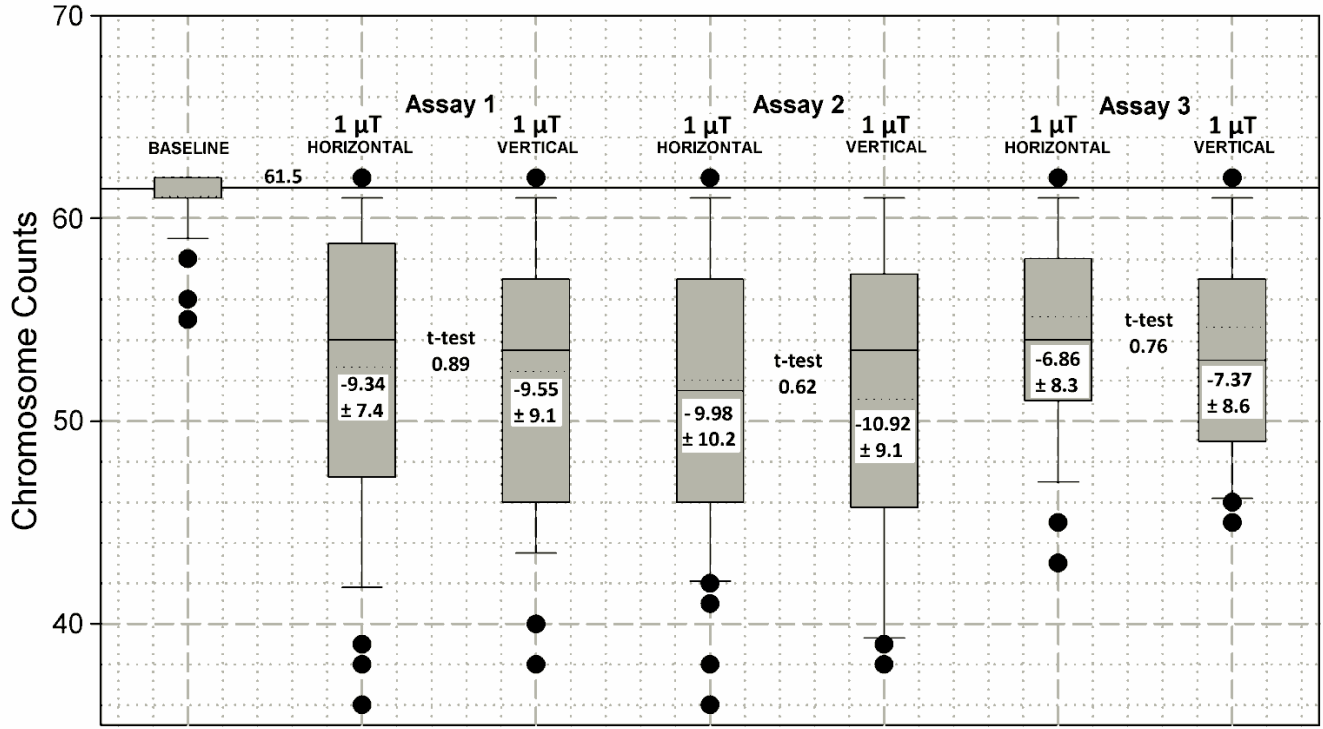

**Fig. 3.13:** **K562 chromosome counts produced by 60-Hz, 1 µT MFs applied either horizontally or vertically in three 6-day assays. The baseline T-12 culture vessel contains anoxic cells at less than 4 nT (60-Hz), with an average of 61.5 chromosomes (horizontal line), and a very narrow distribution (at left). Box plots show median (solid), average (dotted), 25 and 75 % limits (box), 10 and 90 % limits (whiskers), and outside values (dots). 56 (Assay 1), 50 (Assay 2) and 51 (Assay 3) metaphases karyotyped in each orientation. Inside the box plots are average chromosome losses, with standard deviation. The Student's t-test results quantify the probability that the horizontal and vertical results are identical.**

We assumed direct MF, rather than induced current action on the basis that variations of current density by a factor of 6 did not affect the results. But it should be kept in mind that the same observations would result if induced currents had a very flat dose-response, already saturated at the lower current. Further, direct MF action on KC does not preclude that other effects of MFs may depend on induced currents.

We are confident that the highest MFs applied to our cell cultures (5 µT) do not produce temperature rises larger than 0.1 K in the medium, and are in all likelihood actually much smaller. Temperatures elevations would be readily detectable in our assays, as K562 is a good thermal sentinel. Hyperthermia manifests by larger cell sizes starting at + 0.5 K, while + 1 K seriously impairs proliferation, and + 2 K over a few days is lethal.

The larger cell sizes observed under mild hyperthermia (+ 0.5 K) suggest chromosome number increases, as the cells probably use such a strategy to meet thermal challenges. Cell size and chromosome numbers often increase together in a given cell type. As expected, our measurement of the K562 karyotype at 38°C (+ 1 K) yielded an increase in chromosome number from 61.5 to a mean of 68.8 (n = 20). This means that any thermal effects from the application of MFs, if they were present, would tend to attenuate the KCs induced by MFs. This opposition of effects is not unexpected, as MFs are metabolic restrictors, whereas temperature rises stimulate metabolism. These measurements confirmed by biological means the athermal context of our MF results.

### 3.4.2 Static Magnetic Fields

Certain theories on MF biological effects have invoked a relation between static (Earth) and power-frequency MFs (Liboff, 1985). Although logic would suggest that living organisms be resistant to static fields, since life evolved in them, an interaction between the two is not easy to discount.

The *naïve* cells for our experiments are kept inside magnetic shields. Within our 10 sets of shields, we measured different static fields, comparable to Earth's in magnitude, that varied over small distances (~ 3 cm), and with seemingly random orientations. As we assayed various levels of 60-Hz MFs for KC, different shield sets were used, but we detected no influence of static fields on our results.

According to the research of Russian physicists (Semikhina and Kiselev, 1981), the effects of static MFs on the structure of water, and consequently on ATPS function (section 4.3.1),  are negligible unless the Earth's MF (57.8 µT) is reduced below 34 nT. As will be discussed in more depth in section 4.3.1, the measurements of Russian physicists on water are entirely compatible with our experimental observations on cancer cells exposed to both static and alternating current magnetic fields.

### 3.4.3 Dose-Response

Fig. 3.14 shows the chromosome losses experienced by naïve K562 cells after 6-day exposures in various magnetic flux densities. The graph covers time-averaged MFs representing different environments. 0 to 0.2 µT represents domestic, 0.07 to 0.5 µT commercial, while 0.1 to 1 µT is typical of occupational exposures. In the baseline at left, 75 % of the cells have either 62 or 61 chromosomes. Under any MF exposure, the few karyotypes of the baseline expand to a variety of chromosome counts, and there are substantial KCs across all MFs.

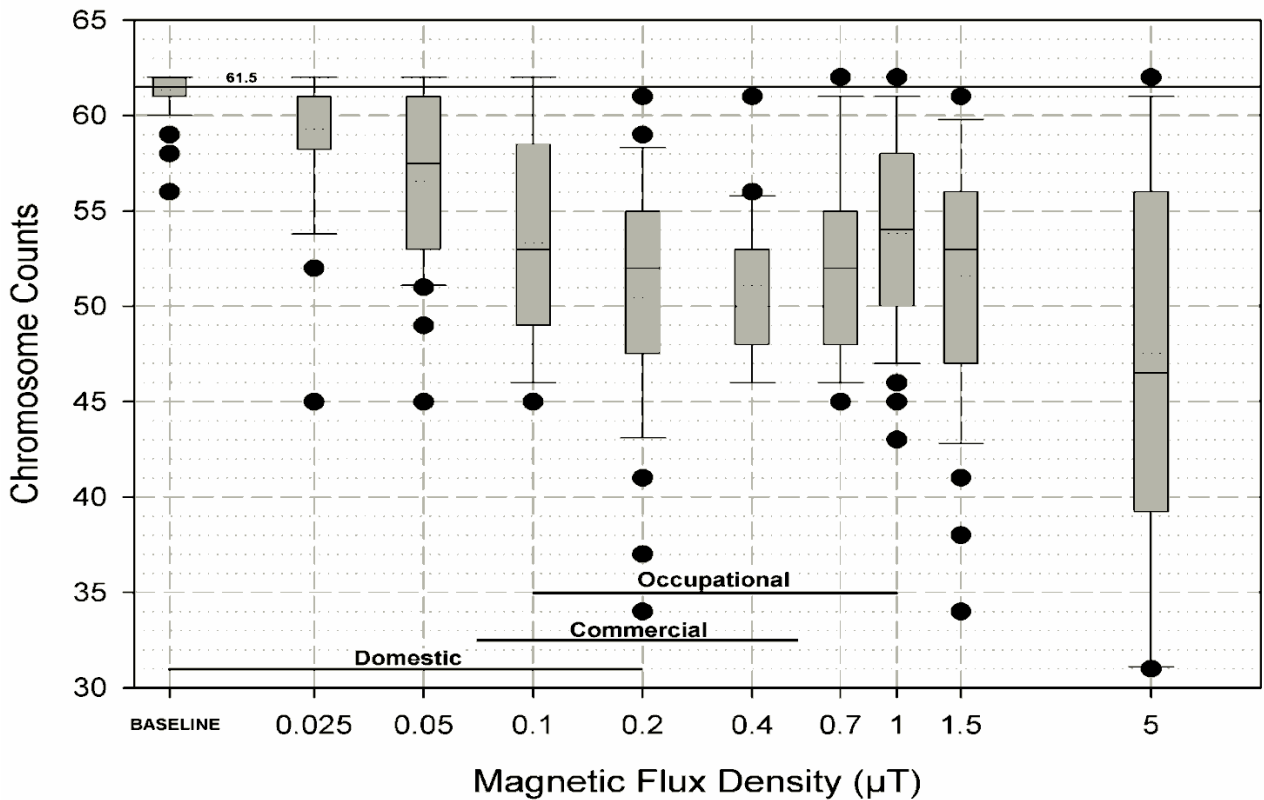

**Fig. 3.14: K562 chromosome counts as a function of 60-Hz Magnetic Flux Density applied for 6 days to naïve cells. In sequence, 65, 28, 50, 77, 46, 33, 65, 102, 56 and 50 metaphases. Approximate ranges for typical domestic, commercial and occupational exposures are indicated.**

Three features are of importance. First, a no-effect-level lower than 25 nT. Second, a progression of chromosomes losses between 0 and 0.4 µT. Third, the relatively flat dose-response between 0.1 and 1.5 µT.

Other cancer cell lines show even flatter dose-responses than K562. Over two orders of magnitude of the MF, erythro-leukemia (HEL 92.1.7), breast (MCF7) and lung (NCI-H460) cancer cells lose between 8 and 13 chromosomes (Fig. 3.15). HEL, our second erythro-leukemia cell line, shows a rise at lower fields similar to that of K562.

Classical toxicology and epidemiology, where smoothly climbing doses-responses are justified by binding chemistry and the central tendency theorem , do not expect the flat dose-responses over two orders of magnitude observed in Fig. 3.14 and Fig. 3.15. Also, the effects found for different cell types are strikingly similar, with parallel low-field deviations in the two erythro-leukemia cells, suggesting common, basic mechanisms.

**Fig. 3.15: Average Chromosome Losses in Erythro-Leukemia, Breast, Lung and Colon cancer cells as a function of 60-Hz Magnetic Flux Density applied for 6 days to naïve cells. The references ("0") for naïve cells (< 4 nT) are: 66 (HEL), 74 (MCF7), 57 (NCI-H460) and 54 (COLO 320DM) chromosomes. In sequence, 32, 22, 29, 32; 19, 22, 19, 21; 29, 22, 24; 22, 34 and 46 metaphases. HEL, NCI-H460 and COL 320DM assays used 21 % oxygen, rather than anoxic conditions, as some anoxic karyotype modes are too close to 46.**

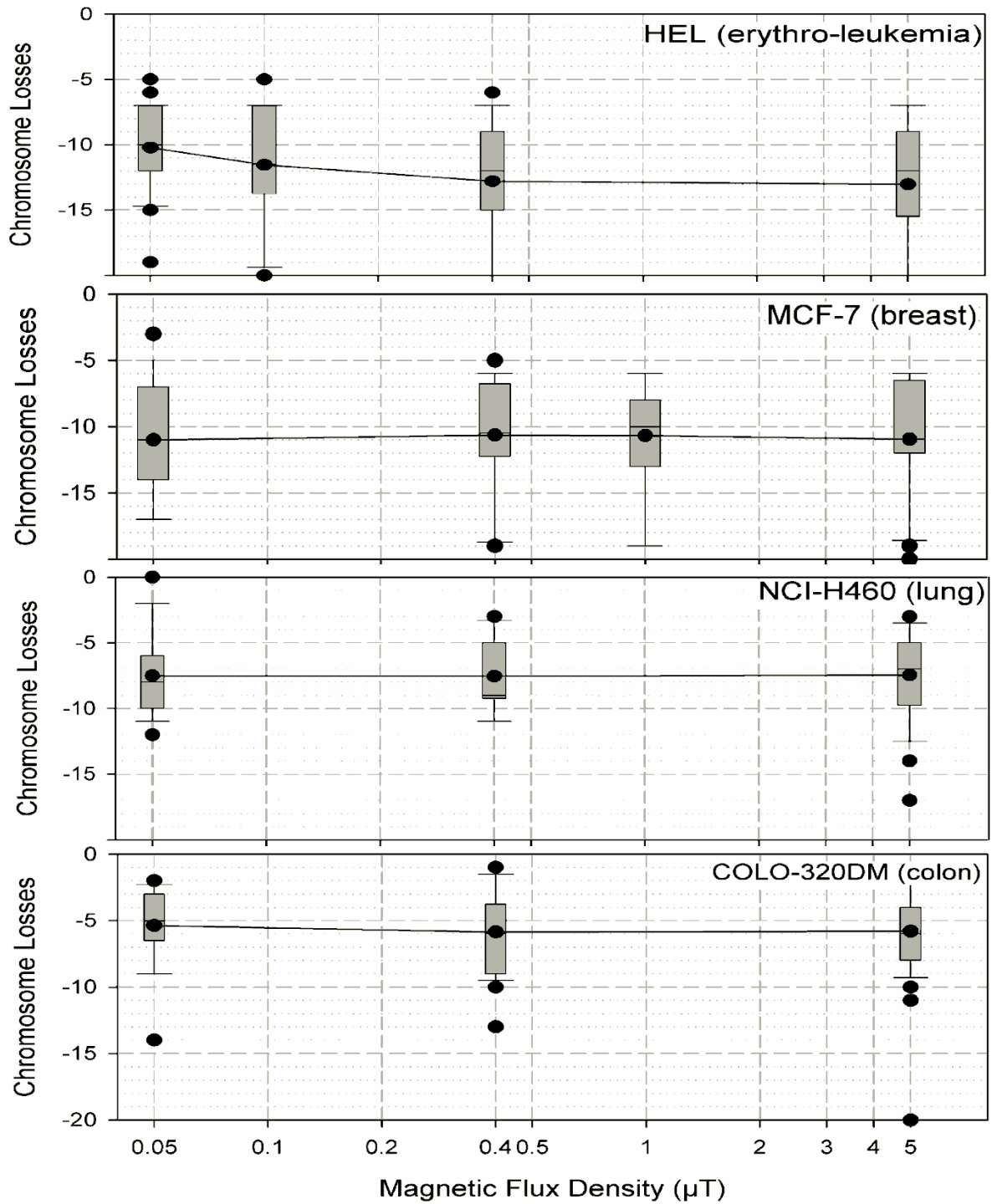

### 3.4.4 Differential Action

K562 cells with magnetically-lowered chromosome counts such as those displayed in Fig. 3.14 progressively recover their original chromosome counts after 3 weeks, even as the MF is maintained at a constant level (Fig. 3.16). It is interesting to note that chromosome count is restored earlier than chromosome count dispersion.

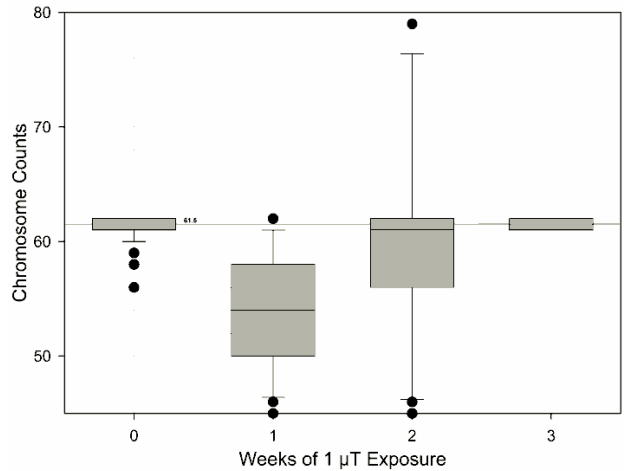

**Fig. 3.16: K562 chromosome counts return to baseline after 3 weeks of continuous 1 µT MF exposure. 65, 102, 50 and 37 metaphases.**

Thereafter, if the original field is altered by a small percentage of the original value, either positively or negatively, KCs are again observed, as shown in Fig. 3.17. 0.1 µT and 1 µT were chosen here because they cover one order of magnitude, and because they are enviromentally representative. Starting from low (0.1 µT) or high (1 µT) baselines, symmetrical chromosome losses are observed as the fields are slightly augmented or reduced. KC is also observed when fields are reduced from 50 nT to < 4 nT. This 3-week adaptation and bilateral sensitivity to changes is unforeseen by conventional toxicological analysis.

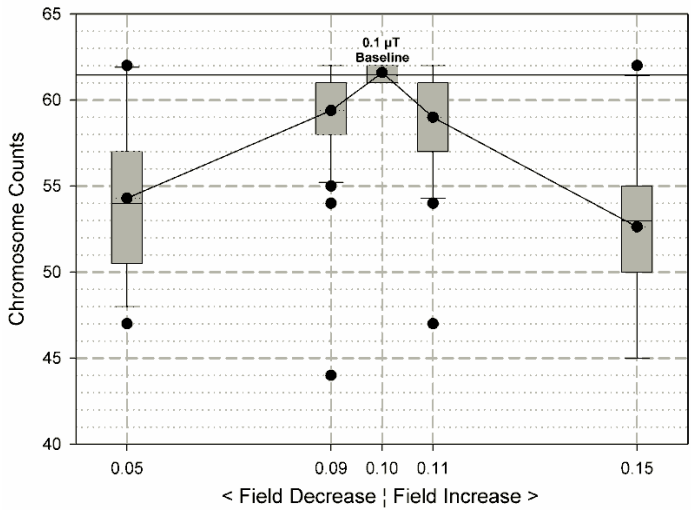

**Fig. 3.17:  K562 Chromosome counts obtained after 6 days by altering baseline MFs of 0.1 µT and 1 µT.**
**For 0.1 µT, 20, 31, 37 (baseline), 31, 35 metaphases.**
**For 1 µT, 28, 28, 37 (baseline), 28, and 28 metaphases.**
**Although the symmetry of the chromosome counts is strong, there is more cell decay in the cultures with increased than with reduced fields.**

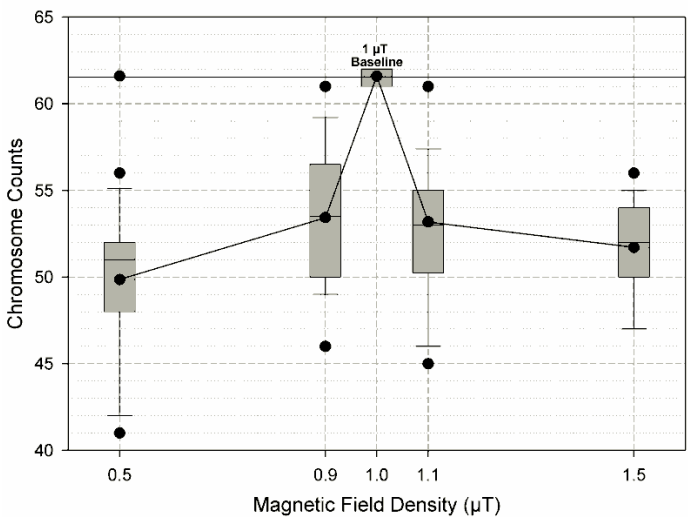

The chromosome losses will be interpreted below as caused by magnetically-induced changes in intra-cellular ATP levels. These results cast doubt on the stability of cancer cell models housed in incubators with MFs that are highly variable over space and time (Mild et al., 2009).

In view of these interesting results, we performed a number of verifications to insure that the KCs observed were not specific to the synthetic medium, RSF1. First, we verified that the KC observed under the anoxic transition (only) for K562 also occurred under RPMI-1640 plus 10 % FBS.  Next, we verified that the KCs recorded at 0.05, 0.4 and 5 µT MF for anoxic MCF7 also occurred under RPMI-1640 plus 10 % FBS.  Finally, we verified that the KCs recorded at 0.05, 0.4 and 5 µT MF for atmoxic NCI-H460 also occurred under RPMI-1640 plus 10 % FBS.

### 3.4.5 MF>ATPS>AMPK Assays

Previous experiments had show a link in all our cancer cell lines between metabolic restriction and KC. Anoxia alone induced partial KC, 6 to 8 chromosomes lost, depending on the cell type. Deeper contractions, almost to normalization of the karyotypes to 46, were produced by $IC_{50}$ doses (allowing 50 % of the normal cell division rate) of the metabolic restrictors oligomycin and imatinib. Similar KCs were produced by physiological levels of melatonin and vitamin C together. We believed that comparing the characteristics of metabolically restricted cultures with MF-exposed cultures may provide clues on action mechanisms, as the different restrictors have different sites of action.

A K562 culture subjected to a MF field of 0.4 µT, very effective for KC, was compared to those of various metabolic restrictors. Fig. 3.18 displays the close similarity between two of seven K562 assays, one exposed to 0.4 µT MF, and the second to a sub-toxic level ($IC_{50}$) of oligomycin. Oligomycin and the 0.4 µT MF stand apart from all others in having smaller cell diameters and lower ratios of cells-to-objects below 11 µm, the decay particles and apobodies. For comparison, we supply the characteristics of a pristine "baseline" culture with very few particles, and those of other metabolic restrictors. The similarity between these two cell size histograms suggests that MFs and oligomycin share a common mode of action.

In spite of the close similarity between the oligomycin and 0.4 µT assay results in Fig. 3.18, oligomycin is faster-acting than the 0.4 µT MF: the changes in cell size, revealing of KC, are visible at 1 day, earlier than for the MF. But more efficant MFs, such as 5 µT at 60-Hz or 1 µT at 120-Hz, reveal similar effects earlier.

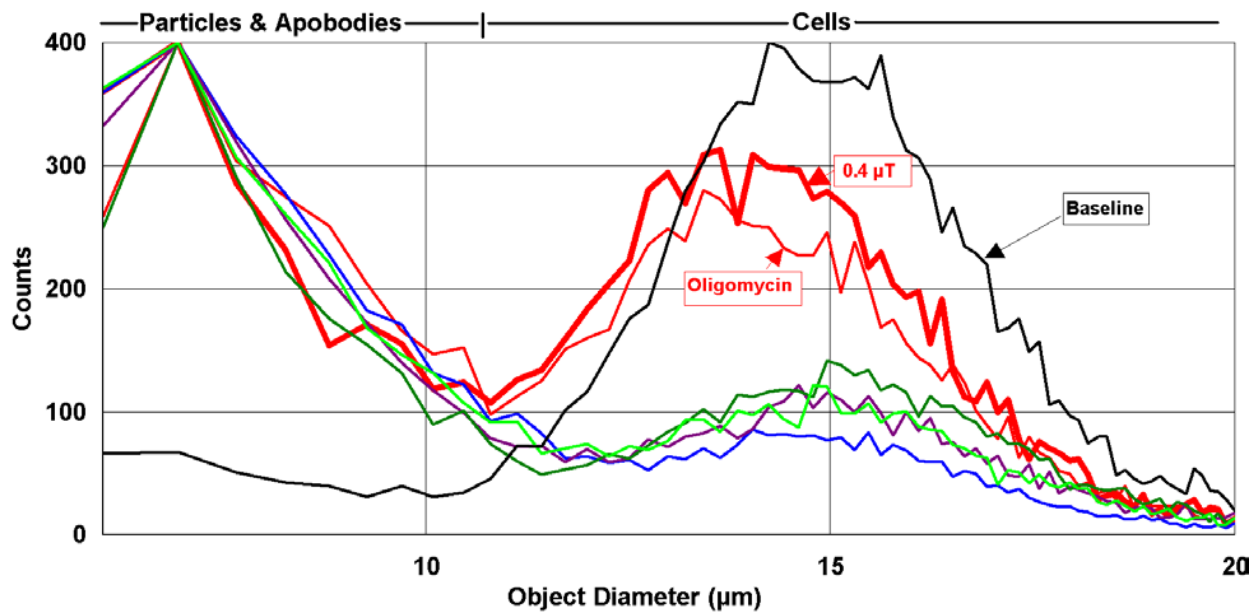

**Fig. 3.18: Object Diameter histograms for 6-day anoxic exposures of K562 cultures to 0.4 μT MF at 60-Hz and oligomycin at 2.5 ng/ml. The lower 4 curves are: imatinib (0.04 μg/ml) in blue, resistin (40 ng/ml) in violet, metformin (0.01 mg/ml) in light green and melatonin-vitamin C (0.3 μg/ml, 26 μg/ml) in dark green. Exposed cultures are adjusted to a common small particle count maximum.**

The similarity between 0.4 μT and oligomycin suggests that the MF may be an inhibitor of ATPS, as oligomycin is a highly specific inhibitor of ATPS. If this were the case, inhibition of mitochondrial ATPS by MFs would activate AMPK, because healthy cells must maintain a high level of phosphorylation capacity (ATP:ADP ≈ 10) to function well (Hardie and Hawley, 2001). AMPK is a sensitive regulator that switches *on* catabolic pathways and *off* many ATP-consuming processes, both acutely and chronically, through gene expression. AMPK's regulation could explain the adaptation of cancer cells to fixed MF levels (Fig. 3.16), as well as the differential action of MFs on chromosome reduction (Fig. 3.17).

The MF>ATPS>AMPK pathway was investigated using metformin and resistin. Metformin is a diabetes drug that activates peripheral AMPK, leading to reduced glucose production in the liver, and reduced insulin resistance in muscle. Metformin usually causes weight loss and reduced appetite, and is considered an attractive anti-aging drug.

Resistin, a product of the RSTN gene, is a 9.9 kDa protein containing 93 amino acid residues which at 20 ng/ml or more inhibits AMPK, and interferes with phosphorylation of Akt (serine/threonin protein kinase), active in multiple cellular processes such as glucose metabolism, cell proliferation, apoptosis, transcription and cell migration.

| Agent | Concentration/Intensity | Chromosomes Lost |
|---|---|---|
| Metformin (activator) | 0.01 mg/l | - 9 |
| Resistin (inhibitor) | 40  ng/l | - 10 |
| Metformin + MF | 0.01 mg/l + 1 µT | - 11 ± 0.34 |
| Resistin+ MF | 1 µT | - 4, ±0.46, |
| MF | 1 µT | - 7.5 |

**Table 6: K562 Karyotype Contractions under action of AMPK Modulators and MFs.**

Metformin (0.01 mg/l) and resistin (40 ng/l) alone induce average KCs of 9 and 10 chromosomes respectively, in K562. When a 1 µT MF is added to metformin, even larger KCs are observed (9 becomes 11, with a standard deviation of  0.34). When a 1 µT MF is added to resistin, the 10 chromosome KC of resistin falls to 4 chromosomes, with a standard deviation of 0.46, also less than the KC of 1 µT MF alone, at 7.5.

The conclusion is that MFs enhance the action of metformin, but neutralize the effect of resistin, supporting the MF>ATPS>AMPK pathway.

### 3.4.6 NCI-H460 Proliferation

Beyond effects on cancer cells karyotypes, MFs also impact proliferation rate, adhesion and cell shape, which are not reported in detail here. Some prominent effects may not be solely related to ATPS interference, since they are strongly dependant on MF intensity, and disappear with the addition of serum. For example, the cell counts of lung cancer cells (NCI-H460) after 4 days in our synthetic medium at 50 nT, 400 nT and 5 µT are 8, 9.2 and 14.8 times larger than those of naïve cells. Naïve NCI-H460 do not attach in our synthetic medium, but do so under any MF exposure.

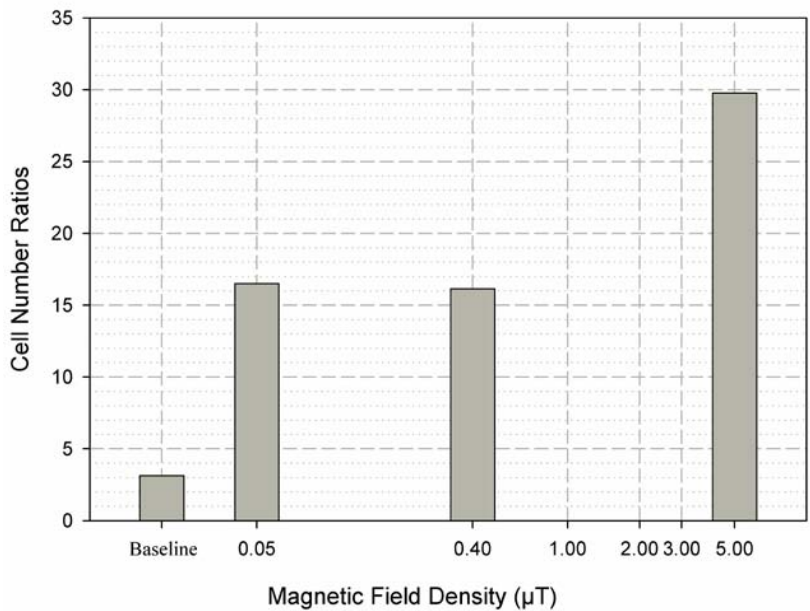

**Fig. 3.19:  NCI-H460 cell number ratios between initial and 4-day counts under Baseline (<4 nT), 0.05, 0.4 and 5 µT MFs.**

# 4.0 DISCUSSION

# 4.1 Metabolism, EndoReDuplication and Evolution

Researchers have been aware of the intimate relation between cancer and metabolism for a long time (Warburg, 1956). The identification of specific cancer mutations and the success of imatinib therapy have inspired on-going work both on the metabolic consequences of cancer lesions (King et al., 2006; Menon and Manning, 2009) and on cancer therapies based on metabolic control (Chen et al., 2007; Kim et al., 2011; Duivenvoorde et al., 2011). Studies have also been performed on the evolution over time of cancer metabolism (Hu et al, 2011;  Alfarouk et al., 2011), and even on reversing cancer phenotypes and genetic instability using anti-oxidants and nitric oxide inhibitors (Martinez-Outschoorn et al., 2010).

It is well accepted that cancer karyotypes can evolve over time, and be influenced by culture conditions (Polianskaia and Miu, 1988; Jin et al., 1993; Wenger et al., 2004). Even normal human embryonic stem cells, often incubated under low oxygen conditions, are vulnerable to karyotype instability, and it is recommended that chromosomal status be checked at intervals of 5 passages to monitor possible translocations and aneuploidies (Moralli et al., 1990). But, to our knowledge, there has been no previous work on devolution of tumor karyotypes using metabolic restriction.

## 4.1.1 Cancer Expression Depends on Metabolism

The classical view of pleomorphism in CML is that many cell populations with different proliferation and adhesion properties are observed, but that the progression from chronic phase through acceleration and blast crisis *seems* to be driven by the acquisition of new chromosomal abnormalities. CML patients typically show normal or slowly climbing chromosome counts over time, even in the acute phase (average of 46.4 in 31 patients) (Haas et.al., 1984), as acquisition of larger chromosome counts is a late characteristic of the disease. The exact influence of chromosome changes in CML and other cancers is an open question.

In the data we reported, anoxic and atmoxic K562 cells both share the same BCR:ABL lesion, but exhibit clearly different phenotypes and genotypes. Phenotypically, the anoxic cells, even with increased proliferation rate, exhibit more stable characteristics, as previously described. Genotypically, the reduction in markers, an exceptionally narrow karyo-histogram and the speed with which chromosomes are lost in the anoxic transition all point to a reduction of CIN under anoxia. Whatever the mechanisms, a basic observation remains: the instability of K562 can be quenched and reversed both at the phenotype and genotype levels by suppressed metabolism, confirming karyotype variation, in particular, as a secondary rather than primary characteristic of cancer.

Looking back in history, the K562 cells harvested by Lozzio at 45-46 chromosomes in 1971 may have increased their chromosome counts as a result of the standard atmoxic cell culture conditions. All evidence points to an influence of oxygen gradients over space and time on cell culture karyotype diversity and pleomorphism even under standard atmoxic cell culture conditions. When transported *in vivo*, the influences of oxygen and metabolism can explain the diverse phenotypes and karyotypes of tumors (Nowell, 1976; Lengauer et al., 1998). Further, the high level of CIN in tumors, although rooted in the initial cancer lesion, is likely controlled by the strong oxygen and metabolic gradients of tumor structure.

BCR:ABL and CIN-END seem mechanistically separated. While BCR:ABL unbridles tyrosine kinase and multiple downstream effects in metabolism, CIN-END, which lump together both chromosome addition and loss (Yoon et al., 2002) are linked with extra-mitotic chromosome duplication (END) or to defects in mitotic operations.

Similar irregularities in mechanisms of chromosome duplication occur even in normal tissues. For example, chromosome addition by persistence of kinetochore-microtubule misattachment caused by a defective mitotic spindle checkpoint is a documented cause of aneuploidy in 0.1 to 0.8 % of normal blood lymphocytes (Cimini and Degrassi, 2005). Chromosome loss has been reported for mitotically unstable cancer cells (Rajagopalan and Lengauer, 2004), and it is not uncommon for hybridomas (Kessler et al., 1993) to shed chromosomes in culture.

A further argument can be found in the fact that the other erythro-leukemia cell which we investigated, HEL, displays the same sensitivity and chromosome loss to anoxia, but is BCR: ABL negative (Honma et al., 1995).

In our opinion, rather than BCR:ABL or other specific lesion, the lead actor in the control of CIN and END is metabolic rate.

It is interesting that the disparate genetic mechanisms of cancer somehow come together to create tumors with relatively similar characteristics. This suggests that these diverse lesions may come to a common ground by interfering with a physiologically large target, which we propose to be connected to oxygen, but also to general metabolism. In this view, it is heightened metabolism that is the immediate that cause of CIN-END, determining the ultimate karyo-histogram of K562. We submit that metabolic rate supersedes the basic cancer lesion in controlling the expression of carcinogenicity.

## 4.1.2 Status of EndoReDuplication

Although clonal expansion may play some role in our oxic transition tests, the rapid changes in chromosome counts in our cells, and probably also in tumors, can be more adequately explained by the connection between oxygen changes and CIN-END. END has primarily been studied in plants, and until now, only a modest role has been documented for it, related to a list of drugs, physical agents, and a mutant cell line defective in DNA repair (Cortés et. al, 2004). Our results suggests that END should have its role expanded beyond an "unusual finding" (Bottura and Ferrari, 1963), and should be described as a dormant hyperploidy reflex triggerable by hyperoxia or enhanced metabolism, with an important role in cancer, as suggested by Larizza and Schirrmacher (Larizza and Schirrmacher, 1984).

The atmoxic transitions described here could be an adequate simulation of a "karyotype variation front", helping tumors, through CIN-END, to deal with chemotherapeutic drugs and the immune system (Walenta et al., 2001). Resistance to imatinib, for example, arises partly from amplifications and chromosome duplications in chromosomally unstable CML cells (Rajagopalan and Lengauer, 2004). Rapid adaptations of cancer cells to changes in their microenvironment, similar to the ones described here for oxygen, may be the cause of the amazingly rapid replacement of tumor cells with variants that are resistant to therapy and the immune system (Rajagopalan and Lengauer, 2004; Duesberg et al., 2007).

It may be that the stem cells of primitive tumor cores are only fit for the anoxic niche, thus explaining their poor relation with oxygen. These cores would support hypoxic growth, while peripheral cells with expanded karyotypes and suppressed apoptosis would invade oxygen-rich areas, spurred by the richer blood supply. Perhaps these distinct populations are responsible for residual disease in CML. The rapid adaptation of cell karyotypes to oxygen levels described in this paper has implications for our understanding of cancer. Chromosome count is rapidly and drastically changed by oxygen, providing, as a minimum, a powerful environmental mechanism to accelerate genomic evolution beyond random mutations in tumors.

### 4.1.3 Tumors and Evolution

Although tumor evolution is traditionally believed to occur by slow mechanisms of clonal expansion (Höckel and Vaupel, 2001), our data indicates that cells can use END to rapidly alter karyotypes and phenotypes *in vitro* when metabolic resources are variable.

CIN-END is a powerful mechanism to accelerate genetic evolution beyond random mutations. Tumour evolution may be a model of Evolution itself. The manifestation of END as a tumor stress response implies that stuttering is not so onerous to biological systems. It appears that redundancies at the level of the gene and chromosome are affordable, creating opportunities for variation. Gains or losses in one chromosome steps are more likely to be compatible with cell viability than the larger chromosome count changes associated with mitotic apparatus defects (Rasnick et al., 1999). Darwin described natural selection driven by random mutations. But the deliberate genome expansion of END may be a conserved trait that increases competitiveness of the biota as a whole. The CIN-END mechanisms may have played the role of evolutionary autopilots, driving the *expansion of threatened code*.

# 4.2 Metabolic Restriction and Cancer

## 4.2.1 Cancer and the Niche

Stem cells divide through mitosis, and differentiate into diverse specialized cell types. K562 falls into the category of adult stem cells. Although it is not our intention to debate the various theories of cancer, one leading theory is that cancer is a disease of stem cells (Gupta et al., 2009). One of the characteristics of stem cells is that they reside for protection in hypoxic niches (Li and Neaves, 2006).

Our data provides two strong arguments to support K562's origin in a low-oxygen niche.

The first is our observations on the improved phenotype of anoxic K562, which are supported by observations of other authors, who underline the importance of low oxygen in preserving the properties of germinal lines and stem cells (Cipolleschi et al., 1993).

The second is the deep KC caused by the anti-oxidants melatonin and ascorbic acid, used at physiological medullary levels, which roll K562 almost back to the original Lozzio karyotype.

When carcinogenic lesions such as K562's BCR-ABL translocation occur, stem cells expand beyond the niche because of (1) niche structure failure or (2) expansion out of the hypoxic niche. Stem K562 with an enhanced metabolism then finds itself in a hyperoxic environment and exploiting anaerobic metabolism, used in the niche, even in the presence of oxygen.

This picture is fairly close to the classic illustrations of the "Warburg effect", where leukemic and lung cancer cells process glucose anaerobically into lactate, even with oxygen present (Warburg, 1956), a mismatch between where the cell is, and where it metabolically "thinks" it is. Cancer lesions may be diverse, but they seem to share a vulnerability to oxygen which is expressed in the phenotype and karyotype alterations we documented.

## 4.2.2 Metabolic Restriction Mechanics

The basic cancer lesion enhances cell metabolism, allowing for faster growth. This growth drives down local oxygen concentrations because micro-vascular organization does not accommodate the heightened metabolism and cell numbers. But cancer cells can readily adapt to this reduction because of the unrestricted use of anaerobic metabolism. Their heightened glycolytic rate favors the evolution of a group of hypoxic cancer cells into a tumor core. Metabolic restriction further allows the cells to dispense with the detoxification mechanisms associated with oxygen exposure, such as glutathione-S-transferase and CYP3A4 expression (Nagai et al., 2004), and to concentrate on bio-synthesis (proliferation). Smaller karyotypes are maintained under metabolic restriction, contributing to tumor core expansion, as fewer chromosomes can be more rapidly duplicated.

Our data shows that the removal of oxygen from cancer stem cells is supportive of the development of tumor core structure by enhancing proliferation and self-adhesion, and reducing inflammation. We have observed as well (Table 4) smoother cell borders, smaller cell footprints, and reduced macrophage numbers of increased activity in anoxic K562. The survival of the tumor core is thus enhanced by metabolic restriction.

The tumor core must inevitably slow in its development towards anoxic stability, and as it does so, it is inevitably surrounded by a more oxygenated and perfused tumoral cortex. The more highly oxygenated and perfused cortex is less stable (Warburg, 1956; Li and Neaves, 2006; Walenta et al., 2001), but is ground for the expression of END. Increased chromosomes numbers in the tumor cortex provide a karyotype and phenotype variation front. The chromosomes added by END interface with normal tissues by diluting the initial cancer lesion, slowing down cell division because of the increased chromosome count, and increasing detoxification metabolism, similar to hyperploid liver cells. The cortex will also shed metastatic cells to new sites, which will ultimately revert to hypoxic states as the metastatic sites expand.

Our data supports the view that tumors stabilize their core under metabolic restriction, while their cortex diversifies karyotypes, and sets the stage for metastasis. Therefore, standard *atmoxic* cell cultures *in vitro* may be more relevant to metastasis, while anoxic, hypoxic and restricted metabolic states are more desirable for tumor eradication studies.

### 4.2.3 Metabolic Restriction and Promotion

It has been repeatedly confirmed that cancer cells become more malignant under hypoxia (Höckel et Vaupel, 2001; Hill et al., 2009; Lash et al., 2002; Jögi et. al., 2003) *in vitro* (Anderson et al., 1989; Cuvier et al., 1997) and in the clinic (Brizel et al., 1996, Nordsmark et al., 1996), to the point where it has become a central issue in tumor physiology and treatment (Cuvier et al. 1997; Höckel and Vaupel, 2001). Since our data ties the metabolic restriction of cells to KC, it is logical to conclude that KC caused by metabolic restriction is an indicator of meta-genetic cancer promotion.

### 4.2.4 Anti-Oxidants as Cancer Promoters?

Free radicals and reactive oxygen species can randomly destroy molecules. An apparent corollary is that anti-oxidants protect biological materials, attenuating the carcinogenic action of toxicants and of normal metabolism. Unfortunately, epidemiological evidence following administration of many anti-oxidants does not show the expected benefits, and may point more to cancer rate increases. Two large epidemiological studies found negative impacts of vitamin A on lung cancer rates in smokers (Research Group, 1996; Omenn et al., 1996). The results could not be explained, as vitamin A at low doses has no known toxicity.

We ran a brief series of experiments to investigate this problem. When we applied 25 µM vitamin A, as retinyl acetate, to lung cancer cells (NCI-H460), they lost 7 chromosomes (from 57) after 1 day; K562 (2.5 µM) lost 13 chromosomes (from 69) after 1 day.

KC provides a possible explanation for these epidemiological observations. While anti-oxidants may reduce general cellular damage by controlling free radicals, their ability to contract the karyotypes of *existing cancers* may ultimately *increase* malignancy. This mechanism would be particularly relevant to the penetrating anti-oxidants able to reach tumor cores. At the metabolic level, anti-oxidants reduce oxygen use and enhance glycolysis, a change that cancer cells are well equipped to handle.

 Since these initial epidemiological studies, a number of other investigators have confirmed the paradoxical result that anti-oxidants favor rather than suppress cancer. Our in vitro data supports the idea that certain anti-oxidants can increase cancer rates by inducing KC in pre-existing cancer cells.

# 4.3 Biological Effects of Magnetic Fields

Robust KCs in 5 cancer cells lines suggest that MFs at all intensities are powerful inhibitors of ATP production. A critical difference between molecule-based metabolic restrictors and MFs is that, while all chemicals undergo some form of control before access to inner cellular compartments, MFs benefit from unchecked penetration, making MF action fundamentally different from that of any other agent.

## 4.3.1 Site of Action of Magnetic Fields

Oligomycin inhibits ATPS by binding to its Fo segment. The Fo δ subunit is also named *oligomycin sensitivity conferral protein*. The structure of ATPS is documented in detail (Boyer, 2002) as a rotating motor-alternator structure activated by the trickle of high-density protons from the inter-membrane space into the mitochondrial matrix. Proton diffusion along the 15 nm thick inter-membrane space is not a rate-limiting step in proton translocation across the membrane (1-2 µs) (Procopio and Fornés, 1997). Protons enter the Fo of ATPS along an *entry half-channel* made of four hydrophilic α- helices, to reach a rotating helix. With rotation, protons flow out through a similar *exit half-channel*. The rotation is used by the $F_1$ segment of ATPS to produce ATP (Sasada and Marcey, 2010).

Power-frequency MFs influence the flow of protons through the half-channels of ATPS. These hydrophilic pockets (Fillingame et al. 2003) provide a high density of hydrogen bonds, while the mitochondrial inter-membrane space feeds a high-density of protons. The high-density protons (pH 1, (Procopio and Fornés, 1997)) are driven through the half-channels by a 180 kV/cm electric field (Zorov et al. 2009) across the inner membrane (Mitchell, 1966).

Semikhina at al. (1988) have documented by electrical dissipation factor (ωRC, also known in electrical engineering as *tg δ*) and optical measurements (the dimerization of dilute rhodamine 6G solutions) that alternating MFs in the range 25 nT- 879 µT alter the arrangement of water molecules, particularly under *high concentrations of hydrogen bonds and protons*. It is notable that the effect's reported threshold (25 nT) corresponds to the results in Fig. 3.14.

The dose-responses of Figs. 3.14 and Fig. 3.15 are determined by rising proton impedance (decreased soliton tunneling) through ATPS half-channels. This tunneling of protons exploited by ATPS has also been observed as double wells in neutron Compton scattering studies performed on nanotubes (Reiter et al. 2011).

Semikhina et al (1988) observed a progressive inception of MF effects on water over 5 hours, and dissipation over 2 hours after the field is turned off. The effects are absent above 40-50°C, as water structure changes (Semikhina and Kiselev, 1981). The maximum effect was detected at 156.2-Hz and 15.45 µT for 7°C pure water. It was observed by Semikhina and Kiselev (1981) that a narrow resonance in water is easily broadened by the presence of even small levels of impurities. To investigate the presence of a ratio between frequency and field intensity as reported by Semikhina et al (1988), we measured in anoxic K562 (5 % carbon dioxide and 37°C) 6-day tests at 1 µT the average KCs over frequency as follows: -3.6 ±0.79 at 50-Hz , -9.36 ±1.06 at 60-Hz, -12.71 ±1.82 at 120-Hz and -9.8 ±1.31 at 155-Hz. A polynomial fit predicts maximum MF effect on ATPS at 113 Hz for 1 µT. The ATPS resonance at 1 µT we documented is indeed wider than that reported by Semikhina for pure water.

For many cancer cell types, the dose-response of chromosome contraction vs MFs is remarkably flat (Fig. 3.15). The deviation from flatness in erythro-leukemia cells (Fig. 3.14 and Fig. 3.15) is due, we suspect, to extra-mitochondrial ATP secretion in the cell membrane (Arakaki et al. 2003), a probable feature of this cell type (Das et al. 1994).

## Static Magnetic Field

The influence of the static MF on K562 was investigated using a small cylindrical shield 6.5 cm in inner diameter and 38 cm long comprising 10 layers of 0.4 mm Nickel-Iron-Molybdenum alloy (NIM) (ASTM A753 Type 4) spaced 2.7 mm apart by a neoprene membrane. Such a NIM shield would be highly effective against static as well as ELF MFs, as each layer is computed to have a static MF attenuation of 200, only two of them being theoretically required to attain a 1 nT static MF floor. The NIM shield should reduce both static and ELF MFs to unmeasurable values, lower than the threshold of 34 nT below which the structure of water is assumed to have its optimal molecular arrangement (Semikhina and Kiselev, 1981). According to Russian physicists, proton tunneling through water is reduced by ELF fields higher than 25 nT, but improved by static fields lower than 34 nT. This would be expected to increase the efficiency of ATPS function.

When K562 cells from our < 4 nT at 60-Hz structural steel shields were cultured in the NIM shield for 6 days, they showed signs of a transition by exhibiting loss of apoptotic activity, as revealed by Scepter measurements of cell size distribution, but most importantly, cell numbers increased by a factor of 2.05 ± 0.13 (standard deviation) after 4 days over cells kept in structural steel shields. This increase in proliferation rate is comparable to that observed between 21 % oxygen and anoxic cultures (Fig. 3.1).  Consequently, our observations on cellular metabolism fall in line with the predictions of Semikhina and Kiselev for proton tunneling both for static and ELF fields.

### 4.3.2 Karyotype Contraction and AMPK

The connection between metabolic restrictors, including MFs, and KC can be explained by AMPK action. Perturbations of ATP concentrations trigger AMPK, which activates p53 and reduces both ATP consumption and DNA synthesis (Jones et al. 2005; Motoshima et al. 2006). The suppression of DNA synthesis, part of AMPK's catabolic control, leads to KCs through suppression of chromosome endo-reduplication, the mechanism probably responsible for rapid chromosome count increases in cancer cells (Li et al., 2011).

Two unusual aspects of MF action, adaptation to a stable field over three weeks (Fig. 3.16), and shorter-term sensitivity to small MF changes (Fig. 3.17 and Fig. 3.18) are also explainable by AMPK physiology. AMPK controls long-term dynamic adaptation in muscle (Winder et al. 2000), but is easily triggered by small changes in ATP levels (Hardie and Hawley, 2001).

That small variations in MFs and both increases and decreases in MFs (Fig. 3.17 and Fig. 3.18) alter chromosome counts is unusual. As far as we know, this is the first example of an agent presenting this kind of symmetry, making it possible to sustain KCs indefinitely by judicious selection of MF sequences.

### 4.3.3 Magnetic Field and Diabetes

The MF>ATPS>AMPK pathway is easily detectable in cancer cells because of chromosome instability, but there is no reason to think that the ATPS of normal cells is spared. A major regulator of metabolism (Liu et al. 2006), AMPK modulates insulin secretion by pancreatic beta-cells (Winder and Hardie, 1999), and is investigated for the treatment of diabetes (Viollet et al. 2009). Under continued stimulation, AMPK may facilitate an oxidative as opposed to a glycolytic energy metabolism (Winder and Hardie, 1999). The effects of interference with AMPK function may be wide. When the activity of brain AMPK was pharmacologically inhibited, mice ate less and lost weight. When AMPK activity was pharmacologically raised, mice ate more and gained weight (Kim et al. 2004), a possible link with metabolic syndrome. AMPK activation is also found to attenuate immune cell behaviour (Kanellis et al. 2006), a possible link with the hygiene hypothesis.

## 4.3.4 Karyotype Contractions and Cancer

Cancer cells depend on glycolysis and significantly upregulate it when respiration is inhibited. The Warburg effect manifests as increased glycolysis and reduced mitochondrial respiration (Wu et al. 2006; Jezek et al. 2010). These capabilities of cancer cells allow growth under metabolic restriction by concentration of their efforts on bio-synthesis through the elimination of detoxification mechanisms associated with oxygen exposure, as described in section 2.2.2. The smaller karyotypes maintained under metabolic restriction contribute to tumour core expansion, as fewer chromosomes can be more rapidly duplicated. The survival of tumors could thus be enhanced by chronic metabolic restrictions of hypoxia, oligomycin or MFs (section 4.2.3).

## 4.3.5 Magnetic Fields and Cancer Epidemiology

If KC is indeed a marker of increased malignancy, there is a possibility of carcinogenicity from MF exposures. In such a case, the phenomenon would not be easy to document through epidemiology. First, the threshold for the effect (25 nT) is very low, which means that *all* the population is "exposed". Second, the dose-response is unusually flat (Fig. 3.15), such that useful low and high exposure groups with otherwise similar characteristics would be difficult to assemble. Third, the differential action of MFs may confuse conventional exposure analysis.

Occupational studies are often at the forefront of epidemiological discovery because of their higher and better documented exposures. According to Fig. 3.14, occupational populations of low (0.1 µT) and high exposures (1 µT) have between them a KC difference of "1 chromosome". Domestic MF epidemiology on leukemia may have been successful (Ahlbom et al. 2000; Svendsen et al. 2007) because it benefited from a KC of "10 chromosomes" between 0 and 0.4 µT (Fig. 3.14).

The increased proliferation rates reported above for lung cancer cultures may also be important. Lung cancer was pointed in at least four studies related to EMFs (Vågerö and Olin, 1983; Armstrong et al. 1994; Miller et al. 1996).

-------------------------------------------------------------------------------------------------------

# 5.0 CONCLUSIONS & FUTURE DIRECTIONS

-------------------------------------------------------------------------------------------------------

Metabolic restriction and stimulation contract and expand the chromosome counts of hyperploid cancer cell lines. Metabolic restriction, even in cancer cells not displaying chromosome count changes, may signal increased malignancy. KC caused by metabolic restrictors is therefore an indicator of meta-genetic cancer promotion. According to our models, lipid-soluble anti-oxidants as a means of reducing cancer incidence, particularly in older individuals more likely to harbour tumours, should be viewed with caution.

MFs induce KCs, and we propose that they do so through ATPS inhibition. The first argument is

from physics: MFs in the presence of high concentrations of protons and hydrophilic bonds

change the properties of water, and presumably ATPS's proton flow. The second is from biology:

changes in cell culture characteristics (Fig. 3.18) induced by MFs match closely those of a

specific ATPS inhibitor, oligomycin. A third is the data set relating MFs with KCs across 5

cancer cells and two orders of magnitude of the MF, a threshold predicted by Russian physicists,

and characteristic of a disruption with little sensitivity to MF intensity or to the particularities of cell metabolism: the knockout of a biological enzyme by physics. The fourth is the finding of widened resonances at specific frequency-amplitude combinations, compatible with Semikhina's observations on water structure. The fifth is in the enhanced metabolism of K562 under zero static field, which was predicted by Russian studies on water structure.

What this article proposes is that environmental MFs, through ATPS, act on the core of human metabolism. Past evaluations of MF bio-effects were at a serious disadvantage because of traditional toxicological and epidemiological assumptions, that large responses correspond to large exposures. The controls of *in vitro* scientists were already randomly exposed by the MFs of their incubators. The flatness of MFs' dose-response impaired epidemiological work, as most studies, except for domestic leukemia, worked with tainted controls (Milham, 2010) .

The interaction between living cells and power-frequency MFs may have lied undetected for a long time, because of these unexpected characteristics.

Some diseases appear to have strengthened with no clear causation as more advanced technology, in great part based on electricity, has expanded. Chronic diseases that increased or decreased in the last century, and that are connected to ATP metabolism, should be examined for a link with MFs. Impacts of MFs on ATPS are predictable, as a direct physical effect of MFs on water structure, but our understanding of AMPK and metabolism is incomplete (Jones and Thompson, 2009), making a link between MFs and a specific disease such as diabetes uncertain. And the fact that MF is a physiological agonist of metformin suggests that MF exposure may have had a role in the increased lifespan observed in developed countries in the last century.

# APPENDIX A:
# VB Program Code used to Generate Fig. 2.10

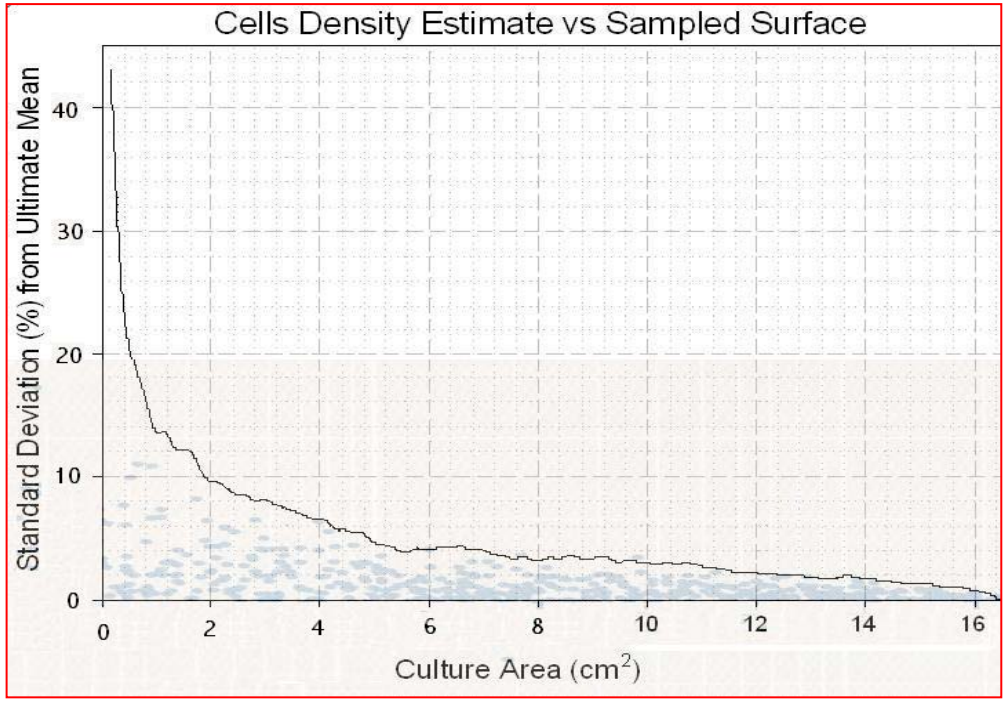

**Fig. 2.10 Accuracy of Cell Density estimate for K562 at 10,000 cells/cm2, as a function of sampled surface. Up to 100 images from a set of 168 images are randomly picked to simulate under-sampling. The standard deviation as a function of the number of images sampled is plotted. 34 images (5.5 cm2) reduces the standard deviation on cell counts to less than 4 %.**

```
Dim CelConts(10, 201) As Long
Dim StaArray(40, 100) As Single
Dim StaArray1(40, 100) As Single

Private Sub CMDEnd_Click()
End
End Sub

Private Sub CmdLogDeriva_Click()
 For N = 1 To 199
  CelConts(9, N) = 1000 * (CelConts(4, N + 1) - CelConts(4, N)) / CelConts(4, N)
  Debug.Print CelConts(5, N)
 Next N
 CelConts(10, 1) = CelConts(9, 1)
 CelConts(10, 2) = (CelConts(9, 1) + CelConts(9, 2) + CelConts(9, 3)) / 3
```

```
 CelConts(10, 3) = (CelConts(9, 1) + CelConts(9, 2) + CelConts(9, 3) + CelConts(9, 4) +
CelConts(9, 5)) / 5
  For M = 4 To 196
    CelConts(10, M) = (CelConts(9, M - 3) + CelConts(9, M - 2) + CelConts(9, M - 1) +
CelConts(9, M) + CelConts(9, M + 1) + CelConts(9, M + 2) + CelConts(9, M + 3)) / 7
  Next M
  bobo = 1
  For X = 1 To 196
  If CelConts(10, X) > bobo Then
  bobo = CelConts(10, X)
  End If
  Next X
  Text1.Text = CStr(bobo)
With Form1.MSChart1
MSChart1.chartType = VtChChartType2dLine
    .ColumnCount = 2
    .RowCount = 201
   For j = 1 To 2
      For I = 1 To 201
        .Column = j
        .Row = I
        .Data = CelConts(j + 8, I)
      Next I
      Next j
End With
End Sub

Public Sub cmdReadFile_Click()
ChDrive "C"
ChDir "C:\Ying\vb practise"
Open "Ying Growth00.txt" For Input As #1
j = 1
Do While Not EOF(1)
Line Input #1, Invalue
Rank = InStr(Invalue, vbTab)
CelConts(0, j) = Val(Left(Invalue, Rank - 1))
CelConts(1, j) = Right(Invalue, Len(Invalue) - Rank)
Debug.Print CelConts(0, j)
Debug.Print CelConts(1, j)
j = j + 1
Loop
NumofReadings = j - 1
Debug.Print NumofReadings
End Sub

Private Sub cmdAssArray_Click()
```

```
Dim A(9) As Long
Dim N As Integer
N = CInt(CmbstpFactor.Text)
If N = 1 Then
  CelConts(2, 1) = CelConts(1, 1)
  CelConts(3, 1) = 200 * Log(CelConts(1, 1))
  CelConts(4, 1) = 200 * Log(CelConts(2, 1))
  For j = 2 To 200
    CelConts(2, j) = (CelConts(1, j - 1) + CelConts(1, j) + CelConts(1, j + 1)) / 3
    CelConts(3, j) = 200 * Log(CelConts(1, j))
    CelConts(4, j) = 200 * Log(CelConts(2, j))
  Next j
  CelConts(2, 201) = CelConts(1, 201)
  CelConts(3, 201) = 200 * Log(CelConts(1, 201))
  CelConts(4, 201) = 200 * Log(CelConts(2, 201))
End If
If N = 2 Then
  CelConts(2, 1) = CelConts(1, 1)
  CelConts(3, 1) = 200 * Log(CelConts(1, 1))
  CelConts(4, 1) = 200 * Log(CelConts(2, 1))
  CelConts(2, 2) = (CelConts(1, 1) + CelConts(1, 2) + CelConts(1, 3)) / 3
  CelConts(3, 2) = 200 * Log(CelConts(1, 1))
  CelConts(4, 2) = 200 * Log(CelConts(2, 1))
  For j = 3 To 199
    CelConts(2, j) = (CelConts(1, j - 2) + CelConts(1, j - 1) + CelConts(1, j) + CelConts(1, j + 1) +
CelConts(1, j + 2)) / 5
    CelConts(3, j) = 200 * Log(CelConts(1, j))
    CelConts(4, j) = 200 * Log(CelConts(2, j))
  Next j
  CelConts(2, 200) = (CelConts(1, 199) + CelConts(1, 200) + CelConts(1, 201)) / 3
  CelConts(3, 200) = 200 * Log(CelConts(1, 200))
  CelConts(4, 200) = 200 * Log(CelConts(2, 200))
  CelConts(2, 201) = CelConts(1, 201)
  CelConts(3, 201) = 200 * Log(CelConts(1, 201))
  CelConts(4, 201) = 200 * Log(CelConts(2, 201))
End If
If N = 3 Then
  CelConts(2, 1) = CelConts(1, 1)
  CelConts(3, 1) = 200 * Log(CelConts(1, 1))
  CelConts(4, 1) = 200 * Log(CelConts(2, 1))
  CelConts(2, 2) = (CelConts(1, 1) + CelConts(1, 2) + CelConts(1, 3)) / 3
  CelConts(3, 2) = 200 * Log(CelConts(1, 2))
  CelConts(4, 2) = 200 * Log(CelConts(2, 2))
  CelConts(2, 3) = (CelConts(1, 1) + CelConts(1, 2) + CelConts(1, 3) + CelConts(1, 4) +
CelConts(1, 5)) / 5
  CelConts(3, 3) = 200 * Log(CelConts(1, 3))
```

```vba
  CelConts(4, 3) = 200 * Log(CelConts(2, 3))
 For j = 4 To 198
    CelConts(2, j) = (CelConts(1, j - 3) + CelConts(1, j - 2) + CelConts(1, j - 1) + CelConts(1, j) +
CelConts(1, j + 1) + CelConts(1, j + 2) + CelConts(1, j + 3)) / 7
    CelConts(3, j) = 200 * Log(CelConts(1, j))
    CelConts(4, j) = 200 * Log(CelConts(2, j))
    Debug.Print j, CelConts(2, j)
 Next j
 CelConts(2, 199) = (CelConts(1, 197) + CelConts(1, 198) + CelConts(1, 199) + CelConts(1,
200) + CelConts(1, 201)) / 5
 CelConts(3, 199) = 200 * Log(CelConts(1, 199))
 CelConts(4, 199) = 200 * Log(CelConts(2, 199))
 CelConts(2, 200) = (CelConts(1, 199) + CelConts(1, 200) + CelConts(1, 201)) / 3
 CelConts(3, 200) = 200 * Log(CelConts(1, 200))
 CelConts(4, 200) = 200 * Log(CelConts(2, 200))
 CelConts(2, 201) = CelConts(1, 201)
 CelConts(3, 201) = 200 * Log(CelConts(1, 201))
 CelConts(4, 201) = 200 * Log(CelConts(2, 201))
End If
If N = 4 Then
 CelConts(2, 1) = CelConts(1, 1)
 CelConts(3, 1) = 200 * Log(CelConts(1, 1))
 CelConts(4, 1) = 200 * Log(CelConts(2, 1))
 CelConts(2, 2) = (CelConts(1, 1) + CelConts(1, 2) + CelConts(1, 3)) / 3
 CelConts(3, 2) = 200 * Log(CelConts(1, 2))
 CelConts(4, 2) = 200 * Log(CelConts(2, 2))
 CelConts(2, 3) = (CelConts(1, 1) + CelConts(1, 2) + CelConts(1, 3) + CelConts(1, 4) +
CelConts(1, 5)) / 5
 CelConts(3, 3) = 200 * Log(CelConts(1, 3))
 CelConts(4, 3) = 200 * Log(CelConts(2, 3))
 CelConts(2, 4) = (CelConts(1, 1) + CelConts(1, 2) + CelConts(1, 3) + CelConts(1, 4) +
CelConts(1, 5) + CelConts(1, 6) + CelConts(1, 7)) / 7
 CelConts(3, 4) = 200 * Log(CelConts(1, 4))
 CelConts(4, 4) = 200 * Log(CelConts(2, 4))
 For j = 5 To 197
    CelConts(2, j) = (CelConts(1, j - 4) + CelConts(1, j - 3) + CelConts(1, j - 2) + CelConts(1, j - 1)
+ CelConts(1, j) + CelConts(1, j + 1) + CelConts(1, j + 2) + CelConts(1, j + 3) + CelConts(1, j +
4)) / 9
    CelConts(3, j) = 200 * Log(CelConts(1, j))
    CelConts(4, j) = 200 * Log(CelConts(2, j))
 Next j
 CelConts(2, 198) = (CelConts(1, 195) + CelConts(1, 196) + CelConts(1, 197) + CelConts(1,
198) + CelConts(1, 199) + CelConts(1, 200) + CelConts(1, 201)) / 7
 CelConts(3, 198) = 200 * Log(CelConts(1, 198))
 CelConts(4, 198) = 200 * Log(CelConts(2, 198))
```

```
CelConts(2, 199) = (CelConts(1, 197) + CelConts(1, 198) + CelConts(1, 199) + CelConts(1, 200) + CelConts(1, 201)) / 5
CelConts(3, 199) = 200 * Log(CelConts(1, 199))
CelConts(4, 199) = 200 * Log(CelConts(2, 199))
CelConts(2, 200) = (CelConts(1, 199) + CelConts(1, 200) + CelConts(1, 201)) / 3
CelConts(3, 200) = 200 * Log(CelConts(1, 200))
CelConts(4, 200) = 200 * Log(CelConts(2, 200))
CelConts(2, 201) = CelConts(1, 201)
CelConts(3, 201) = 200 * Log(CelConts(1, 201))
CelConts(4, 201) = 200 * Log(CelConts(2, 201))
End If
Debug.Print j, CelConts(2, j)
  For N = 1 To 199
  CelConts(5, N) = 1000 * (CelConts(1, N + 1) - CelConts(1, N)) / CelConts(1, N)
  Debug.Print CelConts(5, N)
  Next N
CelConts(6, 1) = CelConts(5, 1)
  For M = 2 To 196
CelConts(6, M) = (CelConts(5, M - 1) + CelConts(5, M) + CelConts(5, M + 1)) / 3
  Next M
CelConts(7, 1) = CelConts(6, 1)
CelConts(7, 2) = CelConts(6, 2)
CelConts(7, 3) = (CelConts(5, 1) + CelConts(5, 2) + CelConts(5, 3) + CelConts(5, 4) + CelConts(5, 5)) / 5
CelConts(7, 4) = (CelConts(5, 1) + CelConts(5, 2) + CelConts(5, 3) + CelConts(5, 4) + CelConts(5, 5) + CelConts(5, 6) + CelConts(5, 7)) / 7
  For M = 5 To 196
   CelConts(7, M) = (CelConts(5, M - 4) + CelConts(5, M - 3) + CelConts(5, M - 2) + CelConts(5, M - 1) + CelConts(5, M) + CelConts(5, M + 1) + CelConts(5, M + 2) + CelConts(5, M + 3) + CelConts(5, M + 4)) / 9
   CelConts(8, M) = (CelConts(5, M - 5) + CelConts(5, M - 4) + CelConts(5, M - 3) + CelConts(5, M - 2) + CelConts(5, M - 1) + CelConts(5, M) + CelConts(5, M + 1) + CelConts(5, M + 2) + CelConts(5, M + 3) + CelConts(5, M + 4) + CelConts(5, M + 5)) / 11
  Next M
  bobo = 1
  For X = 1 To 196
  If CelConts(6, X) > bobo Then
  bobo = CelConts(6, X)
  End If
  Next X
  Text1.Text = CStr(bobo)
End Sub

Private Sub CmdOrgn_Click()
With Form1.MSChart1
MSChart1.chartType = VtChChartType2dLine
```

```
        .ColumnCount = 8
        .RowCount = 201
        For j = 1 To 8
           For I = 1 To 201
              .Column = j
              .Row = I
              .Data = CelConts(1, I)
           Next I
        Next j
End With
End Sub

Private Sub CmdAverage_Click()
With Form1.MSChart1
MSChart1.chartType = VtChChartType2dLine
        .ColumnCount = 8
        .RowCount = 201
        For j = 1 To 8
           For I = 1 To 201
              .Column = j
              .Row = I
              .Data = CelConts(2, I)
           Next I
        Next j
End With
End Sub

Private Sub CmdLogOrig_Click()
With Form1.MSChart1
MSChart1.chartType = VtChChartType2dLine
        .ColumnCount = 8
        .RowCount = 201
        For j = 1 To 8
           For I = 1 To 201
              .Column = j
              .Row = I
              .Data = CelConts(3, I)
           Next I
        Next j
End With
End Sub

Private Sub CmdLogAver_Click()
With Form1.MSChart1
MSChart1.chartType = VtChChartType2dLine
        .ColumnCount = 8
```

```vb
        .RowCount = 201
      For j = 1 To 8
        For I = 1 To 201
           .Column = j
           .Row = I
           .Data = CelConts(4, I)
        Next I
      Next j
End With
End Sub

Private Sub CmdAll_Click()
With Form1.MSChart1
     .ColumnCount = 4
  .RowCount = 201
MSChart1.chartType = VtChChartType2dLine

      For j = 1 To 4
        For I = 1 To 201
           .Column = j
           .Row = I
           .Data = CelConts(j, I)
        Next I
      Next j
End With
End Sub

Private Sub CmdDxdt_Click()
With Form1.MSChart1
MSChart1.chartType = VtChChartType2dLine
      .ColumnCount = 3
      .RowCount = 201
     For j = 1 To 3
        For I = 1 To 201
           .Column = j
           .Row = I
           .Data = CelConts(j + 4, I)
        Next I
        Next j
End With
End Sub

Private Sub cmdStaCurve_Click()
ChDrive "C"
ChDir "C:\Ying\vb practise"
Open "08m09i16.txt" For Input As #1
```

```
j = 1
Do While Not EOF(1)
Line Input #1, Invalue
Rank = InStr(Invalue, vbTab)
Dim CellStat(2, 168) As Long
CellStat(0, j) = Val(Left(Invalue, Rank - 1))
CellStat(1, j) = Right(Invalue, Len(Invalue) - Rank)
j = j + 1
Loop
NumofReadings = j - 1
Dim RNDArray(100) As Long
With Form1.MSChart1
MSChart1.chartType = VtChChartType2dLine
        .ColumnCount = 30
        .RowCount = 100
For X = 1 To 30
    Dim Sum
    Dim Average
    Average = 0
    Sum = 0
      For N = 1 To 100
        Dim MyValue

        MyValue = Int((168 * Rnd) + 1)
        RNDArray(N) = CellStat(1, MyValue)
        Sum = Sum + RNDArray(N)
      Next N
      Average = Sum / 100
       'Debug.Print Average
  For I = 1 To 100
    If I = 1 Then
    StaArray(X, 1) = RNDArray(1)
    Else
  StaArray(X, I) = (StaArray(X, I - 1) * (I - 1) + RNDArray(I)) / I
    End If
    StaArray1(X, I) = StaArray(X, I) / Average
    'Debug.Print StaArray(X, I) / Average
            .Column = X
            .Row = I
            .Data = StaArray(X, I)
        Next I
Next X
End With
End Sub

Private Sub Cndsigma_Click()
```

```
ChDrive "C"
ChDir "C:\Ying\vb practise"
Open "08m09i16.txt" For Input As #1
j = 1
Do While Not EOF(1)
Line Input #1, Invalue
Rank = InStr(Invalue, vbTab)
Dim CellStat(2, 168) As Long
CellStat(0, j) = Val(Left(Invalue, Rank - 1))
CellStat(1, j) = Right(Invalue, Len(Invalue) - Rank)
j = j + 1
Loop
NumofReadings = j - 1
Dim RNDArray(100) As Long
For X = 1 To 40
   Dim Sum
   Dim Average
   Average = 0
   Sum = 0
     For N = 1 To 100
       Dim MyValue
       MyValue = Int((168 * Rnd) + 1)
       RNDArray(N) = CellStat(1, MyValue)
       Sum = Sum + RNDArray(N)
     Next N
     Average = Sum / 100
      'Debug.Print Average
  For I = 1 To 100
    If I = 1 Then
    StaArray(X, 1) = RNDArray(1)
    Else
   StaArray(X, I) = (StaArray(X, I - 1) * (I - 1) + RNDArray(I)) / I
    End If
     StaArray1(X, I) = StaArray(X, I) / Average
      Next I
Next X
Dim s(100) As Single
With Form1.MSChart1
MSChart1.chartType = VtChChartType2dLine
      .ColumnCount = 8
      .RowCount = 100
For j = 1 To 8
For I = 1 To 100
    Sum = 0
    Average = 0
    Sum1 = 0
```

```
    For X = 1 To 40
        Sum = Sum + StaArray1(X, I)
    Next X
     Average = Sum / 40
    Debug.Print Average
    For X = 1 To 40
     Sum1 = Sum1 + (StaArray1(X, I) - Average) ^ 2
    Next X
    s(I) = (Sum1 / 40) ^ 0.5
            .Column = j
            .Row = I
            .Data = s(I)
    Next I
Next j
End With
End Sub

Private Sub MSChart1_Click()
End Sub
```

# APPENDIX B:
# VB Program Code used to Automatically Assess Cell Cultures

```
Public CNumber As Integer
Public Allnumber As Integer
Public Anumber  As Integer
Dim Roundness(2000) As Single

Private Sub CmdCalculate_Click(Index As Integer)
Dim ExpDen As Integer
Dim CelNum As Single
ExpDen = CInt(ExpDenTxt.Text)
Call cellnumber_assessment
TxTClCnt.Text = CStr(CNumber)
CelNum = CNumber / 0.0025
Text6.Text = CStr(CelNum)
TXTPlCnt.Text = CStr(Allnumber - CNumber)
If Form1.CombContainer.Text = "T12" Then
Text3.Text = CStr(3.75 * ExpDen / CelNum)
Text7.Text = CStr(3.75)
End If
If Form1.CombContainer.Text = "T25" Then
Text3.Text = CStr(7.5 * ExpDen / CelNum)
Text7.Text = CStr(7.5)
End If

If Form1.CombContainer.Text = "96-well Quartz Plate" Then
Text3.Text = CStr(7.5 * ExpDen * 96 * 0.317 / (25 * CelNum))
Text7.Text = CStr(0.44 * 96)
End If
End Sub

Private Sub cmdCntRound_Click(Index As Integer)
Call cell_roundness
'ReDim Roundness(Anumber - 1) As Single
x = 0
For I = 1 To Anumber - 1
x = x + Roundness(I)
Next I
y = x / I
TxtRndness.Text = CStr(y)
End Sub
```

```
Private Sub Command1_Click()
End
End Sub

Public Sub cellnumber_assessment()
    ret = IpWsLoad("R:\cellnumbers.jpg", "jpg")
    ret = IpWsConvertImage(IMC_GRAY12, CONV_SCALE, 0, 0, 0, 0)
    ret = IpSegSetRange(0, 0, 1000)
    ret = IpSegPreview(ALL_W_B)
    ret = IpSegShow(1)
    ret = IpSegSetAttr(SETCURSEL, 0)
    ret = IpSegSetAttr(Channel, 0)
    ret = IpSegPreview(ALL_W_B)
    ret = IpSegCreateMask(5, 0, 1)
    ipICal(0) = 0
    ipICal(1) = 4095
    ret = IpBlbMultiRanges(ipICal(0), 1)
    ret = IpSegShow(0)
    ret = IpBlbShow(1)
    ret = IpBlbEnableMeas(BLBM_AREA, 1)
    ret = IpBlbEnableMeas(BLBM_ROUNDNESS, 1)
    ret = IpBlbSetAttr(BLOB_BRIGHTOBJ, 0)
    ret = IpBlbSetAttr(BLOB_MINAREA, 1)
     ret = IpBlbEnableMeas(BLBM_AREA, 1)
     ret = IpBlbEnableMeas(BLBM_ROUNDNESS, 1)
    ret = IpBlbSetFilterRange(BLBM_AREA, 90, 1000)
    ret = IpBlbLoadSetting("C:\Ying\vb practise\cell assessment\Cells.ENV")
    Cnum = IpBlbCount()
    CNumber = Int(Cnum)
    ret = IpBlbSetFilterRange(BLBM_AREA, 5, 1000)
    ret = IpBlbSaveSetting("C:\Ying\vb practise\cell assessment\All.ENV")
    'Perform Count AND analysis of all objects [particles AND cells]
    Allnum = IpBlbCount()
    Allnumber = Int(Allnum)

    If ChkNumbers.Value = 1 Then
    ret = IpBlbSetAttr(BLOB_LABELMODE, 1) 'With numbers
    Else
    ret = IpBlbSetAttr(BLOB_LABELMODE, 0) 'No numbers
    End If
End Sub

Private Sub Form_Load()
Form1.Top = 0
```

```
Form1.Left = 0
ret = Shell("C:\IPWin4\ipwin32.EXE", 1) ' Run Image-pro Plus
ret = IpAppMaximize()
ret = IpAppSize(678, 645)
ret = IpAppMove(353, 0)
End Sub

Public Sub cell_roundness()
    ret = IpWsLoad("R:\cellroundness.jpg", "jpg")
    ret = IpWsConvertImage(IMC_GRAY, CONV_SCALE, 0, 0, 0, 0)
    ret = IpBlbSetAttr(BLOB_MEASUREOBJECTS, 1)
    ret = IpBlbSetAttr(BLOB_OUTLINEMODE, 3)
    ret = IpBlbSetAttr(BLOB_FILLHOLES, 1)
    ret = IpBlbSetFilterRange(BLBM_AREA, 90#, 10000000#)
    ret = IpBlbEnableMeas(BLBM_ROUNDNESS, 1)
    ret = IpBlbLoadSetting("C:\Ying\vb practise\cell assessment\Cells.ENV")
    Anumber1 = IpBlbCount()
    Anumber = Int(Anumber1)
    ret = IpBlbUpdate(0)
ReDim Roundness(Anumber - 1) As Single
'Get the data from images
ret = IpBlbData(BLBM_ROUNDNESS, 0, Anumber - 1, Roundness(0))
End Sub
```

**APPENDIX C:**
**Li Y, Héroux P,  Kyrychenko I.**
**Cytotoxicity Testing with Anoxic K-562.**
**XII International Congress of Toxicology, 2010, P201-009:**
**In Vitro Testing Methods.**

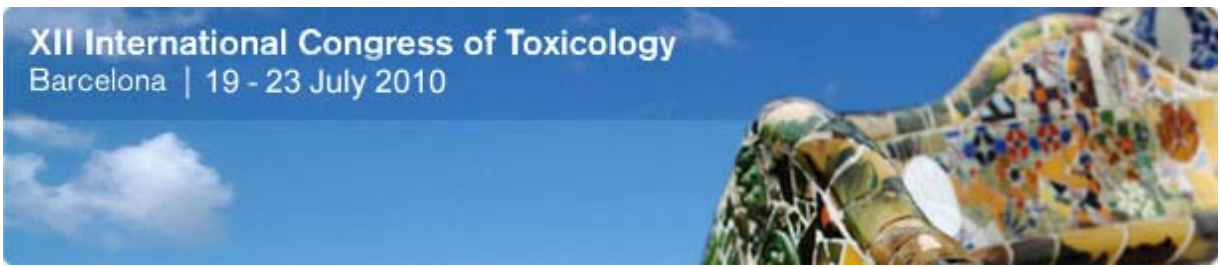

**CYTOTOXICITY TESTING WITH ANOXIC K-562**

An *in vitro* cellular model is developed to investigate the cytotoxicity of industrial metals. Based on the erythroleukemia cell K-562 in a synthetic medium, the model is supported by automated data acquisition and computer vision. Since Reactive Oxygen Species (ROS) are central to metal toxicities, we investigated the effect of free radicals by growing the cells in various levels of oxygen. From culturing K-562 in 0, 2, 5 and 21 % oxygen, we found that oxygen levels influence proliferation rate, footprint, roundness and self-adhesion. Somewhat surprisingly, K-562 proliferated 1.45 times faster under anoxia than under oxic (21 % oxygen) conditions. Phenotype measurements also showed a more normal behavior and uniform cell anatomy under anoxia. We find that the K-562 chromosome count reduces under anoxia, from the pseudo-triploid (69) K-562 oxic karyotype. The anoxic karyotype is closer to the ones found in clinical chronic myeloid leukemia studies, and may result either from stem cells occurring below usual detection levels or from chromosome count instability. Because anoxic K-562 grows faster and has a most regular shape, we propose that K-562 originates from a very anoxic medullary niche in the body, and culturing it under anoxic conditions is most representative of the original cancer lesion. Anoxia also displays a higher apoptosis rate than oxic culture, which is compatible with previous reports of oxygen as an inhibitor of apoptosis. We conclude that anoxia produces a most sensitive K-562 model for metals cytotoxicity to stem cells because it (1) is physiological, (2) has a larger sensitivity to apoptosis, allowing detection of stem cell chronic depletion, (3) has a low baseline ROS and therefore sensitivity to metals-generated ROS, and (4) has a lower chromosome number.

**APPENDIX D:**
**Li Y, Héroux P,  Kyrychenko I.**
**Metabolic restriction of cancer cells in vitro causes**
**karyotype contraction — an indicator of cancer promotion?**
*Tumor Biology* **33(1), pp. 195-205, 2012.**

RESEARCH ARTICLE

# Metabolic restriction of cancer cells in vitro causes karyotype contraction—an indicator of cancer promotion?

**Ying Li · Paul Héroux · Igor Kyrychenko**

**Abstract** The metabolism of cultured cancer cells is stimulated by 21% oxygen and generous nutrition, while real tumors grow in oxygen and nutrient-restricted environments. The effect of these contrasted conditions was studied in five hyperploid (54–69) cancer cell lines. When grown under anoxia and in the presence of antioxidant metabolic restrictors, the cell lines quickly reverted to almost normal chromosome numbers (47–49). The stepped withdrawal of oxygen over K562 showed progressive increases in proliferation rate and the acquisition of a stable, stem phenotype. In genetic studies, hyperploid cancer cells adjusted their chromosome numbers up or down to match their micro-environment through rapid mechanisms of endo-reduplication or chromosome loss. These fast reactions may explain the surprising adaptability of tumor cells to oncologic interventions. Furthermore, karyotype contraction may provide a basis for the previously observed carcinogenic influence of the administration of some antioxidants in human populations.

## Background

The growth of cancer cells is widely believed to be limited by oxygen and nutrients, occulting their truly critical property, the ability to thrive in diverse sites. Because of their metabolic flexibility, cancer cells survive both in the highly hypoxic tumor cores and in the blood. Considering that 82% of oxygen readings taken in solid tumors are less than 0.33% [1], compared to up to 21% in standard in vitro culture, relatively few oncological studies are based on hypoxic or metabolically restricted models. The work available on cancer cell hypoxia typically reaches down to 1% oxygen, and there are scant studies on true anoxia. One interesting study on a wide range of bone marrow samples compared proliferation at 19%, 3%, and 1% oxygen, showing that the large majority of samples generated more cells and increased colony-forming units under increased hypoxia [2]. Cell survival and proliferation rates reported by these deep hypoxia studies vary widely [3–7]. The oxygen and nutrient gradients [8, 9] between the core and cortex of a tumor influence both the phenotype and genotype of cancer cells. One prominent characteristic is chromosome number. Although variable and increased chromosome numbers (above 46) are routinely observed in the majority of human tumors, little physiological meaning has been associated with this variable. If human tumor cells are cultured ex vivo, chromosome numbers may increase even further [10], thus displaying a basic cancer characteristic termed chromosome instability (CIN). Therefore, in vivo as well as in vitro, the majority of cancer cells display hyperploid as well as unstable karyotypes.

Y. Li · P. Héroux
InVitroPlus Laboratory, Department of Surgery,
Royal Victoria Hospital,
Montreal, Canada

Y. Li · P. Héroux
Department of Epidemiology,
Biostatistics and Occupational Health, McGill University,
Montreal, Canada

I. Kyrychenko
Cell Culture Technologies,
Toronto, Ontario, Canada

P. Héroux (✉)
Faculty of Medicine, McGill University,
1110 Pine Ave West,
Montreal, Canada H3A 1A3
e-mail: paul.heroux@mcgill.ca

 Springer

Some perspective on the meaning of chromosome number increases can be gained by reviewing the four classes of genetic alterations classically recognized in neoplasms: gene amplification, sequence changes, translocations (such as BCR:ABL in K562), and *chromosome number alterations* [11]. These classes testify to a plurality of mechanisms in cancer, but all create hyperplasia, suggesting that these lesions find a common ground in metabolic stimulation [12]. Gains or losses of chromosomes can be associated with cancer initiation, but the majority view is that these changes occur together with more basic genetic changes, for example, trisomy of chromosome 7 with a mutation in MET in hereditary papillary renal carcinoma [13] and loss of 10q with alteration of the PTEN gene in glioblastoma multiforme [14]. On the other hand, added chromosomes mark cells for greater cancer risk [15] and poor clinical outcomes in leukemias [16].

The experiments described here illustrate how metabolic modulation can reversibly add or remove whole chromosomes in a cell culture, but without altering the original cancer lesion, showing CIN as a post-initiation contributor to karyotype evolution. In this context, anoxia and metabolic restriction may be useful techniques to identify the critical early mutations [17, 18] that cause cancer.

To investigate the effects of metabolic restriction on chromosome duplication, we studied karyotype changes in five cell lines: K562 and HEL 92.1.7 (erythro-leukemias), NCI-H460 (lung cancer), COLO 320DM (colon cancer), and MCF7 (breast cancer). The cell lines are challenged by five *metabolic restrictors* that impair oxygen metabolism, ATP synthesis, or ATP use: an environmental chemical (anoxia), a neuro-hormone (melatonin), a dietary antioxidant (vitamin C), a metabolic inhibitor (oligomycin), and a cancer drug (imatinib).

Cellular metabolism has two important segments, glycolysis and mitochondrial oxidation, which can be controlled independently. Anoxia is of great interest because it essentially shuts down mitochondrial metabolism by enhancing glycolysis (the Warburg effect [19]) and driving pyruvate to lactate. It is also a condition that is perfectly well tolerated by tumors, a flexibility absent in most normal cells. Two of the five metabolic restrictors used in this study suppress mitochondrial metabolism [20]. Anoxia eliminates oxidative phosphorylation in mitochondria, and oligomycin inhibits the $F_0$ complex of ATP synthase (ATPS), the bottleneck enzyme allowing mitochondrial production of ATP [21]. The antioxidants melatonin and vitamin C reduce oxygen consumption and thus perturb metabolic activity. Melatonin is a direct scavenger of $O_2$ [22], while vitamin C is an electron donor for eight different enzymes [23]. The last, imatinib, is a competitive inhibitor of the BCR–ABL enzyme's ATP-binding site and is used to treat chronic myeloid leukemia (CML) and other malignancies [24].

When administered, each of the five metabolic restrictors elicits cellular reactions aimed at maintaining homeostasis. The clearest example is oligomycin, which inhibits ATPS [21] and triggers AMP-activated protein kinase (AMPK), a sensitive regulator of ATP levels [8, 12, 25–28]. The activation of AMPK suppresses anabolism and stimulates catabolism [29–31], which we believe lead to the chromosome losses observed in our experiments.

Cancer cells displaying CIN are ideal test objects to assess perturbations in metabolism, as these perturbations, through the regulation of AMPK, are reflected in the following cell divisions as altered chromosome numbers. This article uses chromosome number as a reporter of metabolic state and focuses on the rapid and reversible losses of chromosomes resulting from metabolic restriction in hyperploid cells.

## Materials and methods

### Cells and culture conditions

The K562, HEL 92.1.7, NCI-H460, COLO 320DM, and MCF7 cell lines were obtained from the ATCC and maintained at 5% $CO_2$, 90% humidity, and 0% (anoxia), 2%, 5%, 10%, or 21% (atmoxia) oxygen as needed. The culture medium is RPMI-1640 with L-glutamine (Sigma 61-030-RM), sodium selenite 20 nM (Sigma S-5261), bovine insulin 1 mg/l (Sigma I5500), iron-saturated bovine transferrin 25 mg/l (Sigma T1408), sodium bicarbonate 2 g/l (Sigma S-6014), and bovine serum albumin 4 g/l (Sigma A3311). For proliferation, inflammation, and adhesion measurements, cells were incubated in vented T-25 flasks and transferred to T-25 s (Sarstedt 83.1810.502) or T-12 s (Falcon 353018) for experiments.

### Imaging

A series of images for proliferation, inflammation, and adhesion measurements were produced using Diavert (Leitz) or DMIRB (Leica) microscopes, in brightfield illumination, at magnifications of ×20 to ×2.5 with an Infinity Lite (1.5 Mpixels) CMOS camera (Lumenera). For cytogenetics, ×100 oil immersion, a Laborlux D (Leitz) microscope, and an Infinity X (21 Mpixels) CMOS camera (Lumenera) were used. Object recognition functions for serial images and image enhancement for cytogenetics were provided by Visual Basic control of Media Cybernetics ImagePro Plus macro language [32]. Object recognition for serial images was facilitated by the acquisition of pairs of images with offset focal planes. A focused image (maximized JPG size) and a second image offset by 25 μm at ×10 are subtracted to improve contrast. The resulting image is then enhanced, and the objects are recognized using conventional imaging procedures.

Chromosome number, cell shape, and cell grouping

Metaphase preparation and cytogenetic analysis were performed according to standard cytogenetic procedures on the five cancer cell lines with a trypsin-Giemsa banding technique. Spectral karyotyping (Applied Spectral Imaging, Sky Vision) and FISH protocols of anoxic K562 cells were performed on metaphase spreads [33] at the Banque de Cellules Leucémiques du Québec (Centre de Recherche, Hôpital Maisonneuve-Rosemont, Montreal, Quebec, Canada).

Measurements of proliferation, *Roundness* and *Hex-distance*, are derived from a large series of cell culture micrographs automatically acquired over days. Culture vessels, T-12 s or T-25 s, are placed on the motorized stage (Marzhauser MCL) of a microscope contained in an incubator. Stage acceleration is limited to 0.34 cm/s², producing barely detectable vibration of settled cells in a T-12 with 3 mm of medium. Stage control, image acquisition, analysis, and first line data compilation procedures are realized by custom Visual Basic software developed in our laboratory [32]. An advantageous configuration is the joining of four T-12 s together (edge epoxy), which are then seeded with culture aliquots, and identically treated both mechanically and environmentally. In the transition tests, oxygen concentrations are changed without test interruption by feeding the desired oxygen concentrations through four flexible tubes connecting the vented caps of T-12 s (volume of 25 ml) to 2 liter gas reservoirs.

Roundness is the dimensionless ratio between the imaged cell perimeter and a circle of identical area. A circle has a Roundness of 1; all other shapes yield higher numbers. For suspension cells, which are normally close to spherical, Roundness quantifies taxis of the sub-population of macrophage-like cells within the culture.

Hex-distance is 360° divided by the sum of the angles shadowed on a cell's surface by its six nearest neighbors. In ideal two-dimensional packing of identical spheres (Fig. 5), dimensionless Hex-distance is equal to 1. It rates 5 to 7 for cells newly seeded at 9,000 to 7,000/cm².

$$Hex - distance = \frac{\pi}{\sum_1^6 \sin^{-1} \frac{1}{D_i} \sqrt{\frac{A_i}{\pi}}} \qquad (1)$$

where $D_i$ is the center-to-center distances between one cell and its six nearest neighbors of area $A_i$.

## Results

### Karyotype contractions

Karyotype contractions on five hyperploid cancer cell lines are presented in Table 1. The "Atmoxia baseline" column lists the chromosome numbers observed under standard culture conditions, for comparison with the other columns reporting the chromosome losses sustained by cell cultures under four strong metabolic restrictions. Anoxia represents the deep hypoxia of tumor cores. Oligomycin and imatinib were used at sub-toxic levels: serial dilutions allowed the determination of concentra-

**Table 1** Karyotype contractions in cancer cell lines after 3-day metabolic restrictions

| Cell | Type | Atmoxia baseline | Anoxia alone | Atmoxia | | |
|------|------|------------------|--------------|---------|---|---|
| | | Mode (80% range) | Mode (80% range) | Oligomycin 0.1 μM[a] Mode (80% range) | Imatinib 0.08 μM[b] Mode (80% range) | Melatonin–vit C 0.3 μM, 150 μM[c] Mode (80% range) |
| K-562[d] | Erythro-leukemia | 69 (64–70) | 62 (58–62) | 48 (46–53) | 47 (45–51) | 48 (45–52) |
| HEL 92.1.7[e] | Erythro-leukemia | 66 (62–67) | 59 (57–60) | 47 (46–51) | 48 (47–53) | 49 (46–52) |
| NCI-H460 | Large cell lung cancer | 57[f] (53–65) | 51 (45–52) | 47 (46–49) | 47 (46–50) | 47 (45–51) |
| COLO 320DM | Colo-rectal adenocarcinoma | 54[f] (49–61) | 48 (46–49) | 46 (46–48) | 46 (45–48) | 47 (45–49) |
| MCF7 | Breast adenocarcinoma | 82[f] (66–87) | 74 (61–75) | 64 (59–66) | 65 (61–68) | 63 (59–65) |

Number of metaphases for each determination: 122, 108, 30, 25, 25; 50, 50, 30, 25, 20; ATCC, 20, 20, 20, 20; ATCC, 50, 20, 20, 20; ATCC, 35, 20, 25, 30

[a] Proliferation IC$_{50}$ of 0.0125 μM for K562 and HEL IC$_{50}$ of 0.1 μM for other types

[b] K562 proliferation IC$_{50}$

[c] The melatonin–vitamin C concentrations were optimized for K562 chromosome number normalization

[d] BCR-ABL positive

[e] BCR-ABL negative

[f] ATCC data

tions suppressing proliferation to 50% of normal ($IC_{50}$), still adequate for karyotyping. Melatonin–vitamin C were optimized for maximum chromosome drop through serial dilutions, and these levels were subsequently found to be physiological, as they matched those in bone marrow [34] and plasma.

"Anoxia alone" (Table 1) resulted in partial karyotype contractions, which means that six to eight chromosomes were lost in the five cell lines, bringing their totals closer to 46. This contraction remains as long as anoxia persists, but is reversed over a few weeks if atmoxia is restored.

The metabolic restrictors of the last three columns of Table 1 contract karyotypes further than anoxia in all cell lines, whether starting from atmoxia or anoxia (not shown). NCI-H460 and COLO 320DM chromosome numbers are almost normalized by anoxia alone, possibly indicating a deeper dependence on oxygen metabolism. The apparent failure to reach peri-normal numbers in MCF7 may be due to the induction of cell fusions by the metabolic restrictors [35].

K562 chromosome number ranges can be expanded beyond the values in Table 1. Higher melatonin levels contracted them below 46, while hyperoxia (50% and 95% oxygen) expanded the usually small "2S" (~138 chromosomes) population. Since anoxia alone reduces by six to eight chromosomes, while other metabolic restrictors return chromosome numbers close to normal, except for MCF7, it can be concluded that hyperploidy in cancer cells is not *essential*, but is circumstantially connected with enhanced metabolism.

## K562 chromosomes lost in anoxic transition

The karyo-histogram of anoxic K562 (black in Fig. 2) is remarkably narrower compared to that of any atmoxic cancer cell culture [36], or even of normal human tissue, such as the brain [37]. The stability of the 62-karyotype (56% in Fig. 2) created an opportunity for a chromosome-by-chromosome comparison with a documented K562 atmoxic 67-karyotype published by Naumann et al. [38].

G-banding and spectral karyotyping (Fig. 1) detected the whole human genome in the anoxic 62-karyotype. Chromosome 14 shows a Robertsonian translocation, common in acute myeloid leukemia and CML. The lost satellite of chromosome 14 is redundant with micro-satellites of 13, 15, 21, and 22. There is also a 20% incidence of a normal chromosome 14 in the culture. A single normal chromosome 17, reported by Lozzio and Lozzio [10], is conserved in both anoxic and atmoxic karyotypes. Fluorescence in situ hybridization showed that the BCR:ABL fusion on chromosome 22, part of the standard atmoxic K562 line, is conserved in anoxia. Comparison of the two karyotypes reveals that 11 chromosomes move closer to normality, while two become more abnormal. Many markers of the 67-karyotype are eliminated in the 62-karyotype. These observations, together with a low rate of apoptosis in anoxic transitions (see below), suggest that anoxia contracts karyotypes by methods that are not degenerative or random, but selective.

## Oxygen and karyotypes

Anoxic cell cultures, which have no oxygen concentration inhomogeneity, show low chromosome number ranges. All cell lines in Table 1 have larger atmoxic than anoxic number ranges. In Fig. 2, the left tail of the histograms is larger for atmoxic (white, 65 to 67) than anoxic (black, 58 to 60) distributions. These two observations can be explained if oxygen increases not only chromosome number, but also chromosome number range.

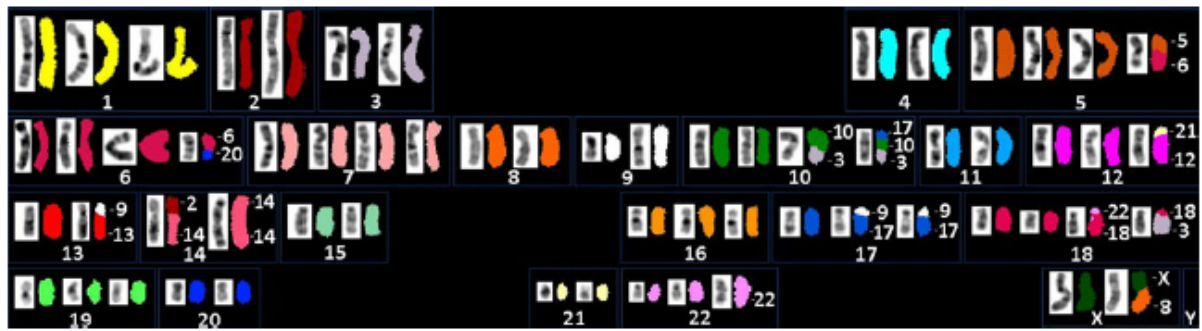

**Fig. 1** G-banding and SKY results from 20 anoxic K562 cells, detected in all but four (absence of i(14)(q10): normal chromosome 14). Chromosomal abnormalities are relate to the 3n (69 chromosomes) level, according to ISCN 2009: −2, dup(2)(q?): duplicated long arm segment localized on p, −3, −4, +der(5)t(5;6)(q?;p?), dup(6)(p?), +der(6)t (6;20), +inv(7)(p?p?), −8, −9, del(9)(p12), dup(9)(q?): duplicated long arm segment localized on 9p, der(10)t(3;10), +der(10)t(3;10;17), −11, der (12)t(12;21), −13, der(13)t(9;13), −14, der(14)t(2;14), i(14)(q10), −15, der(17)t(9;17)×2, ?del(18)(q?), der(18)t(18;22), der(18)t(3;18), −20, −21, dup(22q?q?), −X, der(X)t(X;8), dup(X)(q?): duplicated long arm segment localized on Xp. From *Plate-Forme de Cytogénétique*, Maisonneuve-Rosemont Hospital

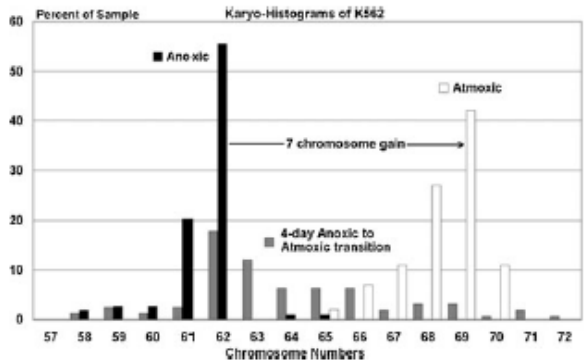

**Fig. 2** K562 karyo-histograms of anoxic cultures (*black*, 108 metaphases) and atmoxic cultures (*white*, 50 metaphases). Similar atmoxic data are published by ATCC and Schultz and Chan [20]. The *gray* 4-day transition data include 157 metaphases from three different tests. Some atmoxic transition karyotypes lie below (22%) and above (11%) the graph, while such outliers are almost entirely absent from the anoxic transition (not shown)

Oxygen diffuses poorly in culture media, so concentrations at the bottom of a dish depend on depth and cell density [39]. Growth patterns also influence karyo-histograms under oxygen, as a cell's position within a group impacts its oxygen exposure. This view is supported in Table 1 by the smaller atmoxic chromosome ranges for suspension cells (the two erythro-leukemia types) compared to the adherent lines.

Re-seeding of cancer cell cultures at 21% oxygen is likely to increase chromosome numbers and range, because fewer and evenly scattered cells are better oxygenated. Different karyo-histograms have been reported for atmoxic K562 [40], supporting that different karyotypes result from different culture conditions. Looking back in history, the K562 cells harvested by Lozzio at 45–46 chromosomes in 1971 may have increased their chromosome numbers as a result of the standard atmoxic cell culture conditions. All evidence points to an influence of oxygen gradients over space and time on cell culture karyotype diversity and pleomorphism under standard atmoxic cell culture conditions.

When transferred in vivo, the influences of oxygen and metabolism can explain the diverse phenotypes and karyotypes of tumors [41, 42]. Furthermore, the high level of CIN in tumors, although rooted in the initial cancer lesion, is likely controlled by the strong oxygen and metabolic gradients of tumor structure.

### Oxic transitions

Anoxic K562 is not only an ideal model to detect metabolic disturbances, as revealed by chromosome number changes, but can also be used to investigate the mechanisms of CIN. We performed *oxic transition assays*, where stable anoxic and atmoxic cultures are transferred to their opposite oxygen level (21%>0% and 0%>21%). By karyotyping at various time intervals following the transitions, it was noticed that chromosome losses after anoxia are instantaneous, while gains after atmoxia need many cell cycles to complete.

Figure 2 compares three K562 karyo-histograms: anoxic, atmoxic as well as an intermediate, the 4-day anoxic to atmoxic transition. Four days is an auspicious time to investigate the atmoxic transition (Fig. 2, gray), as intermediate karyotypes (63 to 67) bridge the gap between the anoxic mode of 62 and the atmoxic mode of 69.

The traditional mechanism of tumor adaptation, clonal expansion [41], strains to explain the rapid evolution of karyotypes observed. How can the 63 chromosome karyotype, undetectable under anoxia (black in Fig. 2), appear at 12% only 4 days later (gray 63) and disappear afterwards from the long-term atmoxic signature (no white 63)? The results of Fig. 2 imply, rather, that chromosome number instability (CIN) [11, 42] allows chromosome additions to occur with high probability at each cell division.

Chromosome numbers could increase in the atmoxic transition as a result of endo-reduplication (END) [43], a mechanism of unscheduled chromosome duplication. Disturbances of chromosome segregation or cytokinesis can also cause chromosome gains or losses. Mitotic disturbances such as nondisjunction (weakened mitotic checkpoint) or merotelic attachment (kinetochore attached to both mitotic spindles) [44] produce a high probability of both lowered and increased chromosome counts in daughter cells. More complex spindle aberrations, such as multipolar divisions, cause severe chromosome missegregation, leading to lethality [45]. We compared in a computer model the relative importance of non-disjunction with that of END, based on the assumption that failed segregation contributes symmetrically around a mode, while END contributes a simple chromosome gain. Starting from the anoxic karyo-histogram, the computer simulation modifies karyotypes according to various probabilities of the competing mechanisms, to match the 4-day transition karyo-histogram in Fig. 2. Results show that the dominant mechanism is END (21% probability per division), rather than asymmetric segregation (1.5% probability per division). The simulations choose END to dominate asymmetric segregation because the transition karyo-histogram needs to drift right from the anoxic karyo-histogram.

In the opposite transition (21%>0%), only overwhelming CIN can explain the rapid chromosomes losses observed, where within 1 day, two thirds of metaphases are already within the long-term anoxic envelope (not shown). The speed at which these transitions occur far outpaces clonal expansion following random mutations. The mechanism supporting the rapid chromosome losses is uncertain. It is possible that under altered metabolism, END chromosomes are not or are incompletely synthesized. An alternate mechanism would involve a small probability that chromosomes do not reach the mitotic plate and thereafter decay.

Apoptosis was measured [32] in the transition assays at 0.4% to 0.15%/h for about 2 h, followed by stable incidence rates of about 0.15%/h (not shown). But over long exposures, apoptosis is strongly inhibited by atmoxia, to typical levels of only 0.01%/h. Rises in apoptosis rates coincide with rapid karyotype changes, but neither karyotype contraction nor expansion creates substantial debris within the culture. Karyotype contractions produce few highly hyperploid (>72) karyotypes: at 4 days, the proportion is only 2.4% with oligomycin and undetected with imatinib, vitamin C–melatonin, and baseline cultures.

Our observations suggest that tumors can contract and expand their karyotypes with surprising speed as oxygen levels and metabolic rates are altered. From cultures in transition, it was observed that chromosome changes were shadowed by phenotype alterations and that the superior karyotype stability under anoxia was mirrored in the cells' phenotypes.

Proliferation, inflammation, and adhesion

The hypoxic center of tumors is widely perceived as in need of more perfusion, because anoxia in viewed as an inherently pathological state. Measurements below present a different picture.

First, Fig. 3 shows the proliferation rate of K562 increasing smoothly with deepening hypoxia. Although

ATP metabolism is more efficient with oxygen, anoxic cancer cells simply increase glycolysis to achieve higher mitotic rates.

Second, anoxic cultures to which oxygen is rapidly applied were assessed. A number of changes occur soon after such transitions. The cell borders become less smooth, the proportion of macrophage-like cells (larger than 300 μm²) increases steadily with 0%, 2%, 5%, 10%, and 21% oxygen, and so does the number of necrotic cell bodies (not shown).

Figure 4 shows an especially robust phenomenon, the appearance of the "Roundness 2" cell sub-population in an anoxic culture suddenly challenged by 21% oxygen. The figure documents the change of cell *shapes* over time. The near edge of the figure is a stable anoxic culture, and at time zero, the anoxic gas of the culture is replaced by atmoxic gas. Low *Roundness* cells, at the extreme left, are the most numerous at any time. But within as little as 4 h, the "Roundness 2" population appears, solidifies in the 40 h documented by the graph, and is sustained indefinitely, displaying the initial stages of an inflammation response.

Third, we investigated K562 self-adhesion in quadruple simultaneous experiments, where anoxic and atmoxic aliquots are simultaneously passaged into identical or opposite oxygen environments. In Fig. 5, four cultures are plotted against cell density, rather than time, to allow a fair comparison of cell clustering. Small values of *Hex-distance* correspond to tighter packing of cells. Newly seeded cells are randomly dispersed, showing Hex-distance values of 5 to 7 on the vertical axis, compatible with their seeding density.

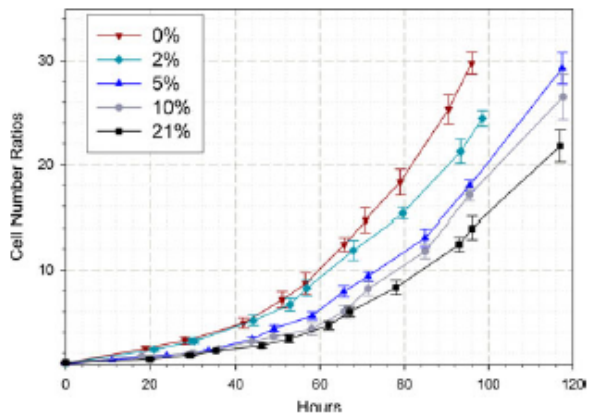

**Fig. 3** Cell numbers on K562 at various oxygen levels. *Bars* represent ±95% CI. Proliferation increased as oxygen diminishes from atmoxia (23-h doubling time) to anoxia (15.8-h doubling time)

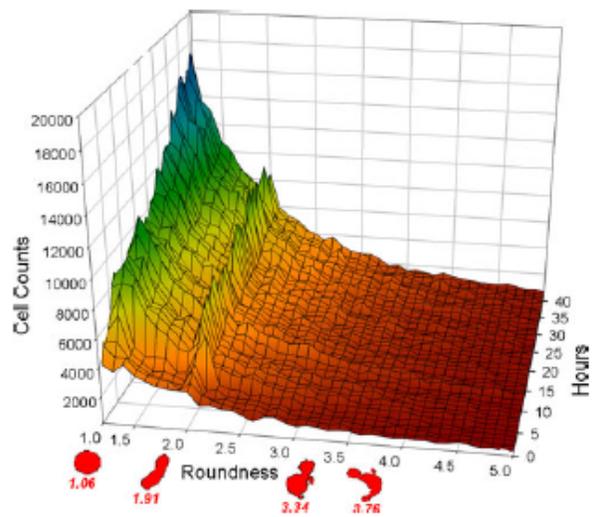

**Fig. 4** Oxygen produces inflammation in K562, illustrated by time-series histograms of cell shape, quantified as *Roundness*. The anoxic gas phase is replaced by atmoxic gas at time zero, triggering the appearance of the "Roundness 2" peak. The total cell number climbs from 22,000 to 46,000 from the near to the far edge of the graph

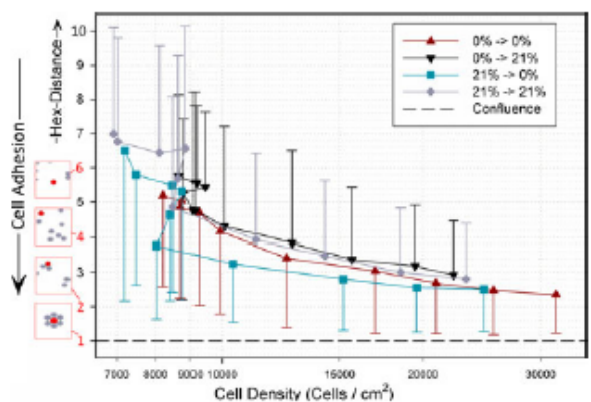

**Fig. 5** Self-adhesion of K562 under various oxic conditions. The vertical axis icons illustrate *Hex-distance*, the average distance between a cell and its six closest neighbors, expressed in cell diameters. Each data point is a distribution of 5,000 to 32,000 Hex-distance individual measurements, as averages±1σ, compiled at 1, 4, 7, 10, 18, 29, 40, 51, 62, and 73 h. Cyto-adhesion is strongest for anoxic and anoxic transition cells, while atmoxic cells are more easily shed

How Hex-distance values decay towards 1 (= confluence) reveals the level of self-adhesion.

To the right of the seeding phase ending at 10,000 cells/cm², we observe more adhesion in anoxic than atmoxic culture. But the strongest cell adhesion is in cells passaged from atmoxic to anoxic medium. Lowered adhesion is relevant to metastasis and increased adhesion to formation of tumors. Here, we show that cells can lose and regain self-adhesion, according to oxic conditions and suggesting that cells on the more oxygenated periphery of a tumor are more easily shed.

The three phenotype measurements above converge to present a picture of anoxic K562 as a more dynamic (increased proliferation), stable (less inflammation), and cohesive (more self-adhesion) tissue. The picture suggested by these measurements is that bringing the unregulated metabolism of cancer cells under control by metabolic restriction not only produces fewer pathological chromosomes, but that cancer cells also increase their proliferation rate and acquire the phenotype of a stable, growing tissue.

## Discussion

Researchers have been aware of the intimate relation between cancer and metabolism for a long time [19]. The identification of specific cancer mutations and the success of imatinib therapy have inspired on-going work both on the metabolic consequences of cancer lesions [46, 47] and on cancer therapies based on metabolic control [48–50]. Studies have also been performed on the evolution over time of cancer metabolism [51, 52], and even on reversing cancer phenotypes and genetic instability using antioxidants and nitric oxide inhibitors [53].

It is well accepted that cancer karyotypes can evolve over time and be influenced by culture conditions [54–56]. Even normal human embryonic stem cells, often incubated under low-oxygen conditions, are vulnerable to karyotype instability, and it is recommended that chromosomal status be checked at intervals of five passages to monitor possible translocations and aneuploidies [57]. But to our knowledge, there has been no previous work on devolution of tumor karyotypes using metabolic restriction.

### Endo-reduplication, fast karyotype changes, and tumor evolution

The traditional view of tumor evolution is a slow clonal expansion based on genetic instabilities arising from the initial cancer lesion [1]. END, specifically, is only acknowledged in the literature in a modest role [58–62]. But our in vitro experiments suggest that cells can use END to rapidly alter karyotypes and phenotypes when metabolic resources are available. The rapid adaptations of cells to changes in their micro-environment that we observed may be an important mechanism to support the amazingly rapid replacement of tumor cells with variants that are resistant to chemotherapy and the immune system [63]. It is acknowledged, for example, that resistance to imatinib arises partly from amplifications and chromosome duplications in chromosomally unstable CML cells [42]. In our view, END should be recognized as a dormant hyperploidy reflex available to cancer cells in tumors, as suggested by Larizza and Schirmacher [64].

CIN-END are powerful mechanisms to accelerate genetic evolution beyond random mutations. Tumor evolution may be a model of evolution itself. The manifestation of END as a tumor stress response implies that stuttering is not so onerous to biological systems. It appears that redundancies at the level of the gene and chromosome are affordable, creating opportunities for variation. Gains or losses in one chromosome steps are more likely to be compatible with cell viability than the larger chromosome number changes associated with mitotic apparatus defects [65]. Darwin described natural selection driven by random mutations. But the deliberate expansion of the genome by END mechanisms may be a conserved trait that increases competitiveness of the biota as a whole. CIN-END may have played the role of evolutionary autopilots, driving the expansion of threatened code.

### Metabolic restriction promotes cancer

Cancer lesions enhance cell metabolism, allowing faster growth. This growth drives down oxygen concentrations,

but cancer cells can readily accommodate this reduction because of their flexible metabolism. Enhanced glycolysis favors the evolution of a group of hypoxic cancer cells into a tumor core. Metabolic restriction further allows the cells to dispense with the detoxification mechanisms associated with oxygen exposure, such as glutathione-$S$-transferase and CYP3A4 expression [66], and to concentrate on bio-synthesis. The smaller karyotypes maintained under metabolic restriction contribute to tumor core expansion, as fewer chromosomes are more rapidly duplicated.

Our data show that hypoxia is supportive of the development of tumor core structure by enhancing proliferation and self-adhesion and reducing inflammation. We observed as well (not shown) smoother cell borders, smaller cell footprints, and reduced macrophage numbers of increased activity in anoxic K562. Metabolic restriction seems to favor survival of the tumor core. As the tumor core slows in its development towards anoxic stability, it is inevitably surrounded by a more oxygenated and perfused tumoral cortex. This configuration is reminiscent of the niches [67] of germinal lines and stem cells [4].

The more highly oxygenated and perfused cortex is less stable [68, 69], but is ground for the expression of END. Increased chromosome numbers in the tumor cortex provide a karyotype and phenotype variation front. The chromosomes added by the END interface with normal tissues by diluting the initial cancer lesion, slowing down cell division, and increasing detoxification metabolism, similar to what is displayed in hyperploid liver cells. The cortex will also shed metastatic cells, which will ultimately revert to hypoxic states as they proliferate.

In our view, tumors stabilize their core with metabolic restriction, while their cortex diversifies karyotypes and sets the stage for metastasis. Therefore, standard atmoxic cell cultures in vitro may be more relevant to metastasis studies, while anoxic, hypoxic, and restricted metabolic states are more desirable for tumor eradication studies.

It has been repeatedly confirmed that cancer cells become more malignant under hypoxia [67, 70, 71] in vitro [3, 5] and in the clinic [72, 73], to the point where it has become a central issue in tumor physiology and treatment [1]. Since our data tie the metabolic restriction of cells to karyotype contraction, it is logical to conclude that karyotype contraction is an indicator of meta-genetic cancer promotion.

Antioxidants as cancer promoters?

Free radicals and reactive oxygen species can randomly destroy molecules. An apparent corollary is that antioxidants protect biological materials, attenuating the carcinogenic action of toxicants and of normal metabolism [74]. Unfortunately, epidemiological evidence following administration of many antioxidants does not show the expected benefits and may point more to cancer rate increases [75]. Two large epidemiological studies found negative impacts of vitamin A on lung cancer rates in smokers [76, 77]. The results could not be explained, as vitamin A at low doses has no known toxicity. When 25 μM vitamin A, as retinyl acetate, was applied to lung cancer cells (NCI-H460), they lost seven chromosomes (from 57) after 1 day; K562 (2.5 μM) lost 13 chromosomes (from 69) after 1 day.

Karyotype contraction provides a possible explanation for these epidemiological observations. While antioxidants may reduce general cellular damage by controlling free radicals, their ability to contract the karyotypes of *existing cancers* may ultimately *increase* malignancy. This mechanism would be particularly relevant to the penetrating antioxidants able to reach tumor cores. At the metabolic level, antioxidants reduce oxygen use but enhance glycolysis, a change that cancer cells are well equipped to handle.

Since these initial epidemiological studies, a number of other investigators have confirmed the paradoxical result that antioxidants favor rather than suppress cancer [76]. Our in vitro data support the idea that antioxidants may increase cancer rates by inducing karyotype contraction in pre-existing cancer cells.

Conclusions

Metabolic restriction and stimulation contract and expand the chromosome numbers of hyperploid cancer cell lines. Metabolic restriction, even in cancer cells not displaying chromosome number changes, may signal increased malignancy. Karyotype contraction caused by metabolic restrictors may be an indicator of meta-genetic cancer promotion. According to our models, lipid-soluble antioxidants as a means of reducing cancer incidence, particularly in older individuals more likely to harbor tumors, should be viewed with caution.

**Acknowledgments**   We are grateful to Janet Moir and Lorne Beckman of the Royal Victoria Hospital. We are indebted to Josée Hébert, Sylvie Lavallée, and Claude Rondeau, Immunology and Cancer Research Institute of Maisonneuve-Rosemont Hospital. We thank Drs. Zhang Guo Chen and Ju Yan for reviewing the paper. The work was supported by Royal Victoria Hospital Research Institute Fund 65891.

**Conflicts of interest**   None.